\documentclass[twocolumn,preprintnumbers,prd,longbibliography,nofootinbib]{revtex4-1}

\usepackage{graphicx}
\usepackage{dcolumn}
\usepackage{bm}
 \usepackage{amstext}
 \usepackage{amssymb}
 \usepackage{amsmath}
 \usepackage{graphicx}
 \usepackage{color}
 \usepackage{delimset} 

\usepackage[usenames,dvipsnames]{xcolor}
\xdefinecolor{refcolor}{rgb}{0.5,0,0}
\usepackage[colorlinks=true, filecolor=refcolor, citecolor=refcolor, linkcolor=refcolor, urlcolor=refcolor, menucolor=refcolor,
]{hyperref}

\makeatletter
\newlength{\apb@width}
\newcommand{\autoparbox}[2][c]{\settowidth{\apb@width}{#2}\parbox[#1]{\apb@width}{#2}}
\newcommand{\includegraphicsbox}[2][]{\autoparbox{\includegraphics[#1]{#2}}}
\makeatother

\makeatletter
\def\mr@ignsp#1 {\ifx\:#1\@empty\else #1\expandafter\mr@ignsp\fi}%
\newcommand{\multiref}[1]{\begingroup
\xdef\mr@no@sparg{\expandafter\mr@ignsp#1 \: }%
\def\mr@comma{}%
\@for\mr@refs:=\mr@no@sparg\do{\mr@comma\def\mr@comma{,}\ref{\mr@refs}}%
\endgroup}
\makeatother

\newcommand{\secref}[1]{Section~\multiref{#1}}
\newcommand{\Secref}[1]{section~\multiref{#1}}
\newcommand{\Appref}[1]{appendix~\multiref{#1}}

\renewcommand{\eqref}[1]{(\multiref{#1})}

\newcommand{\sfrac}[2]{{\textstyle\frac{#1}{#2}}}
\newcommand{\half}{\sfrac{1}{2}}

\newcommand{\atopfrac}[2]{{{#1}\above0pt{#2}}}

\newcommand{\be}{\begin{equation}}
\newcommand{\ee}{\end{equation}}
\newcommand{\ba}{\begin{align}}
\newcommand{\ea}{\end{align}}
\newcommand{\eqn}[1]{(\ref{#1})}

\newcommand{\gen}[1]{\mathrm{#1}}
\newcommand{\levo}[1]{ \gen{\widehat #1}}
\newcommand{\order}[1]{\mathcal{O}(#1)}
\newcommand{\mbf}[1]{\mathbf{#1}}

\newcommand{\dd}{\mathrm{d}}

\newcommand{\Eval}{s} 
\allowdisplaybreaks

\usepackage{feynmp-auto}
\usepackage{feynmp} 
\begin{document}

\preprint{
HU-EP-20/44,
SAGEX-20-30-E
}

\title{Three-Body Effective Potential in General Relativity at Second \\
Post-Minkowskian Order and Resulting Post-Newtonian Contributions}
\author{Florian Loebbert}
\email{florian.loebbert@physik.hu-berlin.de} 
\author{Jan Plefka} 
\email{jan.plefka@hu-berlin.de}

\author{Canxin Shi} 
\email{canxin.shi@physik.hu-berlin.de}

\author{Tianheng Wang} 
\email{tianheng.wang@physik.hu-berlin.de}

\affiliation{%
Institut f\"ur Physik und IRIS Adlershof, Humboldt-Universi\"at zu Berlin,
Zum Gro{\ss}en Windkanal 6, 12489 Berlin, Germany
}



\date{\today}

\begin{abstract}
We study the Post-Minkowskian (PM) and Post-Newtonian (PN) expansions of the gravitational three-body effective potential. At order 2PM a formal result is given in terms of a differential operator acting on the maximal generalized cut of the one-loop triangle integral. We compute the integral in all kinematic regions and show that the leading terms in the PN expansion are reproduced. 
We then perform the PN expansion unambiguously at the level of the integrand. Finding agreement with the 2PN three-body potential after integration, we explicitly present new $G^2v^4$-contributions at order 3PN and outline the generalization to $G^2v^{2n}$. The integrals that represent the essential input for these results are obtained by applying the recent Yangian bootstrap directly to their $\epsilon$-expansion around three dimensions. The coordinate space Yangian generator that we employ to obtain these integrals can be understood as a special conformal symmetry in a dual momentum space. 
\end{abstract}

                              
\maketitle

\section{Introduction}
\label{sec:intro}

The three-body problem in Newtonian gravity has been a source of inspiration in mathematics
and physics since the time of Newton himself. Families of special solutions are known and 
tied to names such as Euler, Lagrange and Poincar\'e \cite{Poincare:1890}. This
system of non-integrable differential equations poses a challenge to the theory of non-linear systems and numerical approaches to date. They are of clear importance for celestial
mechanics and space-flight, and 
have even been inspirational for science fiction \cite{liu}. In general relativity the problem is more challenging,
as there are now genuine \emph{$N$-body interactions} going beyond the Newtonian 2-body potential. As observations
indicate that many galaxies, including our own, contain supermassive black holes in their core, 
these $N$-body
interactions might be important for the dynamics of multiple-star systems in their vicinity \cite{Will:2013cza}. With the advent of gravitational wave astronomy \cite{Abbott:2016blz,TheLIGOScientific:2017qsa,LIGOScientific:2018mvr} the gravitational
radiation emitted by mergers of compact \emph{binaries} is now observable. It
is an interesting question whether genuine three-body systems, such as hierarchical systems where a
black-hole binary is traversed by a third lighter compact object, will be observable in
the future as well~\cite{Asada:2009qs,Galaviz:2010te,Meiron:2016ipr,Bonetti:2017hnb,Lim:2020cvm}.

In the non-relativistic (post-Newtonian) limit of general relativity the leading three-body interactions are due to Einstein, Infeld and Hoffmann \cite{Einstein:1938yz,eddington1938problem} and arise from the effective potential
\cite{Landau:1987gn}
\be\label{start}
V_{\text{3-body}}^{1\text{PN}} = -  \sum_{i=1}^3 \sum_{\atopfrac{j=1}{j\neq i}}^3 \sum_{\atopfrac{k=1}{k\neq i}}^3 \frac{G^{2}}{2} \frac{m_{i}m_{j}m_{k}}{r_{ij}r_{ik}}\, ,
\ee
with $r_{ij}$ denoting the spatial distance of the two massive bodies $m_{i}$ and $m_{j}$,
$G$ is Newton's constant and we set $c=1$. In the nomenclature of the two-body problem this is
the first post-Newtonian (1PN) contribution to the effective potential 
in which the velocity squared $v^{2}$ and the coupling term $Gm/r$ are of the same order due to the virial theorem. The velocity dependent 1PN terms contributing to the potential beyond \eqn{start} are pure two-body interactions of order $Gv^{2}$.
Numerical simulations of the relativistic three-body problem to date have mostly incorporated
general relativity by restricting to the pure two-body PN terms to various orders
\cite{Asada:2009qs,Gultekin:2005fd,Iwasawa:2005zh,Hoffman:2006iq,Gupta:2019unn}, as
the three-body interactions \eqn{start} and beyond are computationally costly, yet relevant
\cite{Will:2013cza,Galaviz:2010te}.
In fact, a number of numerical studies incorporating the three-body interactions up to 
the presently known 2.5PN order exist
\cite{Lousto:2007ji,Galaviz:2010te,Galaviz:2011qb,Naoz:2012bx,Bonetti:2016eif}
demonstrating their relevance for the dynamics.
Simulations of three black holes in full numerical relativity 
\cite{Campanelli:2007ea,Lousto:2007rj,Galaviz:2010mx} are challenging.

The two-body conservative potential for spin-less
compact binaries is known up to 4PN level for the potential 
\cite{Damour:2014jta,Bernard:2015njp,Damour:2016abl,Bernard:2016wrg,Foffa:2016rgu,Porto:2017dgs,Marchand:2017pir,Damour:2017ced,Foffa:2019rdf,Foffa:2019yfl,Blumlein:2020pog,Galley:2015kus}, including parts 
of 5PN \cite{Foffa:2019hrb,Blumlein:2019zku,Bini:2019nra} and 6PN \cite{Blumlein:2020znm,Cheung:2020gyp,Bini:2020nsb,Bini:2020wpo,Bini:2020hmy,Bini:2020uiq}.
The situation for the $N$-body problem is considerably more open. For three bodies the effective potential is known to 2PN order
\cite{Ohta:1974pq,Damour:1985mt,Schafer:1987,Lousto:2007ji,Galaviz:2010te}, generalizing \eqn{start}
by three-body terms of order $G^{3}m^{4}/r^{3}$ as well as  $v^{2}G^{2}m^{3}/r^{2}$
which entered the abovementioned numerical studies \cite{Lousto:2007ji,Galaviz:2010te,Galaviz:2011qb,Bonetti:2016eif}.
The complexity of the three-body 2PN potential already increases considerably, cp.~eq.~\eqn{eq:2PNfinal}.
For $N\ge4$ the effective potential is in fact \emph{unknown} at 2PN in an analytical form due to an unsolved two-loop spatial integral. The \emph{unintegrated}  $N$-body conservative potential at 2PN was presented in
\cite{Chu:2008xm}. 

Turning to the weak gravitational field but arbitrary velocity limit -- known as the  post-Minkowskian (PM) limit, where one only expands in Newton's coupling but
leaves the velocity inert -- a lot of progress has been made on the two-body problem recently.  
Using
methods of scattering amplitudes for perturbative quantum gravity, 
the 2PM \cite{Cheung:2018wkq,Cristofoli:2019neg} and 3PM \cite{Bern:2019nnu,Bern:2019crd,Cheung:2020gyp}
(including radiation reaction effects \cite{Damour:2020tta, DiVecchia:2020ymx}) results for the effective potential
 have been established. A worldline effective field theory formalism for the PM expansion was recently formulated
\cite{Kalin:2020mvi} and has now been successfully applied to order 3PM 
\cite{Kalin:2020fhe}. Earlier worldline-based PM calculations can be found in refs.~\cite{Westpfahl:1985tsl,Bel:1981be,Ledvinka:2008tk,Damour:2016gwp,Blanchet:2018yvb} for the conservative sector.  
The relation between the world-line quantum field theory and the scattering amplitude approach was recently 
clarified in \cite{Mogull:2020sak}.
Despite this progress, for the $N$-body problem nothing is known beyond 1PM order at which there are no genuine higher body interactions \cite{Ledvinka:2008tk}.

It is the aim of this paper to improve on this
and to construct the 2PM effective potential in the three-body case (the essential 2PM formulae straightforwardly generalize to $N$ bodies). This in turn may be employed 
to determine all the velocity dependent contributions to the potential at order $G^{2}$ in the post-Newtonian expansion, i.e.~the terms of order $v^{2n}G^{2}m^{3}/r^{2}$. As the complexity of these contributions grows dramatically
we shall explicitly provide only the  so far unknown $v^{4}G^{2}m^{3}/r^{2}$ terms which contribute to the 3PN
terms in the potential in \Secref{sec:3PN}. The general tools to determine the higher velocity terms 
will be provided.

We employ the PM worldline effective quantum field theory formalism based on \cite{Kalin:2020mvi,Mogull:2020sak}, generalizing the non-relativistic (PN) effective field theory approach  of \cite{Goldberger:2004jt} to general relativity. The three-body 2PM potential essentially follows from a single Feynman diagram
connecting the three-graviton vertices with the world-lines resulting in a one-loop three-point integral with
coordinate space Green's functions~\cite{Westpfahl:1985tsl}. 
This integral features a Yangian level-one symmetry \cite{Chicherin:2017cns,Loebbert:2020hxk} and is related to a generalized cut of the four-point box integral, which has recently been obtained from Yangian bootstrap \cite{Loebbert:2019vcj,Corcoran:2020epz}. Generalizing the calculation of \cite{Westpfahl:1985tsl}, we explicitly show that our three-point integral is indeed proportional to one of the four Yangian invariants found in \cite{Loebbert:2019vcj}. We then demonstrate that the PN expansion is most efficiently performed at the integrand level, which results in a family of three-point integrals in three dimensions with half integer propagator powers. Again, this family of divergent integrals is invariant under a Yangian level-one generator, which allows to bootstrap their expansion in the dimensional regularization parameter $\epsilon$. This level-one symmetry can alternatively be interpreted as a special conformal symmetry in a dual momentum space, cf.\ \cite{Coriano:2013jba,Bzowski:2013sza}, and \cite{Loebbert:2020glj} for the connection between the two symmetries.
We explicitly perform
the PN expansion to NNLO yielding the previously unknown $v^{4}G^{2}m^{3}/r^{2}$ terms at the 3PN level and illustrate the generalization to  $v^{2n}G^{2}m^{3}/r^{2}$.

This paper is organized as follows:  after a general discussion of the worldline effective field theory in the Polyakov
formulation in \Secref{sec:EffField} we construct the 2PM potential in \Secref{sec:2PMPot}. The computation of the emerging three-point key integral in various kinematical regions is relegated to \Appref{app:3delta}.
\secref{sec:1PNExp} discusses the 1PN limit of the 2PM
potential recovering the Einstein--Infeld--Hofmann Lagrangian. In \Secref{sec:PNandBootstrap} we lay out our general approach to integrate
the 2PM potential in the non-relativistic PN expansion at the level of the integrand making use of a level-one Yangian symmetry
for the emerging master integrals. As concrete applications of this procedure we then recover the known 2PN three-body potential up to the
static term in \Secref{sec:2PN}, and in \Secref{sec:3PN} provide all three-body terms at the 3PN order that scale quartically in velocities and show that they reproduce the known results in the two-body limit. 
%


\section{Effective field theory}
\label{sec:EffField}

Consider three massive spinless point particles coupled to Einstein gravity via the action
\begin{equation}
	S = S_{\mathrm{EH}}+  S_{\mathrm{gf}}+ S_{\mathrm{pp}} .
\end{equation}
Here we have defined
\begin{align}
	S_{\mathrm{EH}}=&-\frac{2}{\kappa^{2}} \int d^{4} x \sqrt{-g} R +(\text{GHY term})  \\
	=&-\frac{2}{\kappa^{2}} \int d^{4} x \sqrt{-g}\left[g^{\mu \nu}\left(\Gamma^{\rho}_{\mu \lambda} \Gamma^{\lambda}_{\nu \rho} -\Gamma^{\rho}_{\mu \nu} \Gamma^{\lambda}_{\rho \lambda}\right) \right], \nonumber
\end{align}
with $\kappa^{2}= 32\pi G$, the gravitational coupling and a Gibbons--Hawking--York (GHY) boundary term \cite{York:1972sj,Gibbons:1976ue}. For the point particles we start out with the action
\begin{align}
	S'_{\mathrm{pp}}=&-\sum_{i} m_i \int d \tau_i \, \sqrt{g_{\mu \nu}(x(\tau_{i})) 
	u_i^{\mu}(\tau_{i}) u_i^{\nu}(\tau_{i})}
	\label{eq:Spp}\, ,
\end{align}
with the 4-velocities $u^{\mu}_{i}=\dd x^{\mu}_{i}/\dd\tau_{i}$ integrated along their world-lines. It turns out to be more
advantageous to work with the Polyakov formulation of the point-particle action.
Upon introducing the einbein $e_{i}=e(x(\tau_{i}))$ this action reads
\be \label{eq:Polyakov}
S_{\mathrm{pp}}= - \sum_{i=1}^{3} \frac{m_i}{2} \int d \tau_i e_{i}\left (\, g_{\mu \nu} 
	u_i^{\mu}(\tau_{i}) u_i^{\nu}(\tau_{i}) + \frac{1}{e_{i}^{2}}\right )\, ,
\ee
which preserves reparametrization invariance by the transformation rule for $e_i$.
Solving the algebraic equations of motion for the inverse einbein yields $e_{i}^{-1}=  \sqrt{g_{\mu \nu} 
u_i^{\mu}(\tau_{i}) u_i^{\nu}(\tau_{i})}$ and plugging this back into the action
recovers the original action \eqn{eq:Spp}.

In the weak field expansion of the metric we take $g_{\mu\nu} = \eta_{\mu\nu} + \kappa\,h_{\mu\nu}$, using the mostly minus convention.
We choose the standard de~Donder gauge fixing term $S_{\mathrm{gf}}= \int d^{4} x f_\mu f^\mu$ with
$
	f^{\mu}=
	\partial^{\nu}h^{\mu }{}_{\nu} 
	-  \tfrac{1}{2} \partial^{\mu }h^{\nu}{}_{\nu}
$. 
This yields the graviton Feynman propagator
\begin{equation}
\includegraphicsbox{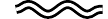} = \frac{i}{2}\frac{P^{\mu\nu\rho\sigma}}{k^2+i\varepsilon},
\end{equation}
with
$
P^{\mu\nu\rho\sigma} = \eta^{\mu\rho} \eta^{\nu\sigma} + \eta^{\mu\sigma} \eta^{\nu\rho} - \eta^{\mu\nu} \eta^{\rho\sigma}
$.
The advantage of the Polyakov formulation \eqn{eq:Polyakov} is that it only gives rise to a single graviton
worldline interaction:
\begin{equation}  \label{FeynmanRules}
\includegraphicsbox{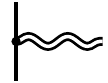} 
 = -i\kappa~ e(\tau) u^\mu(\tau) u^\nu(\tau).
\end{equation}
In the bulk we will only need the three-graviton vertex, which may be found e.g.~in \cite{Sannan:1986tz}.
An important aspect in the construction of the classical effective
action is the causality structure of the propagator as was recently stressed in
\cite{Damour:2020tta}. The Fourier transform to coordinate space
of the graviton Feynman propagator
 reads ($x_{ij}^\mu=x_i^\mu-x_j^\mu$)
 \begin{align}\label{eq:propdef}
\bar D_{ij}
&
=\int \frac{\dd^4 k}{(2\pi)^4} \frac{1}{k^2+i\varepsilon} e^{i k \cdot x_{ij}}
\nonumber\\
&=\frac{1}{4\pi^2}\frac{i}{x_{ij}^2-i \varepsilon}=-\frac{1}{4\pi}\delta(x_{ij}^2)+\frac{i}{ 4\pi^2 x_{ij}^2},
\end{align}
where the famous distributional identity
\be
\lim_{\epsilon\to 0^{+}} \frac{1}{y\pm i\epsilon} = \text{pv}\,\frac{1}{y} \mp i\pi \delta(y)
\ee
was used in the last step, thereby dropping the principal value label.
 In order to construct the classical, conservative action for PM gravity, one should  restrict to the \emph{real} part $D_{ij}$ defined as:
\begin{equation}
D_{ij}=\text{Re} (\bar D_{ij})=-\frac{1}{4\pi}\delta(x_{ij}^2).
\label{eq:DefPropagator}
\end{equation}
This propagator obeys the Green's function identity
\begin{equation}
	\square D_{ij}=-\delta^{(4)}(x_{ij}).
	\label{eq:GreensPropagator}
\end{equation} 
It may also be expressed as
\begin{equation}
	\delta(x^2) = 
	 \frac{\delta(ct- r)}{2  r} + \frac{\delta(ct+ r)}{2  r},
\end{equation}
where $ r = \abs{\bf{x}}$, making manifest that the sum of the retarded and advanced propagator, i.e.~the time symmetric propagator, is the real part of the Feynman propagator. The conservative effective action $S_{\mathrm{eff}}$ may then be obtained upon integrating out the graviton fluctuations.


\section{2PM potential}
\label{sec:2PMPot}
Up to order 2PM the effective action is expanded as
\begin{equation} \label{eq:Seff}
	S_{\mathrm{eff}} = S^{\mathrm{free}} + \kappa^2 S^{\mathrm{1PM}} + \kappa^4 S^{\mathrm{2PM}} 
	+\mathcal{O}(\kappa^6),
\end{equation}
where the free contribution takes the form of \eqref{eq:Polyakov} with $g_{\mu\nu}$ replaced by $\eta_{\mu\nu}$.
Using the Feynman rule \eqref{FeynmanRules}, it is straightforward to compute the 1PM order. It follows from a single graviton exchange between each pair of point masses 
\begin{align}
	\kappa^2 S^{\mathrm{1PM}} &= \sum_{i}\sum_{j\neq i}\raisebox{-.2cm}{\includegraphicsbox[scale=0.8]{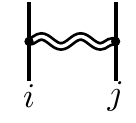}}	 \\
	&= \sum_{i} \sum_{j\neq i} \int {\dd\hat{\tau}_i \dd\hat{\tau}_j}\frac{ \kappa^2 m_i m_j} {32 \pi}
	\brk[s]*{ u_{ij}^2 \!-\! \half u_i^2 u_j^2 } \delta(\!x_{ij}^2\!)\,,\nonumber
\end{align}
with $d\hat{\tau}_i := e_i d \tau_i$, and 
$u_{jk} :=u_j\cdot u_k$. There are no three-body interactions
at this order.
Moving on to 2PM, we find the first genuine three-body interaction in the theory, which arises from a single Feynman diagram.
In coordinate space it reads 
\begin{align}\label{3body}
	\includegraphicsbox[scale=1]{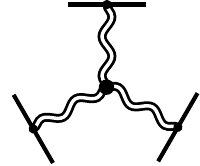} =& \kappa^{4} \int \frac{\dd^3\hat{\tau} }{(4\pi)^3} P(x_1(\tau_1), x_2(\tau_2), x_3(\tau_3) ),
\end{align}
where we have defined $\dd^3\hat{\tau} = \dd\hat{\tau}_1\dd\hat{\tau}_2 \dd \hat{\tau}_3$ as well as
\begin{align}	
	&{8}{(m_1 m_2 m_3)^{-1}}P(x_{i}(\tau_{i})):=\nonumber\\	
	& \pi \big(
      4 u_{12}^2 u_{3}^2 
	- 4 u_{12}^{} u_{13}^{} u_{23}^{}
	- u_{1}^2 u_{2}^2 u_{3}^2 
	\big) \delta(x_{12}^2) \delta(x_{13}^2)
	\nonumber\\
	&+  \big(
      u_{12}^2 u_{3}^{\mu}  u_{3}^{\nu}  
	- \tfrac{1}{2} u_{1}^2 u_{2}^2 u_{3}^{\mu}  u_{3}^{\nu}
    + 2 u_{13}	u_{23}u_{2}^{\mu}  u_{1}^{\nu} \big)
	\partial_{x_1,\mu}\partial_{x_2,\nu} I_{3\delta} 
	\nonumber\\
	&+( \text{cyclic} ),
	\label{eq:PPP}
\end{align}
with the integral $I_{3 \delta}$ further discussed below.
Note that we have discarded all terms proportional to $u_i\cdot \partial_{x_{i}}$, which can be written as derivatives ${\dd}/{\dd\tau_i}$. 
Modulo integration by parts, the $\tau_i$-derivatives act on the $u_i$ and the einbein $e_j$, which also solely depends on the $u_j$ after solving the equations of motion, see \eqref{eq:einbein}. In the end, the $\tau_i$-derivatives yield terms involving accelerations, which, by proper field redefinition of $x_i$, can be replaced by lower order equations of motion that lift these terms to the next order in $\kappa^2$, cf.\ \cite{Damour:1990jh}.
Note that we will employ this mechanism at several points of the paper.

To complete the three-body action we need to include the two-body interactions at 2PM. These can be obtained from \eqref{3body} by identifying two of the three world-lines and multiplying with a symmetry factor ${1}/{2}$. 
The full 2PM three-body action thus becomes 
\begin{equation}
\label{eq:S2PMgen}
	S^{\mathrm{2PM}} = \frac{1}{6} \int \frac{\dd^3\hat{\tau} }{(4\pi)^3} \sum_{i,j,k}^{}\! {}^\prime P(x_i(\tau_1), x_j(\tau_2), x_k(\tau_3) ),
\end{equation}
where the sum ${\sum^\prime_{i,j,k}}$ runs over $i,j,k=1,2,3$ but excludes $i=j=k$. 
Moreover, propagators that have both ends on the same worldline vanish in dimensional regularization. 
We note that in fact this becomes the $N$-body 2PM action if we allow $i,j,k$ to run from $1$ to~$N$.

The central ingredient in the above formula \eqref{eq:PPP} for the three-body contribution to the effective potential is the integral
\begin{equation}
I_{3 \delta} :=\int \dd^{4} x_{0} \,\delta(x_{01}^2)\delta(x_{02}^2)\delta(x_{03}^2)
=
\includegraphicsbox{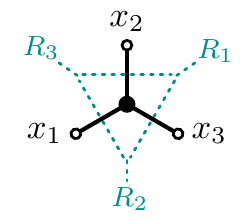},
\label{eq:I3delta}
\end{equation}
which is interesting for various reasons. In the present paper it arises as the one-loop three-point integral in coordinate space (black solid diagram). Alternatively, we can interpret it as the generalized maximal cut of the momentum space triangle integral  where all propagators are put on shell (green dashed diagram), expressed in terms of region momenta $x_j$, which map to the dual momenta $R_j$ via 
\begin{equation}
R_j^\mu :=x_{j+1}^\mu-x_{j-1}^\mu.
\label{eq:DualMomenta}
\end{equation}
Moreover, $I_{3\delta}$ is related to a generalized cut of the four-point (box) integral, in the limit where one point is sent to infinity. The box integral is invariant under a Yangian algebra, an extension of its well known conformal symmetry \cite{Chicherin:2017cns,Chicherin:2017frs}.
As such, in the region $R_j^2<0$ the integral is given by the minimal transcendentally solution of the Yangian constraints found in~\cite{Loebbert:2019vcj} (modulo a piecewise constant):
\begin{equation}
I_{3 \delta} = \frac{C}{\sigma},
\qquad\quad
\sigma^2 :={(R_2 \cdot R_3)^2 - R_2^2 R_3^2}.
\label{eq:I3deltaC}
\end{equation}
Note that due to $R_1+R_2+R_3=0$ this representation is not unique and one may pick any two $R_{i}$'s to define~$\sigma^{2}$.

To obtain $I_{3\delta}$, it is useful to generalize the steps of Westpfahl \cite{Westpfahl:1985tsl}, who evaluated the integral for the retarded propagator. This generalization performed in  \Appref{app:3delta} shows that the value of the integral depends on the sign of $\sigma^2$.
In fact, for $R_j^2<0$ with $j=1,2,3$ the expression \eqref{eq:I3deltaC} can be compared with the result of \cite{Westpfahl:1985tsl} which shows that $C({\sigma^2>0},{R_j^2<0})=\pi/4$ in the above expression.
However, more care is needed to obtain $C$ for generic kinematics. 
The explicit calculation given in \Appref{app:3delta} shows that for $\sigma^2>0$ we have
\begin{equation}
I_{3\delta} = 
	\frac{\pi}{4 \sigma} \Theta (-R_1^2 R_2^2 R_3^2 ).
	\label{eq:I3deltaresultsigmagreater}
\end{equation}
Here $\Theta$ denotes the Heaviside-function as defined in \eqref{eq:DefHeaviside}.
For $\sigma^2<0$ the integral diverges and for $\sigma^2=0$ it is proportional to $\sum_i \delta(R_j^2)$, see \Appref{app:3delta}.


\section{The 1PN Expansion}
\label{sec:1PNExp}

In this section we want to provide a first test of the above expression for the full 2PM effective action against known results for the three-body potential at 1PN order.
For this we first solve the equation of motion ${\delta S}/{\delta e_i} =0 $ for $e_i$ perturbatively up to order $\kappa^2$:
\begin{align}
    e_i = \frac{1}{ \sqrt{u_i^2} } 
    + \sum_{j \neq i} \!\int\! \dd\tau_j 
    \frac{\kappa^2 m_j}{16\pi \sqrt{u_i^6 u_j^2}}
    \left( u_{ij}^2 - \half u_i^2 u_j^2 \right) + \order{\kappa^4}.
    \label{eq:einbein}
\end{align}
Plugging this solution back into \eqref{eq:Seff} and expanding to order $\kappa^4$ yields the 2PM effective action free of the einbein. 
We then consider its non-relativistic limit,
 choosing the convenient gauge $\tau_i = t_i$. Reintroducing the speed of light $c$ such that
\begin{equation}
	u_i^\mu = \left( 1, \frac{\textbf{v}_i}{c} \right), \quad
	\frac{\partial}{\partial x_i^\mu} =\left( \frac{\partial }{c \partial t_i},  \frac{\partial }{\partial \textbf{x}_i} \right), 
    \quad \kappa \rightarrow \frac{\kappa}{ c },
\end{equation}
we see that in $P(x_{i})$ of \eqref{eq:PPP} only the second line  contributes
at leading order in $c^{-1}$:
\be \label{eq:Pleading}
\sum_{i,j,k}{}^\prime P(x_{i}) = - \frac{3 \pi m_{1}m_{2}m_{3}}{8}   \sum_{i}\! \sum_\atopfrac{j\neq i}{k\neq i}\delta(x_{ij}^2) \delta(x_{ik}^2) + \mathcal{O}(c^{-2}).
\ee
Note that we have rewritten the sum by discarding propagators with both ends on the same worldline.
Using the non-relativistic expansion of the propagator \eqn{eq:DefPropagator} 
\begin{align}
	\delta(x_{ij}^2) = &\frac{\delta(t_i- t_j)}{r_{ij}} 
            - \frac{ r_{ij} }{2 c^2} \partial_{t_i} \partial_{t_j} \delta(t_i - t_j) 
            \\
            &+ \frac{r_{ij}^3}{24 c^4} \partial_{t_i}^2 \partial_{t_j}^2 \delta(t_i - t_j)
             + \mathcal{O}(c^{-4}), \nonumber
\end{align}
where $ r_{ij} =\abs{\textbf{r}_{ij} } $ with $ \textbf{r}_{ij} = \textbf{x}_i - \textbf{x}_j $, yields a
localized time integration in the effective action \eqref{eq:S2PMgen}.
After some rearrangements, we find the 1PN three-body effective action\footnote{
Note that in the GR literature the PN action is typically rescaled by a factor of $c^2$.
}
\begin{align}
	S=& \sum_{i} \int dt \bigg[ \!-\! m_i + \frac{1}{c^2} \bigg( \frac{ m_i \textbf{v}_i^2 }{2} + \sum_{j\neq i} \frac{G m_i m_j}{2 r_{ij}}  \bigg)  \nonumber \\
	&+ \frac{1}{c^{4}} \bigg( \frac{ m_i \textbf{v}_i^4 }{8} + \sum_{j\neq i} \frac{G m_i m_j }{4 r_{ij}} \big( 6\, \textbf{v}_i^2 \!-\! (\textbf{n}_{ij} \!\cdot\! \textbf{v}_i) (\textbf{n}_{ij} \!\cdot\! \textbf{v}_j) \nonumber \\
    &\quad - 7\, \mathbf{v}_i\cdot \mathbf{v}_j  \big) -  \sum_{j\neq i} \sum_{k\neq i} \frac{G^2 m_i m_j m_k}{2 r_{ij} r_{ik} } \bigg) \bigg],
\end{align} 
where we abbreviate $\textbf{n}_{ij} :={\textbf{r}_{ij} }/{r_{ij} }$ and
$
G={\kappa^2}/{32\pi}.$
This result agrees with the well known 1PN expression \cite{Landau:1987gn}.


\section{Post-Newtonian Expansion and Integral Bootstrap}
\label{sec:PNandBootstrap}

The 1PN expansion obtained in the previous section merely tests the second line of the three-body contribution  \eqref{eq:PPP} to the effective potential. In order to obtain the expansion at 2PN order, also the third line in \eqref{eq:PPP} has to be taken into account. This includes second derivatives of the three-delta integral, $\partial_j^\mu\partial_k^\nu I_{3\delta}$, cf.\ the $\Theta$-function in \eqref{eq:I3deltaresultsigmagreater}. As outlined in detail in \Appref{sec:DerivativesI3delta}, taking these derivatives leads to lengthy expressions in terms of delta functions and their derivatives which are hard to control.
In fact, it is simpler to perform the non-relativistic expansion directly on the level of the integrand of $I_{3\delta}$ as we will demonstrate in the following.
For convenience of the reader we briefly summarize the below strategy: First, we will show that expanding the integrand of $I_{3\delta}$ leads to the family of key integrals given in \eqref{eq:intfamily}. We will then use the Yangian level-one symmetry of these integrals, i.e.\ invariance under the differential operator \eqref{eq:definitionPhat}, to obtain the differential equations \eqref{eq:PhatI3}. Finally, we explicitly demonstrate how these equations are solved in the form of \eqref{eq:ansatzI3}, which results in the expressions for the dimensionally regularized integrals that enter into the effective potential.
To start we consider the non-relativistic expansion of the propagator of \eqref{eq:propdef}, generalized to $D$ spatial dimensions in the so-called potential region $\omega:=k^{0} \ll |\textbf{k}|$,
writing
\be
\frac{1}{k^{2} }= \frac{1}{\omega^{2}-\textbf{k}^{2}}=-
\sum_{\alpha=1}^{\infty}\frac{\omega^{2\alpha-2}}{(\textbf{k}^{2})^{\alpha}}\, .
\ee
 Inserting this
expansion into the Fourier transformed  expression for the 
time-symmetric propagator yields the common PN-expanded propagator 
\begin{equation}
    \delta(x_{0i}^2)= {4\pi} \int \frac{\dd^D k}{(2\pi)^D} e^{i\textbf{k}\cdot\textbf{x}_{0i}} \sum_{\alpha=0}^\infty \frac{(-1)^{\alpha} \partial_{t_i}^{2\alpha}\delta(t_{0i})}
    {c^{2\alpha}(\textbf{k}^2)^{\alpha+1} } \, ,
    \label{eq:expansiondeltaprop}
\end{equation}
having performed the energy ($\omega$) integral.
Hence, with the expression for the $D$-dimensional Fourier transform of the momentum space propagator,
\begin{equation}
\int \frac{\dd^D k}{(2\pi)^D}\frac{e^{i\textbf{k}\cdot \textbf{x}}}{(\textbf{k}^2)^\alpha}
=
\frac{1}{4^\alpha \pi^{D/2}} \frac{\Gamma_{D/2-\alpha}}{\Gamma_\alpha}r^{2\alpha-D},
\end{equation}
we can write the key integral $I_{3\delta}$ in the PN-expansion for general spatial $D$ as
\begin{align}
&I_{3\delta}=
\sum_{\alpha,\beta,\gamma=0}^\infty 
\frac{(-1)^{\alpha+\beta+\gamma}}{(2c)^{2(\alpha+\beta+\gamma)} \pi^{3(D/2-1)}}
\frac{\Gamma_{\hat \alpha}\Gamma_{\hat \beta}\Gamma_{\hat \gamma}}{ \Gamma_{\alpha+1} \Gamma_{\beta+1} \Gamma_{\gamma+1} }
\nonumber\\
&
\times \int \dd t_0 
\partial_{t_1}^{2\alpha}\delta(t_{01})
\partial_{t_2}^{2\beta} \delta(t_{02})
\partial_{t_3}^{2\gamma}\delta(t_{03})
I_3^D[\hat \alpha,\hat \beta,\hat \gamma] .
\label{eq:PNexpansionI3delta}
\end{align}
Here, $\Gamma_\alpha=\Gamma(\alpha)$ denotes the Gamma-function, we use the shorthand
$\hat\alpha=D/2-\alpha -1 $ 
and we have introduced the following family of (Euclidean) integrals:
\begin{equation}
I_3^D[a_1,a_2,a_3] :=\int \frac{\dd^D\mathbf{x}_0 }{(\mathbf{x}_{01}^2)^{a_1}(\mathbf{x}_{02}^2)^{a_2}(\mathbf{x}_{03}^2)^{a_3}}.
\label{eq:intfamily}
\end{equation}
These integrals represent the central nontrivial input for the above expansion \eqref{eq:PNexpansionI3delta} and we will now discuss how to compute them. Notably, in \cite{Boos:1990rg} the integrals $I_3^D[a_1,a_2,a_3]$ for generic propagator powers $a_j$ and spacetime dimension $D$ have been expressed in terms of Appell hypergeometric functions $F_4$, which converge for small values of the effective ratio variables $r_{12}/r_{13}$ and $r_{23}/r_{13}$. In the present situation we would like to avoid making assumptions on these ratios, which would imply a limited validity of the resulting effective potential. Moreover, note that here we are merely interested in the special case of half integer propagator powers $a_j$ in three dimensions, which satisfy the condition
\begin{equation}
a_1+a_2+a_3\leq \sfrac{D}{2}.
\end{equation}
In particular, this condition implies that the integrals of interest are divergent in strictly three dimensions and we thus consider their $\epsilon$-expansion around $D=3$ in dimensional regularization. Importantly, these integrals are accessible via a bootstrap approach, cf.\ \cite{Loebbert:2019vcj,Loebbert:2020glj}: they feature a non-local Yangian level-one symmetry, i.e.\ they are annihilated by the differential operator
\begin{align}
\label{eq:definitionPhat}
\levo{P}^{\mu} := \frac{i}{2} \sum_{k=1}^3\sum_{j=1}^{k-1} 
	\left(\gen{P}_j^\mu \gen{D}_k \!+\! \gen{P}_{j\nu} \gen{L}_k^{\mu\nu} 
		\!-\! (j\leftrightarrow k)\right) 
	+ \sum_{j=1}^3\! \Eval_j \gen{P}_j^{\mu},
\end{align}
where we have used the following representation of the momentum, Lorentz and dilatation generator of the conformal algebra:
\begin{align}
\gen{P}^{\mu}_j &= -i\, \partial_{x_{j}}^{\mu}, 
\nonumber \\
\gen{L}_j^{\mu \nu} &= i x_j^{ \mu} \partial_{x_{j}}^{ \nu} 
	- ix^{ \nu}_j \partial_{x_{j}}^{\mu}, 
	\label{eq:levzeros} \\
\gen{D}_j &= -i x_{j  \mu} \partial_{x_j}^{\mu} - i. 
\nonumber
\end{align}
The so-called evaluation parameters $s_{j}$ entering the definition of the level-one generator $\levo{P}^\mu$ in \eqn{eq:definitionPhat} take values~\cite{Loebbert:2020hxk}
\begin{equation}
\{s_j\}=\half \{a_2+a_3,a_3-a_1,-a_1-a_2\}.
\end{equation}
Notably, in a dual momentum space, introduced via the transformation \eqref{eq:DualMomenta}, i.e.\ $R_j=x_{j+1}-x_{j-1}$, the level-one generator $\levo{P}$ translates into a representation of the special conformal generator \cite{Loebbert:2020glj}.
Invariance under $\levo{P}^\mu$ implies two independent partial differential equations (cf.\ \cite{Loebbert:2019vcj} for the PDEs in terms of ratio variables)
\begin{equation}
A_1 I_3=0,
\qquad\qquad
A_2I_3=0,
\label{eq:PhatI3}
\end{equation}
with the second order  differential operators
\begin{align}
A_1=
&+r_{12}(\bar w_D-2a_2)\partial_{r_{13}}
-2r_{12}r_{23}\partial_{r_{13}}\partial_{r_{23}}-r_{12}r_{13}\partial_{r_{13}}^2
\nonumber\\
&
+r_{13}(\bar w_D+2a_3)\partial_{r_{12}}
-2r_{12}^2 \partial_{r_{12}}\partial_{r_{13}}-r_{12}r_{13}\partial_{r_{12}}^2,
\nonumber
\\
A_2=&+r_{12}(\bar w_D+2a_1)\partial_{r_{23}}
-r_{12}r_{23}\partial_{r_{23}}^2
\nonumber\\
&-r_{23}(\bar w_D+2a_3)\partial_{r_{12}}
+r_{12}r_{23}\partial_{r_{12}}^2.
\label{eq:Phateq12}
\end{align}
Here, for the conformal weight of the integrals \eqref{eq:intfamily}, we have introduced the  abbreviation
\begin{equation}
w_D=D-2(a_1+a_2+a_3),
\end{equation}
and $\bar w_D=w_D-1$.
For $D=3-2\epsilon$ we make the following ansatz for the $\epsilon$-expansion of the integral $I_3$, which is inspired by \cite{Ohta:1973je}:
\begin{equation}
    \mu^{-2\epsilon}I_3^{3-2\epsilon} =  \frac{A}{2\epsilon} + B + C \log\brk1{\sfrac{r_{12}+r_{13}+r_{23}}{\mu}}+\order{\epsilon}.
\label{eq:ansatzI3}
\end{equation}
Here $\mu$ denotes some mass scale and $A,B,C$ represent polynomials whose form is constrained by the scaling of the integral:
\begin{equation}
X=\sum_{j=0}^{w_3}\sum_{k=0}^{w_3-j} f_{jk}^{(X)} \,r_{12}^j r_{13}^k r_{23}^{w_3-j-k}.
\label{eq:polyansatz}
\end{equation}
For $X\in\{A,B,C\}$ the constant coefficients of the polynomial are denoted by $f_{jk}^{(X)}$. We note that the polynomial $B$ can always be shifted by a term proportional to $C$ via a modification of the mass scale $\mu$. The below results are thus to be understood modulo such a shift.
As the coefficients of $1/\epsilon$ and $\log \mu$ are correlated in the $\epsilon$-expansion of \eqref{eq:ansatzI3}, we must have $A=-C$ which we also find from the bootstrap arguments.

The solution of the homogeneous differential equations \eqref{eq:PhatI3} will depend on some undetermined constants. In general, these can for instance be fixed by comparing a coincident point limit of the solution with the following well known expression for the two-point integral, cf.\ e.g.~\cite{Isaev:2007uy}:
\begin{equation}
\int  \frac{\dd^Dx_0}{x_{01}^{2a_1}x_{02}^{2a_2}}
=\pi^\frac{D}{2} \frac{\Gamma_{a_1+a_2-\frac{D}{2}}\Gamma_{\frac{D}{2}-a_1}\Gamma_{\frac{D}{2}-a_2}}{\Gamma_{a_1}\Gamma_{a_2}\Gamma_{D-a_1-a_2}} r_{12}^{D-2a_1-2a_2}.
\label{eq:twopointint}
\end{equation}
However, for the lower propagator powers considered below, some of the arguments of the Gamma-functions will actually be zero.
It is thus useful to note that the Laplacian acting on leg 1 of the integral generates a recursive structure on the above integrals, e.g.\
\begin{equation}
\Delta_1 I_3[a_1,a_2,a_3]=2a_1(2a_1+2-D)I_3[a_1+1,a_2,a_3],
\label{eq:LaplacianI3}
\end{equation}
and similar for legs 2 and 3. This equation can alternatively be used to relate the undetermined coefficients for integrals with negative propagator powers to the leading-order `seed' integral $I_3[\half,\half,\half]$.

In the following we bootstrap the integrals contributing to the leading terms of the non-relativistic expansion \eqref{eq:PNexpansionI3delta} using the level-one Yangian PDEs~\eqref{eq:Phateq12}. We have compared the expansion of the below results for small ratios $r_{12}/r_{13}$ and $r_{23}/r_{13}$ to the expressions in terms of Appell hypergeometric functions given in \cite{Boos:1990rg} finding full agreement, see also \cite{Loebbert:2020glj} for our conventions.
The following integrals serve as input for the three-body effective potential via \eqref{eq:PNexpansionI3delta} and~\eqref{eq:PPP}. 
\paragraph{Order $c^0$: $I_3[\half,\half,\half]$.}
The leading order contribution to the expansion \eqref{eq:PNexpansionI3delta} is given by propagator powers $a_j=1/2$ for $j=1,2,3$. Using the ansatz \eqref{eq:ansatzI3} it is straightforward to solve the PDEs \eqref{eq:Phateq12} in the $\epsilon$-expansion around $D=3$, which yields
\begin{equation} 
\mu^{-2\epsilon}I_{3}\brk[s]*{\half,\half,\half}=\frac{b_1}{2\epsilon}
-b_2\log(\sfrac{r_{12}+r_{13}+r_{23}}{\mu})+\order{\epsilon},
\label{eq:epsexpI3allhalf}
\end{equation}
for some undetermined constants $b_1$,$b_2$.
The parameters $b_1,b_2$ are fixed by comparison with the two-point integral \eqref{eq:twopointint} to $b_1=b_2=4\pi$. Note that we do not display an additional constant that can be shifted by modification of the mass scale $\mu$.
The above logarithmic result for this integral is already contained in \cite{Ohta:1973je}.

\paragraph{Order $c^{-2}$: $I_3[\half,\half,-\sfrac{1}{2}]$.}
In complete analogy to the above, we find at next-to-leading order the following $\epsilon$-expansion of the single contributing integral:
\begin{align}
\mu^{-2\epsilon}
&I_{3}[\half,\half,-\half]
=
\label{eq:epsexpl3twohalfoneminus}
\\
&-\sfrac{2\pi}{3}\Big[
\sfrac{r_{12}^2-r_{13}^2-r_{23}^2}{2\epsilon}
-(r_{12}-r_{13})(r_{12}-r_{23})
\nonumber\\
&\qquad\quad-(r_{12}^2-r_{13}^2-r_{23}^2)\log(\sfrac{r_{12}+r_{13}+r_{23}}{\mu})
\Big]
+\order{\epsilon}.
\nonumber
\end{align}
Below we will employ this result to obtain new contributions to the three-body effective potential at 3PN, 
which scale as  $v^{4}G^{2}m^{3}/c^{8}r^{2}$.
\paragraph{Order $c^{-4}$: $I_3[-\half,-\half,\half]$ and $I_3[\half,\half,-\sfrac{3}{2}]$.}
To demonstrate that the above bootstrap approach easily generalizes to higher orders, let us also consider the next order. Note, however, that due to its length we will not evaluate the resulting contribution to the effective potential in this paper, see \Secref{sec:3PN} for the previous order.
At order $c^{-4}$ of the non-relativistic expansion \eqref{eq:PNexpansionI3delta} two integrals contribute. With the ansatz \eqref{eq:ansatzI3} we can again solve the above partial differential equations to find solutions of the form
\begin{equation}
\mu^{-2\epsilon}I_3^{3-2\epsilon} =b\brk[s]*{ \frac{A}{2\epsilon} -B- A \log\brk1{\sfrac{r_{12}+r_{13}+r_{23}}{\mu}}}+\order{\epsilon}.
\label{eq:genericform}
\end{equation}
Note again that the polynomial $B$ is only defined modulo a shift by $A$ due to the arbitrariness of the mass scale $\mu$.
Here we have
\begin{align}
A[-\half,-\half,\sfrac{1}{2}]
&=
-\sfrac{1}{3}r_{12}^4+(r_{13}^2-r_{23}^2)^2-\sfrac{2}{3}r_{12}^2(r_{13}^2+r_{23}^2),
\nonumber\\
B[-\half,-\half,\sfrac{1}{2}]
&=\sfrac{r_{12}^4}{9}-  r_{12}\left(r_{13}-r_{23}\right){}^2
   \left(r_{13}+r_{23}\right)
\\
&+\sfrac{1}{3}r_{12}^3 \left(r_{13}+r_{23}\right) - r_{13} r_{23}\left(r_{13}-r_{23}\right){}^2 
   \nonumber\\
   &+\sfrac{1}{9}r_{12}^2\left(5 r_{13}^2-3r_{13} r_{23} +5
   r_{23}^2\right) ,
   \nonumber
\end{align}
as well as
\begin{align}
A[\half,\half,-\sfrac{3}{2}]
&=
- r_{12}^4
- r_{13}^4
- r_{23}^4\nonumber
\\
&+2r_{12}^2\left(r_{13}^2+r_{23}^2\right) 
-\sfrac{2}{3} r_{13}^2 r_{23}^2,
\\
B[\half,\half,-\sfrac{3}{2}]
&=
-\sfrac{4 r_{12}^4}{3}
+r_{13} r_{23}^3-\sfrac{4}{9} r_{13}^2
   r_{23}^2+r_{13}^3 r_{23}\nonumber
\\
&+r_{12}^3\left(r_{13}+r_{23}\right) 
+ r_{12}^2\left(2 r_{13}^2-r_{23}
   r_{13}+2 r_{23}^2\right)
   \nonumber\\
   &-\sfrac{1}{3} r_{12} \left(5 r_{13}^3+3 r_{23} r_{13}^2+3
   r_{23}^2 r_{13}+5 r_{23}^3\right).
   \nonumber
\end{align}
The overall constants $b$ in \eqref{eq:genericform} are fixed by relating
them to the coefficients for the seed integral \eqref{eq:epsexpI3allhalf} via the recursion \eqref{eq:LaplacianI3}:
\begin{equation}
b[-\half,-\half,\half]=-\frac{\pi}{10},
\qquad
b[\half,\half,-\sfrac{3}{2}]=-\frac{3\pi}{10}.
\end{equation}
This provides all the necessary information to generate the 4PN order $G^{2}$ contributions to the effective potential.


\section{The 2PN Expansion}
\label{sec:2PN}
We now proceed to compute the 2PN expansion of the effective action as introduced in \Secref{sec:1PNExp}. 
This also serves as a test for the integral \eqref{eq:epsexpI3allhalf}. 
For the third line of \eqref{eq:PPP}, we decompose the sum as
\begin{equation}
    \sum_{i,j,k} {}^\prime 
    \rightarrow
    \sum_{i} \sum_{j \neq i } \sum_{k\neq i,j}
    + \bigg( \sum_{i} \sum_{j} \Big|_{k=i}
        \!+ (\text{cyclic})\bigg).
\end{equation}
Here we refer to the first term on the right hand side as the three-body interaction and to the remaining terms as the two-body interactions.
When identifying two of the three indices, we encounter a divergence $1/ r_{ij}|_{j=i}$ and an indefinite unit vector $\mbf{n}_{ij}|_{j=i}$. 
In light of the vanishing of propagators with both ends on the same worldline, we propose to regularize the divergences as $1/ r_{ij}|_{j=i} \rightarrow 0$. 
Terms of odd order in $\mbf{n}_{ij}|_{j=i}$ also vanish due to the anti-symmetry in the indices. 
For the quadratic terms in $\mbf{n}_{ij}|_{j=i}$ of the 2PN result we adopt the following limiting prescription:
\begin{equation}
   \left.
    \mbf{n}_{ij} \!\cdot\! \mbf{v}_\alpha \,
    \mbf{n}_{ij} \!\cdot\! \mbf{v}_\beta \right|_{j=i}
    \rightarrow
    \mbf{v}_\alpha \cdot \mbf{v}_\beta .
    \label{eq:vrule1}
  \end{equation}
That is, whenever the identification of two points yields an expression as given on the left hand side, we replace it by the right hand side. This prescription is natural from the perspective of dimensional analysis and symmetry considerations, and it reproduces the correct results as given in the literature.
With regard to the 3PN result to be discussed in \Secref{sec:3PN} we already give the rule
  \begin{align}
&\left.
    \mbf{n}_{ij} \!\cdot\! \mbf{v}_\alpha \,
    \mbf{n}_{ij} \!\cdot\! \mbf{v}_\beta \,
    \mbf{n}_{ij} \!\cdot\! \mbf{v}_\rho \,
    \mbf{n}_{ij} \!\cdot\! \mbf{v}_\sigma \right|_{j=i}
    \label{eq:vrule2}
    \\ & \quad \rightarrow 
    \mbf{v}_\alpha \!\cdot\! \mbf{v}_\beta \,
    \mbf{v}_\rho \!\cdot\! \mbf{v}_\sigma 
    + \mbf{v}_\alpha \!\cdot\! \mbf{v}_\rho \,
    \mbf{v}_\sigma \!\cdot\! \mbf{v}_\beta 
    + \mbf{v}_\alpha \!\cdot\! \mbf{v}_\sigma \,
    \mbf{v}_\beta \!\cdot\! \mbf{v}_\rho. \nonumber
\end{align}
We note that the $1/\epsilon$-term in \eqref{eq:epsexpI3allhalf} naturally drops out in the final expression for the action due to the derivatives that have to be applied. 
Moreover, we expect this property to hold to all orders in the PN expansion.
This is explicitly shown to be true in the 3PN calculation of \Secref{sec:3PN}.
The 2PN effective action reads
\begin{widetext}
\begin{align}
    S^{\text{2PN}} =& \sum_i \int \frac{\dd t}{c^6} \bigg\{
    \frac{ m_i \mbf{v}_i^6 }{16}
    + \sum_{j\neq i} 
    \frac{ G m_i m_j }{ 16 r_{i j} }
    \Big[
    3 (\mbf{n}_{i j} \!\cdot\! \mbf{v}_i )^2 (\mbf{n}_{i j}\!\cdot\! \mbf{v}_j)^2
    - 6 \, \mbf{n}_{i j} \!\cdot\! \mbf{v}_i \, \mbf{n}_{i j}\!\cdot\! \mbf{v}_j \, \mbf{v}_{ij}^2 
        - 2 \left(\mbf{n}_{i j} \!\cdot\! \mbf{v}_j\right)^2 \mbf{v}_i^2
        \nonumber \\ & \quad
        + 3 \, \mbf{v}_i^2 \, \mbf{v}_j^2
        + 2 \left(\mbf{v}_i \!\cdot\! \mbf{v}_j \right)^2 
        - 20 \, \mbf{v}_i^2 \, \mbf{v}_i \!\cdot\! \mbf{v}_j
        + 14 \, \mbf{v}_i^4 \Big]
    + \sum_{j\neq i} 
        \frac{ G^2 m_i m_j^2}{2 r_{i j}^2}
        \left[
        33 \left( \mbf{n}_{ij} \cdot \mbf{v}_{ij} \right)^2 
        - 17 \mbf{v}_{ij}^2 \right]
    \nonumber \\ & 
    + \sum_{j\neq i} \sum_{k\neq i} 
        \frac{G^2 m_i m_j m_k}{8} 
        \bigg[ \frac{ 1 }{ r_{i j} r_{i k} } 
            \big( 4 (\mbf{n}_{i j} \cdot \mbf{v}_j)^2
            + 18 \, \mbf{v}_i^2 - 16 \, \mbf{v}_j^2 
            - 32 \, \mbf{v}_i \cdot \mbf{v}_j
            + 32 \, \mbf{v}_j \cdot \mbf{v}_k
            \big)
        \label{eq:2PNfinal} \\  & \quad
        + \frac{1}{r_{i j}^2}
            \left(
            14 \, \mbf{n}_{i k} \!\cdot\! \mbf{v}_k \,
              \mbf{n}_{i j} \!\cdot\! \mbf{v}_k
            - 12\, \mbf{n}_{i j} \!\cdot\! \mbf{v}_i \,
              \mbf{n}_{i k} \!\cdot\! \mbf{v}_k
            + \mbf{n}_{i j} \!\cdot\! \mbf{n}_{i k} \,
              ( \mbf{n}_{i k} \!\cdot\! \mbf{v}_k )^2 
            - \mbf{n}_{i j} \!\cdot\! \mbf{n}_{i k} \, \mbf{v}_k^2 
            \right)
        \bigg] 
    \nonumber \\ & 
    + \sum_{j\neq i} \sum_{k\neq i,j} 
        G^2 m_i m_j m_k 
        \bigg[
            \frac{ 2 (\mbf{n}_{i j} \!-\! \mbf{n}_{j k} ) \!\cdot\! 
                \mbf{v}_{ij} }
                 { ( r_{i j} + r_{i k} + r_{j k} )^2}
            \big(
                4 \left(\mbf{n}_{ij} + \mbf{n}_{ik} \right) \!\cdot\!   
                  \mbf{v}_{ij}
                + \left(\mbf{n}_{ik} + \mbf{n}_{jk} \right) \!\cdot\!
                  \mbf{v}_{ik} 
            \big)
            \nonumber \\ &\quad
            +\frac{9 \left(\mbf{n}_{i j} \!\cdot\! \mbf{v}_{ij} \right)^2
                - 9\, \mbf{v}_{ij}^2
                + 2 \left( \mbf{n}_{i j} \!\cdot\! \mbf{v}_{ik} \right)^2 
                - 2\, \mbf{v}_{ik}^2 }
            {r_{i j} \left(r_{i j}+r_{i k}+r_{j k}\right)} \bigg]
        \bigg\} + G^{3}\times\text{[static term]}  \, , \nonumber
\end{align}
\end{widetext}
where we define $\mbf{v}_{ij} := \mbf{v}_i - \mbf{v}_j$. Here we have performed a field redefinition to push terms that involve accelerations to higher orders in $G$.
We have checked that our result agrees with the literature \cite{Ohta:1974pq,Damour:1985mt} up to a total derivative. Note that we do not have access
to the static (velocity independent) term at $\mathcal{O}(G^{3})$ in our approach as it stems from a 3PM computation.

\section{New Contributions at 3PN}
\label{sec:3PN}

In this section we explicitly evaluate the contributions to the 3PN three-body effective potential. Limiting the number of point masses to two gives the two-body 3PN effective action, which we checked to agree with \cite{Bernard:2015njp} up to a total derivative. Next to the novel three-point $G^{2}v^{4}$ terms, the below expression contains terms that scale as  $G v^6$, as well as  two-point  terms of order $G^2 v^4$ which have been known before.
The full 3PN action can be written in the form
\begin{align}
    S^{\text{3PN}} =\sum_i \int \frac{\dd t}{c^8} 
    \bigg\{ 
    \frac{5}{128} m_i \mbf{v}_i^8 
        &+ L^{\text{3PN}}_{(A)}
    + L^{\text{3PN}}_{(B)}
    \label{eq:full3PNaction}
   \\
   & + L^{\text{3PN}}_{(C)}
 + L^{\text{3PN}}_{(D)}    \bigg\}
 +\order{G^3}. \nonumber
\end{align} 
Note that the terms at order ${G^3}$ are not given here and require two yet unknown four-point integrals at one and two loops. Moreover, there are additional $G^4$ contributions at 3PN.
In \eqref{eq:full3PNaction} we have ordered the various terms, which are explicitly given in the following, by their power of $G$ and the structure of summations.
Terms from perturbative solutions of the equations of motion for the einbein, cf.\  \eqref{eq:einbein}, contribute at various places.
Explicit expressions for the terms  in \eqref{eq:full3PNaction} are also provided in an ancillary file to this paper.
The term $L^{\text{3PN}}_{(A)}$ originates from the 1PM action and reads
\begin{widetext}
\begin{align} \label{eq:L3PNA}
L^{\text{3PN}}_{(A)}= \sum_{j\neq i}& \frac{G m_i m_j}{32 r_{i j}}
    \Big[
        -\!5 \left(\mbf{n}_{i j}\!\cdot\! \mbf{v}_i\right)^3 \left(\mbf{n}_{i j}\!\cdot\! \mbf{v}_j\right)^3
        +3\, \mbf{n}_{i j} \!\cdot\! \mbf{v}_i 
          \left( \mbf{n}_{i j} \cdot \mbf{v}_j \right){}^2 
          \left( 2\, \mbf{v}_i^2 \mbf{n}_{i j} \!\cdot\! \mbf{v}_j
            + 6\,  \mbf{v}_i^2\, \mbf{n}_{i j} \!\cdot\! \mbf{v}_i
            - 5\, \mbf{v}_i\cdot \mbf{v}_j  \mbf{n}_{i j} \!\cdot\! \mbf{v}_i \right)
        \nonumber \\[-2ex] &
        + \mbf{n}_{i j}\!\cdot\! \mbf{v}_i \, \mbf{n}_{i j}\!\cdot\! \mbf{v}_j
        \left(10 \left(\mbf{v}_i \!\cdot\! \mbf{v}_j\right){}^2
          +8\, \mbf{v}_i^2 \, \mbf{v}_i \!\cdot\! \mbf{v}_j 
          -5\, \mbf{v}_i^2 \, \mbf{v}_j^2 
          -14\, \mbf{v}_i^4 \right)
        +2\, \left(\mbf{n}_{i j}\!\cdot\! \mbf{v}_j\right)^2 \mbf{v}_i^2\,
          \left( 5 \mbf{v}_i\!\cdot\! \mbf{v}_j -3\, \mbf{v}_i^2 \right)
        \nonumber \\ &
        -6\, \mbf{v}_i^2\, \mbf{v}_j^2 \left(\mbf{n}_{i j}\!\cdot\! \mbf{v}_i\right)^2
        +16\, \mbf{v}_i^4\, \mbf{v}_j^2
        +2 \left(\mbf{v}_i\!\cdot\! \mbf{v}_j\right)^3
        +12\, \mbf{v}_i^2 \left(\mbf{v}_i\!\cdot\! \mbf{v}_j\right)^2
        -19\, \mbf{v}_i^2\, \mbf{v}_j^2\, \mbf{v}_i\!\cdot\! \mbf{v}_j
        -34\, \mbf{v}_i^4\, \mbf{v}_i\!\cdot\! \mbf{v}_j
        +22\, \mbf{v}_i^6
    \Big] .
\end{align}
Here we have added a total derivative as given in \eqref{eq:tdacceleration} in \Appref{sec:3PNlong}. 
Again, accelerations have been pushed to the next order in $G$ by means of field redefinitions. 
The next term $L^{\text{3PN}}_{(B)}$ stems from the two-body interactions of the third line of \eqref{eq:PPP} and reads
\begin{align}\label{eq:L3PNB}
L^{\text{3PN}}_{(B)}=& \sum_{j\neq i} \frac{G^2 m_i m_j^2 }{ 4 r_{ij}^2 }
    \Big[
    \left( 
      - 200\, \mbf{v}_i\!\cdot\! \mbf{v}_j
      + 167\, \mbf{v}_i^2
      + 66\, \mbf{v}_j^2 \right)
    \left( \mbf{n}_{i j} \!\cdot\! \mbf{v}_i \right){}^2
    - 2 \left( 99\, \mbf{v}_i^2
      + 64\, \mbf{v}_j^2
      - 130\, \mbf{v}_i \!\cdot\! \mbf{v}_j \right)
    \mbf{n}_{i j}\!\cdot\! \mbf{v}_j \, \mbf{n}_{i j}\!\cdot\! \mbf{v}_i
    \nonumber \\[-1.8ex] & \quad
    -44 \left( \mbf{n}_{i j}\!\cdot\! \mbf{v}_i 
      - \mbf{n}_{i j}\!\cdot\! \mbf{v}_j \right){}^2 
      \left( 2 \left(\mbf{n}_{i j}\!\cdot\! \mbf{v}_i\right){}^2
        +\left( \mbf{n}_{i j} \!\cdot\! \mbf{v}_j \right){}^2 \right)
    +\left(65\, \mbf{v}_i^2 
      + 96\, \mbf{v}_j^2
      - 128\, \mbf{v}_i \!\cdot\! \mbf{v}_j \right)
     \left(\mbf{n}_{i j}\!\cdot\! \mbf{v}_j\right){}^2
     \nonumber \\ & \quad
    -98 \left(\mbf{v}_i\!\cdot\! \mbf{v}_j\right){}^2
    +96\, \mbf{v}_j^2\, \mbf{v}_i\!\cdot\! \mbf{v}_j
    + \mbf{v}_i^2 \left(134\, \mbf{v}_i\!\cdot\! \mbf{v}_j - 49\, \mbf{v}_j^2\right)
    -51\, \mbf{v}_i^4 
    -32\, \mbf{v}_j^4
    \Big].
\end{align}
Moreover, the term $L^{\text{3PN}}_{(C)}$ to \eqref{eq:full3PNaction} receives contributions from the second line of \eqref{eq:PPP} as well as from field redefinitions and total derivatives that we use to remove terms that involve accelerations:
\begin{align}
&L^{\text{3PN}}_{(C)}=
 \sum_{j\neq i} \sum_{k\neq i} \frac{G^2 m_i m_j m_k}{16} \bigg\{
    \frac{1}{r_{ij} r_{ik} }
    \Big[
        2 \left( \mbf{n}_{i j}\!\cdot\! \mbf{v}_j\right){}^2 
          \left( 16\, \mbf{v}_i\!\cdot\! \mbf{v}_j
            -18\, \mbf{v}_i^2
            -32\, \mbf{v}_j\!\cdot\! \mbf{v}_k
            +12\, \mbf{v}_j^2
            -\left(\mbf{n}_{i k}\!\cdot\! \mbf{v}_k\right){}^2\right)
        \nonumber \\[-0.8ex] &\ \ 
        + 64\, \mbf{v}_i\!\cdot\! \mbf{v}_j \left(
          2 \left(\mbf{n}_{i k}\!\cdot\! \mbf{v}_k\right){}^2 
          + \mbf{v}_i\!\cdot\! \mbf{v}_k
          -2\, \mbf{v}_j\!\cdot\! \mbf{v}_k-\mbf{v}_j^2\right)
        + 16\, \mbf{v}_j^2 \left(
          8\, \mbf{v}_j\!\cdot\! \mbf{v}_k
          -2\, \mbf{v}_j^2
          -2\, \mbf{v}_k^2 
          -\left(\mbf{n}_{i k}\!\cdot\! \mbf{v}_k\right){}^2 \right)
        \nonumber \\ & \ \ 
        +16\, \mbf{v}_i^2 \left( 
          3\, \mbf{v}_j^2
          +2\, \mbf{v}_j\!\cdot\! \mbf{v}_k
          -10\, \mbf{v}_i\!\cdot\! \mbf{v}_j\right)
        -6 \left(\mbf{n}_{i j}\!\cdot\! \mbf{v}_j\right){}^4
        +96 \left(\mbf{v}_i\!\cdot\! \mbf{v}_j\right){}^2
        +49\, \mbf{v}_i^4
    \Big]
    + \frac{1}{ 3\, r_{ij}^2 } \Big[
        20 \left(\mbf{n}_{i k}\!\cdot\! \mbf{v}_k\right){}^3\left(
        \mbf{n}_{i j}\!\cdot\! \mbf{v}_i
        -\mbf{n}_{i j}\!\cdot\! \mbf{v}_k\right) 
    \nonumber \\[-0.7ex] & \ \ 
        -3\, \mbf{n}_{i j}\!\cdot\! \mbf{n}_{i k} 
        \left(\left(\mbf{n}_{i k}\!\cdot\! \mbf{v}_k\right){}^2 - \mbf{v}_k^2\right)
        \left( 3 \left(\mbf{n}_{i j}\!\cdot\! \mbf{v}_j\right){}^2
          +8\, \mbf{v}_i\!\cdot\! \mbf{v}_j
          +\left(\mbf{n}_{i k}\!\cdot\! \mbf{v}_k\right){}^2
          +2\, \mbf{n}_{i k}\!\cdot\! \mbf{v}_i \, \mbf{n}_{i k}\!\cdot\! \mbf{v}_k
          +6\, \mbf{v}_i\!\cdot\! \mbf{v}_k
          -5\, \mbf{v}_i^2
          -4\, \mbf{v}_j^2
          -4\, \mbf{v}_k^2 \right)
        \nonumber \\[0.5ex] & \ \ 
        + 6 \left( \left(\mbf{n}_{i k}\!\cdot\! \mbf{v}_k\right){}^2 - \mbf{v}_k^2 \right)
        \big(
           3\, \mbf{n}_{i j}\!\cdot\! \mbf{v}_j \, \mbf{n}_{i k}\!\cdot\! \mbf{v}_i
          -3\, \mbf{n}_{i j}\!\cdot\! \mbf{v}_j \, \mbf{n}_{i k}\!\cdot\! \mbf{v}_j
          -3\, \mbf{n}_{i k}\!\cdot\! \mbf{v}_i \, \mbf{n}_{i j}\!\cdot\! \mbf{v}_k
          +4\, \mbf{n}_{i j}\!\cdot\! \mbf{v}_i \, \mbf{n}_{i k}\!\cdot\! \mbf{v}_j
          -  \mbf{n}_{i j}\!\cdot\! \mbf{v}_i \, \mbf{n}_{i k}\!\cdot\! \mbf{v}_i \big) 
        \nonumber \\[0.5ex] & \ \ 
        +6\, \mbf{n}_{i k}\!\cdot\! \mbf{v}_k  \big[
          \mbf{n}_{i j}\!\cdot\! \mbf{v}_k\left(
            19\, \mbf{v}_i^2
            +28\, \mbf{v}_j^2
            \!-\! 56\, \mbf{v}_i \!\cdot\! \mbf{v}_j
            \!-\! 2\, \mbf{v}_i \!\cdot\! \mbf{v}_k 
            \!-\! 21 \left(\mbf{n}_{i j}\!\cdot\! \mbf{v}_j\right){}^2 \right) 
          +6\, \mbf{n}_{i j}\!\cdot\! \mbf{v}_j \left( 
            6\, \mbf{v}_i \!\cdot\! \mbf{v}_j
            +7\, \mbf{v}_i \!\cdot\! \mbf{v}_k
            \!-\! 6\, \mbf{v}_i^2
            \!-\! 7\, \mbf{v}_j \!\cdot\! \mbf{v}_k\right) 
          \nonumber \\[-0.5ex] & \quad \ 
          +2\, \mbf{n}_{i j} \!\cdot\! \mbf{v}_i \left(
            11\, \mbf{v}_i^2
            -12\, \mbf{v}_j^2
            -23\, \mbf{v}_i\!\cdot\! \mbf{v}_k
            +28\, \mbf{v}_j\!\cdot\! \mbf{v}_k
            +9 \left( \mbf{n}_{i j}\!\cdot\! \mbf{v}_j\right){}^2\right) \big]
        +18\, \mbf{v}_k^2 \, \mbf{n}_{i k}\!\cdot\! \mbf{v}_k \left(
          5\,  \mbf{n}_{i j}\!\cdot\! \mbf{v}_k
          -4\, \mbf{n}_{i j}\!\cdot\! \mbf{v}_i \right) 
    \Big]
    \bigg\} .
    \label{eq:LC3PN}
\end{align}
The term $L^{\text{3PN}}_{(D)}$ contributing to the above action originates from the three-body parts of the third line of \eqref{eq:PPP}
and can be expressed in terms of derivatives that act on the integrals $I_3[\half, \half, \half]$ and $I_3[\half, \half, -\half]$ as given in \eqref{eq:epsexpI3allhalf} and \eqref{eq:epsexpl3twohalfoneminus} of \Secref{sec:PNandBootstrap}:
\begin{align}
    &L^{\text{3PN}}_{(D)} \!=\! \sum_{j\neq i}\! \sum_{k\neq i, j} \!
    \frac{G^2 m_i m_j m_k}{ 4\pi } 
    \Big\{ 
    \Big[ 
         ( 6 \mbf{v}_i^2 \!+\! \mbf{v}_k^2 \!-\! 8 \mbf{v}_i \!\cdot\! \mbf{v}_j ) 
            (\mbf{v}_{ki} \!\cdot\! \partial_{x_i})(\mbf{v}_{kj} \!\cdot\! \partial_{x_j})
       + ( 8 \mbf{v}_{ik}^2 \!-\! 4 \mbf{v}_k^2 ) 
            (\mbf{v}_{ji} \!\cdot\! \partial_{x_i}) 
            (\mbf{v}_{ij} \!\cdot\! \partial_{x_j})
    \Big] I_3[\half, \half, \half] 
    \label{eq:L3PND}
     \\[-0.5ex] &\ \ 
    + \left( \mbf{v}_{k} \!\cdot\! \partial_{x_k} \right)^2
    \Big[	
           (\mbf{v}_{ki} \!\cdot\! \partial_{x_i}) (\mbf{v}_{kj} \!\cdot\! \partial_{x_j})
        \!+\! 2(\mbf{v}_{ik} \!\cdot\! \partial_{x_k}) (\mbf{v}_{ij} \!\cdot\! \partial_{x_j})
        \!+\! 4(\mbf{v}_{ji} \!\cdot\! \partial_{x_i}) (\mbf{v}_{ij} \!\cdot\! \partial_{x_j})
        \!+\! 8(\mbf{v}_{jk} \!\cdot\! \partial_{x_k}) (\mbf{v}_{kj} \!\cdot\! \partial_{x_j})
    \Big] I_3[\half, \half,\! -\half]
    \Big\}. \nonumber
\end{align}
Here the integrals $I_3$ depend on the three external points $i,j,k$ as opposed to \Secref{sec:PNandBootstrap}, where the labels $1,2,3$ were used. 
For convenience we display again the expressions \eqref{eq:epsexpI3allhalf} and \eqref{eq:epsexpl3twohalfoneminus}:
\begin{align}
\mu^{-2\epsilon}I_{3}\brk[s]*{\half,\half,\half}
&=\frac{2\pi}{\epsilon}
-4\pi\log\brk*{\frac{r_{ij}+r_{ik}+r_{jk}}{\mu}}+\order{\epsilon},
\\
\mu^{-2\epsilon}I_{3}[\half,\half,-\half]
&=
-\frac{2\pi}{3}\Big[
\frac{r_{ij}^2-r_{ik}^2-r_{jk}^2}{2\epsilon}
-(r_{ij}-r_{ik})(r_{ij}-r_{jk})
-(r_{ij}^2-r_{ik}^2-r_{jk}^2)\log\brk*{\frac{r_{ij}+r_{ik}+r_{jk}}{\mu}}
\Big]
+\order{\epsilon}.
\end{align}
Note again that by identifying two indices in the above $L^{\text{3PN}}_{(D)}$ and using the prescriptions given in \eqref{eq:vrule1} and \eqref{eq:vrule2} we obtain the corresponding two-body contributions \eqref{eq:L3PNB}.
The $1/\epsilon$-poles and the mass scale $\mu$ in the expressions for the integrals  $I_3$ drop out after taking the derivatives in \eqref{eq:L3PND}. This property persists at least to the 4PN order. The structure of the result for $L^{\text{3PN}}_{(D)}$ after evaluating these derivatives is displayed in \Appref{sec:3PNlong}.

Note that the contribution from the second line of \eqref{eq:PPP}, which is already contained in the above expression \eqref{eq:LC3PN} for $L^{\text{3PN}}_{(C)}$, can also explicitly  be written in the form before taking derivatives 
\begin{align}
    L^{\text{3PN}}_{(\text{second})} =& \sum_{j\neq i} \sum_{k\neq i} 
        G^2 m_i m_j m_k \bigg\{
            \frac{1}{8\, r_{ij} r_{ik}} \Big[
                \mbf{v}_i^2 \left( \mbf{v}_i^2 - 20\, \mbf{v}_j^2 + 16\, \mbf{v}_j \!\cdot\! \mbf{v}_k \right) 
              + 2\, \mbf{v}_j^2 \big( 32\, \mbf{v}_i \!\cdot\! \mbf{v}_k + 16\, \mbf{v}_j \!\cdot\! \mbf{v}_k  
              - 7\, \mbf{v}_j^2 -9\, \mbf{v}_k^2 \big)
                \nonumber \\ & \ 
              + 32\, \mbf{v}_i \!\cdot\! \mbf{v}_j \left( \mbf{v}_i \!\cdot\! \mbf{v}_j
                - \mbf{v}_i \!\cdot\! \mbf{v}_k - 2\, \mbf{v}_j \!\cdot\! \mbf{v}_k \right)
            \Big]
            + \left( \mbf{v}_i^2 - 3\, \mbf{v}_j^2 -3\, \mbf{v}_k^2 +8\, \mbf{v}_j \!\cdot\! \mbf{v}_k \right) 
             (\mbf{v}_k \!\cdot\! \partial_{x_k})^2  \frac{r_{ik}}{2\, r_{ij}}
             \\ & 
            -\left[ (\mbf{v}_k \!\cdot\! \partial_{x_k})^2 (\mbf{v}_j \!\cdot\! \partial_{x_j})^2 
             \frac{r_{ij} r_{ik} }{4} 
            + (\mbf{v}_k \!\cdot\! \partial_{x_k})^4 \frac{r_{ik}^3}{12\, r_{ij} } \right]
        \bigg\}. \nonumber
\end{align}
This completes the details describing the final result \eqref{eq:full3PNaction} for the $G^2$ contributions to the 3PN effective action. 
We note again that in principle we could proceed with the same method to compute higher order contributions to the effective potential of order $G^2v^{2n}$. The two new integrals that contribute to the next order of the expansion were already given in \Secref{sec:PNandBootstrap}. However, due to the length of the above terms at $G^2v^4$, see also the expanded result in \Appref{sec:3PNlong}, we refrain here from explicitly evaluating the contributions at the next order $G^2v^6$. 
\end{widetext}


\section{Conclusions and Outlook}
\label{sec:Conclusions}

In the present paper we have extended the known results for the three-body effective potential in general relativity as follows:
\begin{itemize} 
\item At order 2PM the potential is given by \eqref{eq:S2PMgen} expressed via a differential operator that acts on the three-point integral $I_{3\delta}$ evaluated in \Appref{app:3delta}.
\item At 3PN, new $G^2 v^4$-contributions were obtained in \Secref{sec:3PN} and are explicitly provided in an ancillary file to this paper. 
\item The key integrals contributing to the effective potential at $G^2 v^{2n}$ can be obtained by the bootstrap approach  described in \Secref{sec:PNandBootstrap}. Due to their length, we here refrain  from evaluating the resulting expressions for the effective potential.
\end{itemize}
There are a number of interesting directions that should be further explored. Firstly, it would be important to establish the connection between the above PN results and the direct non-relativistic expansion of the (integrated!) expression \eqref{eq:S2PMgen} at 2PM. Approaching this problem one faces the lengthy distributional expressions given in \Appref{sec:DerivativesI3delta}, whose PN expansion appears to require some regularization of diverging contributions. For this reason we have performed the PN expansion at the level of the 2PM integrand, and then evaluated the integrals. Still, rederiving the PN expansion from the final 2PM expression would represent an important cross check of the result. This should also be useful to understand the interplay of the different kinematical regions for the integral discussed in \Appref{app:3delta}. 

With regard to the higher order PN contributions which scale as  $G^2 v^{2n}$, it would be interesting to bootstrap a closed formula for the $\epsilon$-expansion of the family of integrals \eqref{eq:intfamily} around three dimensions. Exploiting the Yangian level-one symmetry discussed above, this should be feasible along the lines of \cite{Loebbert:2019vcj}. Here it would be great to prove that the observed mechanism which makes the divergent contributions drop out in the final expression for the potential persists to all orders.
Similarly, it should be explored how far these bootstrap methods reach in obtaining integrals at higher orders of the PM or PN expansion.

The Yangian level-one symmetry that we employed can also be understood as a special conformal symmetry in a dual momentum space. Here the dualized momenta relate to the above position space variables via \eqref{eq:DualMomenta} (not via Fourier transform). It would be interesting to investigate the employed symmetry in Fourier space and to see if there is a relation to the curious conformal symmetry of graviton amplitudes observed in \cite{Loebbert:2018xce}.

Finally, one should see if one can feed the above contributions to the effective potential into numerical simulations updating the studies of
\cite{Lousto:2007ji,Galaviz:2010te,Galaviz:2011qb,Bonetti:2016eif}. Eventually it would be fascinating if the effect of the three-body interactions obtained here could be observed in the future.


\begin{acknowledgements}
We would like to thank J.~Bicak, M.~Levi and G.~Sch\"afer for helpful communications and R.~Gonzo for discussions.
This project has received funding from the European Union's Horizon 2020 research and innovation program under the Marie Sklodowska-Curie grant agreement No. 764850. The work of JP and TW is funded by the Deutsche Forschungsgemeinschaft (DFG, German Research Foundation) -- Projektnummer PL457/3-1. The work of FL is funded by the Deutsche Forschungsgemeinschaft (DFG, German Research Foundation) -- Projektnummer 363895012.
\end{acknowledgements}

\appendix
\section{The $3\delta$ Integral }
\label{app:3delta}

Here we present a detailed calculation of the $3\delta$-integral given by
\begin{align}
I_{3 \delta} &=\int \dd^{4} x_{0} \,\delta(x_{01}^2)\delta(x_{02}^2)\delta(x_{03}^2).	
\end{align}
Recall $\sigma^2 ={(R_2 \cdot R_3)^2 - R_2^2 R_3^2}$ as defined in \eqref{eq:I3deltaC}.
Importantly, the quantity
$-\sigma^2$ may be seen as the square of the area of the parallelogram spanned by  $x_1,x_2,x_3$ and thus characterizes the space $\mathcal{M}$ spanned by these three points:
\begin{align*}
&\sigma^2>0, \qquad \mathcal{M} \text{ is 2D Minkowskian},\\
&\sigma^2=0, \qquad \mathcal{M}  \text{ is a 1D straight line,}\\
&\sigma^2<0, \qquad \mathcal{M} \text{ is 2D Euclidean}.
\end{align*}
We now explicitly evaluate the above integral for these three cases, generalizing the computation of Westpfahl \cite{Westpfahl:1985tsl} for the three-point integral with retarded propagators.
\paragraph{The case $\sigma^2>0$.}
We choose four basis vectors $ {R_2^\mu, R_3^\mu, \xi_1^\mu, \xi_2^\mu}$ such that we can express the integration vector as
\begin{align}
x_{01}^\mu = \tau R_2^\mu + \bar\tau R_3^\mu + r (\cos\varphi \xi_1^\mu + \sin \varphi \xi_2^\mu), \ \  r \geq 0.
\end{align}
Here $\{\xi_1, \xi_2\}$ denotes the (orthogonal) unit basis of the perpendicular complement of $\mathcal{M}$:
\begin{align}		
&\xi_i \cdot R_j = 0 , \qquad \xi_i \cdot \xi_j = - \delta_{ij},\quad \text{for} \ i,j = 1,2.
\end{align}
In these coordinates the integration measure reads 
\begin{equation}
\dd^4 x_0 = \half \sigma \dd\tau \dd\bar\tau \dd r^2 \dd\varphi,
\end{equation}
and the integral simplifies to
\begin{equation}
\int \frac{\dd^{4} x_{0}}{4 \sigma^2}\delta\big(r^2 + \sfrac{R_1^2 R_2^2 R_3^2}{4\sigma^2}\big) \delta\left( \tau - \sfrac{R_3^2 R_1\!\cdot\! R_2}{2 \sigma^2}\right) \delta \left( \bar\tau + \sfrac{R_2^2 R_1\!\cdot\! R_3}{2\sigma^2} \right).
\end{equation}
This straightforwardly yields
\begin{equation}
I_{3 \delta}(\sigma^2>0)
= \frac{\pi}{4 \sigma} \Theta \left(-R_1^2 R_2^2 R_3^2 \right). 
\label{eq_withTheta}
\end{equation}
Here $\Theta$ denotes the Heaviside step function defined as
\begin{align}
\Theta(x) = &\left\{ \begin{matrix}
1, \quad \text{for} \quad x > 0,\\
{1 \over 2}, \quad \text{for} \quad x = 0,\\
0, \quad \text{for} \quad x<0.
\end{matrix} \right.
\label{eq:DefHeaviside}
\end{align}
Hence, we conclude that in the region $\sigma^2>0$ the piecewise constant in \eqref{eq:I3deltaC} is given by 
\begin{equation}
C(\sigma^2>0)= \frac{\pi}{4} \Theta \left(-R_1^2 R_2^2 R_3^2 \right).
\end{equation}
Note that since $\delta(x_{01}^{2})$ is the Green's function of the d'Alembertian, see \eqref{eq:GreensPropagator}, the above integral $I_{3\delta}$ satisfies
\begin{align}
\partial_1^2 I_{3\delta} = 4 \pi \delta(x_{12}^2)\delta(x_{13}^2) \label{eq_dAlemberion}.
\end{align} 
In the region $\sigma^2>0$ this is guaranteed by the Heaviside function in \eqref{eq_withTheta}; dropping the $\Theta$-function in \eqn{eq_withTheta}
would yield a vanishing result as $\partial_1^2 \sigma^{-1}=0$.


\paragraph{The case $\sigma^2<0$.}

In the region where $\sigma^2<0$, we can span $x_{01}$ as
\begin{align}
x_{01}^\mu = t T^\mu + \tau R_2^\mu + \bar\tau R_3^\mu + r \xi^\mu.
\end{align}
Here $T^\mu$ and $\xi^\mu $ denote again unit vectors that are orthogonal to each other and to $R_2^\mu, R_3^\mu$, with $T^\mu$ being time-like and $\xi^\mu$ space-like. The volume element in this coordinate system is 
\begin{align}
\dd^4 x_0 = \sqrt{-\sigma^2} \dd t\dd r\dd \tau \dd\bar\tau,
\end{align}
and the integral becomes
\begin{align}
I_{3\delta} = &\int \dd^4 x_0\, \delta\big( t^2 + (\tau R_2^\mu + \bar\tau R_3^\mu)^2 - r^2 \big) 
\nonumber\\
&\qquad \times\delta\big( 2\tau R_2\cdot R_3 + (2\bar \tau+1) R_3^2 \big) 
\nonumber\\
&\qquad \times\delta \big( 2\bar\tau  R_2\cdot R_3 + (2 \tau-1) R_2^2 \big)
\nonumber\\
=& \frac{\sqrt{-\sigma^2}}{-4\sigma^2} \int \dd t \dd r \,\delta\left( t^2 - r^2 - \sfrac{R_1^2 R_2^2 R_3^2}{4\sigma^2} \right)
\nonumber\\
=& \frac{1}{4 \sqrt{- \sigma^2} } \int_{-\infty}^{+\infty}  \frac{\dd r}{\sqrt{r^2 +1 }} \rightarrow \infty.
\end{align}
Hence, for $\sigma^2<0$ the integral diverges. 


\paragraph{The case $\sigma^2=0$.}

Finally, for $\sigma^2=0$ the surface spanned by the vectors connecting $x_1,x_2$ and $x_3$ degenerates into a line. 
We define the unit vector on this line as $R_u^\mu$, and we set $R_i^\mu = \omega_i R_u^\mu$.  Depending on the nature of this line one finds different expressions as follows. For the line being time-like we have
\begin{align}
&x_{01}^\mu = \tau R_u^\mu + r \left( \xi_1^\mu \cos{\theta} + \xi_2^\mu \sin{\theta} \cos{\phi} + \xi_3^\mu \sin{\theta} \sin{\phi} \right), \nonumber\\
&\dd^4 x_0 = r^2 \sin{\theta} \dd\tau \dd r \dd\theta \dd\phi ,
\end{align}
and thus
\begin{align}
 I_{3\delta} &=  \int \dd^4 x_0 \delta\left( \tau^2 - r^2 \right) \delta\left( \omega_3^2 + 2\tau \omega_3 \right) \delta\left( \omega_2^2 - 2\tau \omega_2 \right) \nonumber \\
&
=\begin{cases}
\infty & \omega_1\omega_2\omega_3=0, \\
0 & \text{otherwise.}
\end{cases} 
\end{align}
For a space-like line and with $T \!\cdot\! R_u = 0$, we have
\begin{align}
 x_{01}^\mu &= t T^\mu + \tau R_u^\mu + r \left( \xi_1^\mu \cos{\theta} + \xi_2^\mu \sin{\theta}  \right),
  \nonumber\\
 \dd^4 x_0 &= r \dd t \dd\tau \dd r \dd\theta,
 \end{align}
 which implies
 \begin{align}
 I_{3\delta} &=  \int \dd^4 x_0 \delta\left( t^2 + \tau^2 - r^2 \right) \delta\left( \omega_3^2 + 2\tau \omega_3 \right) \delta\left( \omega_2^2 - 2\tau \omega_2 \right) \nonumber\\
&
=\begin{cases}
\infty& \omega_1\omega_2\omega_3=0,
\\
0 & \text{otherwise.}
\end{cases} 
\end{align}
And finally for a light-like line with $T \!\cdot\! R_u \neq 0$, we obtain
\begin{align}
  x_{01}^\mu &= t T^\mu + \tau R_u^\mu + r \left( \xi_1^\mu \cos{\theta} + \xi_2^\mu \sin{\theta}  \right),
  \nonumber\\
  \dd^4 x_0 &= \sqrt{(T\cdot R_u)^2 - T^2 R_u^2} r \dd t \dd\tau \dd r \dd\theta,  
 \end{align}
 such that
 \begin{align}
 I_{3\delta} &=  \int \dd^4 x_0 \delta\left( t^2 + 2t\tau T \!\cdot\! R_u - r^2 \right)
 \nonumber\\
& \qquad\qquad\times\delta\left( 2t\omega_3 T \!\cdot\! R_u \right) \delta\left( 2 t \omega_2 T \!\cdot\! R_u\right) 
\nonumber \\
& \sim \int \dd\tau \delta(0) = \infty.
\end{align}
Hence, the result for $\sigma^2=0$ may be  summarized as
\begin{align}
	I_{3\delta}(\sigma^2=0) \sim \delta(R_1^2) + \delta(R_2^2) + \delta(R_3^2).
\end{align}
In total we thus conclude that the $3\delta$-integral can be expressed as
\begin{align}
I_{3\delta} = 
\begin{cases}
	\frac{\pi}{4 \sigma} \Theta (-R_1^2 R_2^2 R_3^2 ),&  \sigma^2>0, \\
	\sim \delta(R_1^2) + \delta(R_2^2) + \delta(R_3^2), & \sigma^2 =0, \\
	\infty, & \sigma^2<0.
\end{cases}
\label{eq:I3deltaallcases}
\end{align}
We note that when using the result for $\sigma^2>0$ it can be useful to expand the theta-function according to
\begin{align}
\Theta \left(-R_1^2 R_2^2 R_3^2 \right) =
& +\Theta \left(-R_1^2 \right) \Theta \left(- R_2^2 \right) \Theta \left(-R_3^2 \right)
\nonumber\\
&+\Theta \left(-R_1^2 \right) \Theta \left(+ R_2^2 \right) \Theta \left(+R_3^2 \right)
\nonumber \\
&+ \Theta \left(+R_1^2 \right) \Theta \left(- R_2^2 \right) \Theta \left(+R_3^2 \right)
\nonumber\\ 
&+ \Theta \left(+R_1^2 \right) \Theta \left(+R_2^2 \right) \Theta \left(-R_3^2 \right).
\end{align}
\begin{widetext}
\section{Derivatives of the $3\delta$ Integral}
\label{sec:DerivativesI3delta}

In this appendix we explicitly evaluate the expressions for the second order derivatives of the triple-delta integral $I_{3\delta}$ for $\sigma^2>0$, cf.\ \eqref{eq:I3deltaallcases}. These enter into the three-body effective potential via \eqref{eq:PPP}.
A priori we find four terms
\begin{align}
\partial_1^\mu \partial_2^\nu I_{3\delta}(\sigma^2>0) =
&
+ \left(\partial_1^\mu \partial_2^\nu \frac{\pi}{4 \sigma} \right) \Theta \left(-R_1^2 R_2^2 R_3^2 \right) 
 +\left( \partial_1^\mu  \frac{\pi}{4 \sigma} \right) \partial_{2}^\nu  \Theta \left(-R_1^2 R_2^2 R_3^2 \right) \nonumber \\
&+\left(\partial_2^\nu  \frac{\pi}{4 \sigma} \right)  \partial_{1}^\mu  \Theta \left(-R_1^2 R_2^2 R_3^2 \right) 
+ \frac{\pi}{4 \sigma} \partial_1^\mu \partial_2^\nu \Theta \left(-R_1^2 R_2^2 R_3^2 \right),\label{eq_twoDerivative}
\end{align}
which evaluate to
\begin{align}
\Theta \left(-R_1^2 R_2^2 R_3^2 \right) \partial_1^\mu \partial_2^\nu\frac{\pi}{4 \sigma} 
=& \frac{\pi \Theta \left(-R_1^2 R_2^2 R_3^2 \right)}{4 \sigma^5} \Big[ 3 \left( R_1 \cdot R_2 \right) \left( R_1^\mu R_1^\nu R_2^2 + R_2^\mu R_2^\nu R_1^2 - R_1^\mu R_2^\nu R_1 \!\cdot\! R_2 - R_2^\mu R_1^\nu R_1 \!\cdot\! R_2  \right) 
\nonumber \\
&\qquad\qquad\qquad\qquad + \sigma^2 \left( \eta^{\mu\nu}  R_1 \!\cdot\! R_2 + R_1^\mu R_2^\nu + R_2^\mu R_1^\nu  \right)\Big] ,
\\
\frac{\pi}{4 \sigma} \partial_1^\mu \partial_2^\nu \Theta \left(-R_1^2 R_2^2 R_3^2 \right) 
=&  \frac{\pi}{4 \sigma} \left[ 2\eta^{\mu\nu}  \delta( R_3^2 ) + 4 R_3^\mu R_3^\nu \delta'( R_3^2 )  \right] \mathrm{sgn} (R_1^2 R_2^2) + \frac{8 \pi}{4 \sigma} \Big[ R_2^\mu R_1^\nu \delta( R_1^2 ) \delta( R_2^2 ) \mathrm{sgn}( R_3^2 )  
\nonumber\\
& \qquad - R_2^\mu R_3^\nu \delta( R_2^2 ) \delta( R_3^2 ) \mathrm{sgn}( R_1^2 ) - R_3^\mu R_1^\nu \delta( R_1^2 ) \delta( R_3^2 ) \mathrm{sgn}( R_2^2 ) \Big] ,
\label{eq_twoDTheta}
\\
 \left(\partial_1^\mu \frac{\pi}{4 \sigma} \right) \partial_{2}^\nu  \Theta \left(-R_1^2 R_2^2 R_3^2 \right)
 =& \frac{\pi}{4\sigma} \left[   \frac{4 R_1^\mu R_1^\nu}{R_2^2 - R_3^2}  \delta( R_1^2 ) \mathrm{sgn}( R_2^2 R_3^2 ) \right. \nonumber \\
&\qquad\quad \left.- \frac{4 R_2^\mu R_3^\nu}{R_1^2 - R_2^2}  \delta( R_3^2 ) \mathrm{sgn}( R_1^2 R_2^2 ) + R_3^\mu R_3^\nu \frac{R_1^2 + R_2^2}{\sigma^2}  \delta( R_3^2 ) \mathrm{sgn}( R_1^2 R_2^2 ) \right].
\end{align}
Here the last line also enters into \eqref{eq_twoDerivative} with the labels $1$ and $2$ interchanged.
Note the appearance of the derivative of the delta function in the first line of \eqref{eq_twoDTheta} that one could resolve
using $\delta'(R^{2}_{3})=-\delta(R^{2}_{3})/ R^{2}_{3}$.
Putting these terms together, eqn.~\eqref{eq_twoDerivative} then becomes (ordered by the number of delta functions)
\begin{align}		
&\partial_1^\mu \partial_2^\nu I_{3\delta} =
\nonumber\\
&\frac{\pi \Theta (-R_1^2 R_2^2 R_3^2 )}{4 \sigma^5} \Big[ 3 \left( R_1 \cdot R_2 \right) \left( R_1^\mu R_1^\nu R_2^2 + R_2^\mu R_2^\nu R_1^2 - R_1^\mu R_2^\nu R_1 \!\cdot\! R_2 - R_2^\mu R_1^\nu R_1 \!\cdot\! R_2  \right)  + \sigma^2 \left( \eta^{\mu\nu}  R_1 \!\cdot\! R_2 + R_1^\mu R_2^\nu + R_2^\mu R_1^\nu  \right)\Big]
\nonumber\\
&  + \frac{\pi}{4\sigma} \left[ +  \frac{4 R_1^\mu R_1^\nu}{R_2^2 - R_3^2}  \delta( R_1^2 ) \mathrm{sgn}( R_2^2 R_3^2 ) + \frac{4 R_2^\mu R_2^\nu}{R_1^2 - R_3^2}  \delta( R_2^2 ) \mathrm{sgn}( R_1^2 R_3^2 )\right]
 \nonumber \\
&
 + \frac{\pi}{4\sigma} \left[
  - \frac{4 R_1^\mu R_3^\nu}{R_2^2 - R_1^2}  \delta( R_3^2 ) - \frac{4 R_2^\mu R_3^\nu}{R_1^2 - R_2^2}  \delta( R_3^2 ) 
 + 2R_3^\mu R_3^\nu \frac{R_1^2 + R_2^2}{\sigma^2}  \delta( R_3^2 ) +2\eta^{\mu\nu}  \delta( R_3^2 )  + 4 R_3^\mu R_3^\nu \delta'( R_3^2 )  \right]
 \mathrm{sgn}( R_1^2 R_2^2 ) 
\nonumber \\
&  + \frac{8 \pi}{4 \sigma} \Big[ +R_2^\mu R_1^\nu \delta( R_1^2 ) \delta( R_2^2 ) \mathrm{sgn}( R_3^2 )- R_2^\mu R_3^\nu \delta( R_2^2 ) \delta( R_3^2 ) \mathrm{sgn}( R_1^2 )- R_3^\mu R_1^\nu \delta( R_1^2 ) \delta( R_3^2 ) \mathrm{sgn}( R_2^2 ) \Big].
\end{align}
Performing the PN expansion starting from this expression seems (also conceptually) much harder than working on the level of the integrand of $I_{3\delta}$ in \eqref{eq:PPP}. The latter is demonstrated in \Secref{sec:PNandBootstrap}.

\section{Details on 3PN}
\label{sec:3PNlong}
In the computation of the 3PN potential, we added the following total derivative to remove the dependence on the derivative of accelerations and possible spurious poles for $r_{ij} \rightarrow \infty$:
\begin{equation}
    L^{\text{td}} = \sum_{j\neq i} \frac{ G m_i m_j }{48\, c^8} \frac{\dd}{\dd t} \Big[ 
    r_{i j}(21\, \mathbf{a}_i \!\cdot\! \mathbf{v}_j -18\, \mathbf{a}_i\!\cdot\! \mathbf{v}_i )
    \left(\left(\mathbf{n}_{i j}\!\cdot\! \mathbf{v}_j\right){}^2 +\mathbf{v}_j^2\right)
    + r_{i j}\, \mathbf{n}_{i j} \!\cdot\! \mathbf{a}_i \mathbf{n}_{i j}\!\cdot\! \mathbf{v}_j 
    \left(\left(\mathbf{n}_{i j}\!\cdot\! \mathbf{v}_j\right){}^2-3 \mathbf{v}_j^2\right)
    \Big].
    \label{eq:tdacceleration}
\end{equation}
Due to its length, here we display only an excerpt of the genuine three-body contribution to the 3PN effective potential from the third line of \eqref{eq:PPP}. The full result is given in an ancillary file. The expression below is organized according to the rational functions of the spatial distances, where each function is multiplied by a sum of numerator structures 
that scale as $v^4$. Note that some numerator structures begin with the same terms but they do not agree. Evaluating the derivatives in \eqref{eq:L3PND} yields the  expression 
\begin{align}
&L^{\text{3PN}}_{(D)} =  \sum_{j\neq i} \sum_{k\neq i, j}  G^2 m_i m_j m_k \times 
\nonumber\\
& \bigg\{ \;\; {1 \over {(r_{ij}+r_{jk}+r_{ik})^2}} 
\left(
(\mathbf{n}_{ik}\!\cdot\!\mathbf{v}_i)
\left( {16\over 3} (\mathbf{n}_{ij}\!\cdot\!\mathbf{v}_i)^3
-12 (\mathbf{n}_{ij}\!\cdot\!\mathbf{v}_i)^2 (\mathbf{n}_{ij}\!\cdot\!\mathbf{v}_j) 
+{20\over 3}(\mathbf{n}_{ij}\!\cdot\!\mathbf{v}_i) (\mathbf{n}_{ij}\!\cdot\!\mathbf{v}_j)^2 \right)
+ \text{245 terms}
 \right)\nonumber\\
& \;\; + {1\over {r_{ij} (r_{ij}+r_{jk}+r_{ik})} } 
\left(  \mathbf{v}_i^2\left(
{16\over 3} (\mathbf{n}_{ij}\!\cdot\! \mathbf{v}_i)^2  
- 11 (\mathbf{n}_{ij}\!\cdot\!\mathbf{v}_i ) (\mathbf{n}_{ij}\!\cdot\!\mathbf{v}_j ) 
+ {16\over 3} (\mathbf{n}_{ij}\!\cdot\! \mathbf{v}_j)^2\right)
+ \text{45 terms}
\right)  \nonumber  \\
& \;\; - {r_{ij} \over (r_{ij}+r_{jk}+r_{ik})^3 } \left( 
{8\over 3} (\mathbf{n}_{ij}\!\cdot\! \mathbf{v}_i)^4 
- 6 (\mathbf{n}_{ij}\!\cdot\!\mathbf{v}_i)^3 (\mathbf{n}_{ij}\!\cdot\!\mathbf{v}_j) 
+ {8\over3} (\mathbf{n}_{ij}\!\cdot\!\mathbf{v}_i)^2 (\mathbf{n}_{ij}\!\cdot\!\mathbf{v}_j)^2 +\text{286 terms}
\right) \nonumber\\ 
& \;\; - {r_{ik} r_{jk} \over {r_{ij} (r_{ij}+r_{jk}+r_{ik})^3 }} 
\left( (\mathbf{n}_{ik}\!\cdot\!\mathbf{v}_k)^2 \left(
16 (\mathbf{n}_{ij}\!\cdot\!\mathbf{v}_i)^2 
-36 (\mathbf{n}_{ij}\!\cdot\!\mathbf{v}_i) (\mathbf{n}_{ij}\!\cdot\!\mathbf{v}_j) 
+ 16 (\mathbf{n}_{ij}\!\cdot\!\mathbf{v}_j)^2 \right) +\text{69 terms}
\right) \nonumber\\
& \;\; -{r_{ij}^2 \over {(r_{ij}+r_{jk}+r_{ik})^4} } \Big(
8 (\mathbf{n}_{ij}\!\cdot\!\mathbf{v}_i)^4 
-18 (\mathbf{n}_{ij}\!\cdot\!\mathbf{v}_i)^3 (\mathbf{n}_{ij}\!\cdot\!\mathbf{v}_j)
+ 16 (\mathbf{n}_{ij}\!\cdot\!\mathbf{v}_i)^2 (\mathbf{n}_{ij}\!\cdot\!\mathbf{v}_j)^2 +\text{143 terms} 
\Big) \nonumber\\
& \;\; +{r_{ik} r_{jk} \over {(r_{ij}+r_{jk}+r_{ik})^4} } \left(  (\mathbf{n}_{ik}\!\cdot\!\mathbf{v}_k)^2
\left(
16 (\mathbf{n}_{ij}\!\cdot\!\mathbf{v}_i)^2 
-36 (\mathbf{n}_{ij}\!\cdot\!\mathbf{v}_i) (\mathbf{n}_{ij}\!\cdot\!\mathbf{v}_j)
+ 16 (\mathbf{n}_{ij}\!\cdot\!\mathbf{v}_j)^2\right) + \text{114 terms}
\right) \nonumber\\
& \;\;-\left[ {1\over {r_{ik} (r_{ij}+r_{jk}+r_{ik})}}
\left({4\over3}\mathbf{v}_i^4
-{4\over3}\left((\mathbf{n}_{ik}\!\cdot\!\mathbf{v}_i)^2- (\mathbf{n}_{ik}\!\cdot\!\mathbf{v}_k)^2\right)\mathbf{v}_i^2
+\text{20 terms}\right) 
+ (i\leftrightarrow j)\right] \nonumber\\
&\;\; -\left[ {r_{ik} \over {r_{jk} (r_{ij}+r_{jk}+r_{ik})^2}} 
\left(
 (\mathbf{n}_{jk}\!\cdot\!\mathbf{v}_k)^2 \left({8\over 3} (\mathbf{n}_{ij}\!\cdot\!\mathbf{v}_i)^2  -6 (\mathbf{n}_{ij}\!\cdot\!\mathbf{v}_i) (\mathbf{n}_{ij}\!\cdot\!\mathbf{v}_j) +{8\over 3} (\mathbf{n}_{ij}\!\cdot\!\mathbf{v}_j)^2 \right) 
 + \text{23 terms}\right) 
 + (i\leftrightarrow j)\right] \nonumber\\
& \;\; + \left[ {r_{ij} \over {r_{ik} (r_{ij} + r_{jk} + r_{ik})^2 }} \left(
 (\mathbf{n}_{jk}\!\cdot\!\mathbf{v}_k)^2
 \left(
{8\over 3} (\mathbf{n}_{ij}\!\cdot\!\mathbf{v}_i)^2  -6(\mathbf{n}_{ij}\!\cdot\!\mathbf{v}_i) (\mathbf{n}_{ij}\!\cdot\!\mathbf{v}_j) +{8\over 3} (\mathbf{n}_{ij}\!\cdot\!\mathbf{v}_j)^2
\right)
+ \text{47 terms}
\right) 
+(i\leftrightarrow j)
\right]\nonumber\\
& \;\;- \left[ {r_{ik} \over r_{ij} (r_{ij}+r_{jk}+r_{ik})^2} \left( 8 (\mathbf{n}_{ij}\!\cdot\!\mathbf{v}_i)^4-18(\mathbf{n}_{ij}\!\cdot\!\mathbf{v}_i)^3 (\mathbf{n}_{ij}\!\cdot\!\mathbf{v}_j)+12 (\mathbf{n}_{ij}\!\cdot\!\mathbf{v}_i)^2 (\mathbf{n}_{ij}\!\cdot\!\mathbf{v}_j)^2 +\text{93 terms}\right)+(i\leftrightarrow j)\right] \nonumber\\
& \;\; - \left[ {r_{ik} \over (r_{ij}+r_{jk}+r_{ik})^3} 
\left(
16 (\mathbf{n}_{ij}\!\cdot\!\mathbf{v}_i)^4 
- 36(\mathbf{n}_{ij}\!\cdot\!\mathbf{v}_i)^3 (\mathbf{n}_{ij}\!\cdot\!\mathbf{v}_j)
+ 24 (\mathbf{n}_{ij}\!\cdot\!\mathbf{v}_i)^2 (\mathbf{n}_{ij}\!\cdot\!\mathbf{v}_j)^2 +\text{285 terms}
\right)
+ (i\leftrightarrow j)
\right]\nonumber\\
& \;\; +\left[ {r^2_{ik} \over {r_{jk} (r_{ij}+r_{jk}+r_{ik})^3} } \left(
(\mathbf{n}_{jk}\!\cdot\!\mathbf{v}_k)^2
\left(
{16\over 3} (\mathbf{n}_{ij}\!\cdot\!\mathbf{v}_i)^2 
-12 (\mathbf{n}_{ij}\!\cdot\!\mathbf{v}_i) (\mathbf{n}_{ij}\!\cdot\!\mathbf{v}_j)
+{16\over 3} (\mathbf{n}_{ij}\!\cdot\!\mathbf{v}_j)^2
\right)+\text{46 terms}
 \right)
+(i\leftrightarrow j)
\right] \nonumber\\
& \;\; - \left[
{r_{ij}^2 \over {r_{ik} (r_{ij}+r_{jk}+r_{ik})^3}}
\left(
 (\mathbf{n}_{ik}\!\cdot\!\mathbf{v}_i)^2
 \left(
{8\over 3} (\mathbf{n}_{ij}\!\cdot\!\mathbf{v}_i)^2  -6(\mathbf{n}_{ij}\!\cdot\!\mathbf{v}_i) (\mathbf{n}_{ij}\!\cdot\!\mathbf{v}_j) +{8\over 3} (\mathbf{n}_{ij}\!\cdot\!\mathbf{v}_j)^2
\right) 
+\text{58 terms}
\right) 
+ (i\leftrightarrow j)
\right] \nonumber\\
& \;\; - \left[
{r_{ij} r_{jk} \over {r_{ik} (r_{ij}+r_{jk}+r_{ik})^3}}
\left(
 (\mathbf{n}_{ik}\!\cdot\!\mathbf{v}_i)^2
 \left(
{8\over 3} (\mathbf{n}_{ij}\!\cdot\!\mathbf{v}_i)^2  -6(\mathbf{n}_{ij}\!\cdot\!\mathbf{v}_i) (\mathbf{n}_{ij}\!\cdot\!\mathbf{v}_j) +{8\over 3} (\mathbf{n}_{ij}\!\cdot\!\mathbf{v}_j)^2
\right) 
+\text{72 terms}
\right) 
+ (i\leftrightarrow j)
\right] \nonumber\\
& \;\; - \left[ 
{r^2_{ik} \over {r_{ij} (r_{ij}+r_{jk}+r_{ik})^3}} 
\left(
 (\mathbf{n}_{ik}\!\cdot\!\mathbf{v}_k)^2
 \left(
 {40\over3}  (\mathbf{n}_{ik}\!\cdot\!\mathbf{v}_i)^2
 - 30 (\mathbf{n}_{ik}\!\cdot\!\mathbf{v}_i)  (\mathbf{n}_{ik}\!\cdot\!\mathbf{v}_j)
 +{40\over3}  (\mathbf{n}_{ik}\!\cdot\!\mathbf{v}_j)^2
 \right)
 +\text{75 terms}
\right) 
+(i\leftrightarrow j)
\right] \nonumber\\
& \;\; + \left[
{r_{ik}^2 \over {(r_{ij}+r_{jk}+r_{ik})^4}}
\left(
 (\mathbf{n}_{ik}\!\cdot\!\mathbf{v}_k)^2
\left(
16 (\mathbf{n}_{ij}\!\cdot\!\mathbf{v}_i)^2-36 (\mathbf{n}_{ij}\!\cdot\!\mathbf{v}_i) (\mathbf{n}_{ij}\!\cdot\!\mathbf{v}_j) 
+ 16 (\mathbf{n}_{ij}\!\cdot\!\mathbf{v}_j)^2
\right)
 + \text{109 terms}
\right) 
+(i\leftrightarrow j)
\right] \nonumber\\
& \;\; -\left[
{r_{ij} r_{ik} \over {(r_{ij}+r_{jk}+r_{ik})^4}}
\left(
16 (\mathbf{n}_{ij}\!\cdot\!\mathbf{v}_i)^4 
- 36(\mathbf{n}_{ij}\!\cdot\!\mathbf{v}_i)^3 (\mathbf{n}_{ij}\!\cdot\!\mathbf{v}_j)
+ 24 (\mathbf{n}_{ij}\!\cdot\!\mathbf{v}_i)^2 (\mathbf{n}_{ij}\!\cdot\!\mathbf{v}_j)^2
+\text{174 terms}
\right) 
+ (i\leftrightarrow j)
\right]\nonumber\\
& \;\; -\left[ {r_{ik}^2 (r_{ij}+r_{jk}) \over {r_{ij} r_{jk} (r_{ij}+r_{jk}+r_{ik})^3}} \left(
 (\mathbf{n}_{jk}\!\cdot\!\mathbf{v}_k)^2 \left({8\over 3} (\mathbf{n}_{ij}\!\cdot\!\mathbf{v}_i)^2  -6 (\mathbf{n}_{ij}\!\cdot\!\mathbf{v}_i) (\mathbf{n}_{ij}\!\cdot\!\mathbf{v}_j) +{8\over 3} (\mathbf{n}_{ij}\!\cdot\!\mathbf{v}_j)^2 \right) 
+\text{21 terms}
\right)
+ (i \leftrightarrow j)
\right]\nonumber\\
& \;\; +\left[
{r_{ik}^2(r_{ik}+2r_{jk}) \over  {r_{ij} r_{jk} (r_{ij}+r_{jk}+r_{ik})^4}}
\left(
(\mathbf{n}_{ik}\!\cdot\!\mathbf{v}_k)^2 
\left(
8(\mathbf{n}_{ij}\!\cdot\!\mathbf{v}_i)^2
-36 (\mathbf{n}_{ij}\!\cdot\!\mathbf{v}_i) (\mathbf{n}_{ij}\!\cdot\!\mathbf{v}_j)
+8 (\mathbf{n}_{ij}\!\cdot\!\mathbf{v}_j)^2
\right)+\text{33 terms}
\right)
+(i\leftrightarrow j)
\right] \bigg\}.
\label{eq:3PNlong}
\end{align}
\end{widetext}

\bibliography{ThreeBodyPotential}

\begin{thebibliography}{81}%
\makeatletter
\providecommand \@ifxundefined [1]{%
 \@ifx{#1\undefined}
}%
\providecommand \@ifnum [1]{%
 \ifnum #1\expandafter \@firstoftwo
 \else \expandafter \@secondoftwo
 \fi
}%
\providecommand \@ifx [1]{%
 \ifx #1\expandafter \@firstoftwo
 \else \expandafter \@secondoftwo
 \fi
}%
\providecommand \natexlab [1]{#1}%
\providecommand \enquote  [1]{``#1''}%
\providecommand \bibnamefont  [1]{#1}%
\providecommand \bibfnamefont [1]{#1}%
\providecommand \citenamefont [1]{#1}%
\providecommand \href@noop [0]{\@secondoftwo}%
\providecommand \href [0]{\begingroup \@sanitize@url \@href}%
\providecommand \@href[1]{\@@startlink{#1}\@@href}%
\providecommand \@@href[1]{\endgroup#1\@@endlink}%
\providecommand \@sanitize@url [0]{\catcode `\\12\catcode `\$12\catcode
  `\&12\catcode `\#12\catcode `\^12\catcode `\_12\catcode `\%12\relax}%
\providecommand \@@startlink[1]{}%
\providecommand \@@endlink[0]{}%
\providecommand \url  [0]{\begingroup\@sanitize@url \@url }%
\providecommand \@url [1]{\endgroup\@href {#1}{\urlprefix }}%
\providecommand \urlprefix  [0]{URL }%
\providecommand \Eprint [0]{\href }%
\providecommand \doibase [0]{http://dx.doi.org/}%
\providecommand \selectlanguage [0]{\@gobble}%
\providecommand \bibinfo  [0]{\@secondoftwo}%
\providecommand \bibfield  [0]{\@secondoftwo}%
\providecommand \translation [1]{[#1]}%
\providecommand \BibitemOpen [0]{}%
\providecommand \bibitemStop [0]{}%
\providecommand \bibitemNoStop [0]{.\EOS\space}%
\providecommand \EOS [0]{\spacefactor3000\relax}%
\providecommand \BibitemShut  [1]{\csname bibitem#1\endcsname}%
\let\auto@bib@innerbib\@empty
\bibitem [{\citenamefont {Poincar\'e}(1890)}]{Poincare:1890}%
  \BibitemOpen
  \bibfield  {author} {\bibinfo {author} {\bibfnamefont {Henry}\ \bibnamefont
  {Poincar\'e}},\ }\bibfield  {title} {\enquote {\bibinfo {title} {{Sur le
  probl\`eme des trois corps et les \'equations de la dynamique}},}\ }\href
  {\doibase 10.2307/1968714} {\bibfield  {journal} {\bibinfo  {journal} {Acta
  Mathematica}\ }\textbf {\bibinfo {volume} {13}},\ \bibinfo {pages} {1--270}
  (\bibinfo {year} {1890})}\BibitemShut {NoStop}%
\bibitem [{\citenamefont {Liu}(2007)}]{liu}%
  \BibitemOpen
  \bibfield  {author} {\bibinfo {author} {\bibfnamefont {Cixin}\ \bibnamefont
  {Liu}},\ }\href@noop {} {\emph {\bibinfo {title} {{The three-body
  problem}}}}\ (\bibinfo  {publisher} {Tor Books},\ \bibinfo {year}
  {2007})\BibitemShut {NoStop}%
\bibitem [{\citenamefont {Will}(2014)}]{Will:2013cza}%
  \BibitemOpen
  \bibfield  {author} {\bibinfo {author} {\bibfnamefont {Clifford~M.}\
  \bibnamefont {Will}},\ }\bibfield  {title} {\enquote {\bibinfo {title}
  {{Incorporating post-Newtonian effects in $N$-body dynamics}},}\ }\href
  {\doibase 10.1103/PhysRevD.89.044043} {\bibfield  {journal} {\bibinfo
  {journal} {Phys. Rev. D}\ }\textbf {\bibinfo {volume} {89}},\ \bibinfo
  {pages} {044043} (\bibinfo {year} {2014})},\ \bibinfo {note} {[Erratum:
  Phys.Rev.D 91, 029902 (2015)]},\ \Eprint {http://arxiv.org/abs/1312.1289}
  {arXiv:1312.1289 [astro-ph.GA]} \BibitemShut {NoStop}%
\bibitem [{\citenamefont {Abbott}\ \emph {et~al.}(2016)\citenamefont {Abbott}
  \emph {et~al.}}]{Abbott:2016blz}%
  \BibitemOpen
  \bibfield  {author} {\bibinfo {author} {\bibfnamefont {B.P.}\ \bibnamefont
  {Abbott}} \emph {et~al.} (\bibinfo {collaboration} {LIGO Scientific,
  Virgo}),\ }\bibfield  {title} {\enquote {\bibinfo {title} {{Observation of
  Gravitational Waves from a Binary Black Hole Merger}},}\ }\href {\doibase
  10.1103/PhysRevLett.116.061102} {\bibfield  {journal} {\bibinfo  {journal}
  {Phys. Rev. Lett.}\ }\textbf {\bibinfo {volume} {116}},\ \bibinfo {pages}
  {061102} (\bibinfo {year} {2016})},\ \Eprint
  {http://arxiv.org/abs/1602.03837} {arXiv:1602.03837 [gr-qc]} \BibitemShut
  {NoStop}%
\bibitem [{\citenamefont {Abbott}\ \emph {et~al.}(2017)\citenamefont {Abbott}
  \emph {et~al.}}]{TheLIGOScientific:2017qsa}%
  \BibitemOpen
  \bibfield  {author} {\bibinfo {author} {\bibfnamefont {B.P.}\ \bibnamefont
  {Abbott}} \emph {et~al.} (\bibinfo {collaboration} {LIGO Scientific,
  Virgo}),\ }\bibfield  {title} {\enquote {\bibinfo {title} {{GW170817:
  Observation of Gravitational Waves from a Binary Neutron Star Inspiral}},}\
  }\href {\doibase 10.1103/PhysRevLett.119.161101} {\bibfield  {journal}
  {\bibinfo  {journal} {Phys. Rev. Lett.}\ }\textbf {\bibinfo {volume} {119}},\
  \bibinfo {pages} {161101} (\bibinfo {year} {2017})},\ \Eprint
  {http://arxiv.org/abs/1710.05832} {arXiv:1710.05832 [gr-qc]} \BibitemShut
  {NoStop}%
\bibitem [{\citenamefont {Abbott}\ \emph {et~al.}(2019)\citenamefont {Abbott}
  \emph {et~al.}}]{LIGOScientific:2018mvr}%
  \BibitemOpen
  \bibfield  {author} {\bibinfo {author} {\bibfnamefont {B.P.}\ \bibnamefont
  {Abbott}} \emph {et~al.} (\bibinfo {collaboration} {LIGO Scientific,
  Virgo}),\ }\bibfield  {title} {\enquote {\bibinfo {title} {{GWTC-1: A
  Gravitational-Wave Transient Catalog of Compact Binary Mergers Observed by
  LIGO and Virgo during the First and Second Observing Runs}},}\ }\href
  {\doibase 10.1103/PhysRevX.9.031040} {\bibfield  {journal} {\bibinfo
  {journal} {Phys. Rev. X}\ }\textbf {\bibinfo {volume} {9}},\ \bibinfo {pages}
  {031040} (\bibinfo {year} {2019})},\ \Eprint
  {http://arxiv.org/abs/1811.12907} {arXiv:1811.12907 [astro-ph.HE]}
  \BibitemShut {NoStop}%
\bibitem [{\citenamefont {Asada}(2009)}]{Asada:2009qs}%
  \BibitemOpen
  \bibfield  {author} {\bibinfo {author} {\bibfnamefont {Hideki}\ \bibnamefont
  {Asada}},\ }\bibfield  {title} {\enquote {\bibinfo {title} {{Gravitational
  wave forms for a three-body system in Lagrange's orbit: Parameter
  determinations and a binary source test}},}\ }\href {\doibase
  10.1103/PhysRevD.80.064021} {\bibfield  {journal} {\bibinfo  {journal} {Phys.
  Rev. D}\ }\textbf {\bibinfo {volume} {80}},\ \bibinfo {pages} {064021}
  (\bibinfo {year} {2009})},\ \Eprint {http://arxiv.org/abs/0907.5091}
  {arXiv:0907.5091 [gr-qc]} \BibitemShut {NoStop}%
\bibitem [{\citenamefont {Galaviz}\ and\ \citenamefont
  {Bruegmann}(2011)}]{Galaviz:2010te}%
  \BibitemOpen
  \bibfield  {author} {\bibinfo {author} {\bibfnamefont {Pablo}\ \bibnamefont
  {Galaviz}}\ and\ \bibinfo {author} {\bibfnamefont {Bernd}\ \bibnamefont
  {Bruegmann}},\ }\bibfield  {title} {\enquote {\bibinfo {title}
  {{Characterization of the gravitational wave emission of three black
  holes}},}\ }\href {\doibase 10.1103/PhysRevD.83.084013} {\bibfield  {journal}
  {\bibinfo  {journal} {Phys. Rev. D}\ }\textbf {\bibinfo {volume} {83}},\
  \bibinfo {pages} {084013} (\bibinfo {year} {2011})},\ \Eprint
  {http://arxiv.org/abs/1012.4423} {arXiv:1012.4423 [gr-qc]} \BibitemShut
  {NoStop}%
\bibitem [{\citenamefont {Meiron}\ \emph {et~al.}(2017)\citenamefont {Meiron},
  \citenamefont {Kocsis},\ and\ \citenamefont {Loeb}}]{Meiron:2016ipr}%
  \BibitemOpen
  \bibfield  {author} {\bibinfo {author} {\bibfnamefont {Yohai}\ \bibnamefont
  {Meiron}}, \bibinfo {author} {\bibfnamefont {Bence}\ \bibnamefont {Kocsis}},
  \ and\ \bibinfo {author} {\bibfnamefont {Abraham}\ \bibnamefont {Loeb}},\
  }\bibfield  {title} {\enquote {\bibinfo {title} {{Detecting triple systems
  with gravitational wave observations}},}\ }\href {\doibase
  10.3847/1538-4357/834/2/200} {\bibfield  {journal} {\bibinfo  {journal}
  {Astrophys. J.}\ }\textbf {\bibinfo {volume} {834}},\ \bibinfo {pages} {200}
  (\bibinfo {year} {2017})},\ \Eprint {http://arxiv.org/abs/1604.02148}
  {arXiv:1604.02148 [astro-ph.HE]} \BibitemShut {NoStop}%
\bibitem [{\citenamefont {Bonetti}\ \emph {et~al.}(2017)\citenamefont
  {Bonetti}, \citenamefont {Barausse}, \citenamefont {Faye}, \citenamefont
  {Haardt},\ and\ \citenamefont {Sesana}}]{Bonetti:2017hnb}%
  \BibitemOpen
  \bibfield  {author} {\bibinfo {author} {\bibfnamefont {Matteo}\ \bibnamefont
  {Bonetti}}, \bibinfo {author} {\bibfnamefont {Enrico}\ \bibnamefont
  {Barausse}}, \bibinfo {author} {\bibfnamefont {Guillaume}\ \bibnamefont
  {Faye}}, \bibinfo {author} {\bibfnamefont {Francesco}\ \bibnamefont
  {Haardt}}, \ and\ \bibinfo {author} {\bibfnamefont {Alberto}\ \bibnamefont
  {Sesana}},\ }\bibfield  {title} {\enquote {\bibinfo {title} {{About
  gravitational-wave generation by a three-body system}},}\ }\href {\doibase
  10.1088/1361-6382/aa8da5} {\bibfield  {journal} {\bibinfo  {journal} {Class.
  Quant. Grav.}\ }\textbf {\bibinfo {volume} {34}},\ \bibinfo {pages} {215004}
  (\bibinfo {year} {2017})},\ \Eprint {http://arxiv.org/abs/1707.04902}
  {arXiv:1707.04902 [gr-qc]} \BibitemShut {NoStop}%
\bibitem [{\citenamefont {Lim}\ and\ \citenamefont
  {Rodriguez}(2020)}]{Lim:2020cvm}%
  \BibitemOpen
  \bibfield  {author} {\bibinfo {author} {\bibfnamefont {Halston}\ \bibnamefont
  {Lim}}\ and\ \bibinfo {author} {\bibfnamefont {Carl~L.}\ \bibnamefont
  {Rodriguez}},\ }\bibfield  {title} {\enquote {\bibinfo {title} {{Relativistic
  three-body effects in hierarchical triples}},}\ }\href {\doibase
  10.1103/PhysRevD.102.064033} {\bibfield  {journal} {\bibinfo  {journal}
  {Phys. Rev. D}\ }\textbf {\bibinfo {volume} {102}},\ \bibinfo {pages}
  {064033} (\bibinfo {year} {2020})},\ \Eprint
  {http://arxiv.org/abs/2001.03654} {arXiv:2001.03654 [astro-ph.HE]}
  \BibitemShut {NoStop}%
\bibitem [{\citenamefont {Einstein}\ \emph {et~al.}(1938)\citenamefont
  {Einstein}, \citenamefont {Infeld},\ and\ \citenamefont
  {Hoffmann}}]{Einstein:1938yz}%
  \BibitemOpen
  \bibfield  {author} {\bibinfo {author} {\bibfnamefont {Albert}\ \bibnamefont
  {Einstein}}, \bibinfo {author} {\bibfnamefont {L.}~\bibnamefont {Infeld}}, \
  and\ \bibinfo {author} {\bibfnamefont {B.}~\bibnamefont {Hoffmann}},\
  }\bibfield  {title} {\enquote {\bibinfo {title} {{The Gravitational equations
  and the problem of motion}},}\ }\href {\doibase 10.2307/1968714} {\bibfield
  {journal} {\bibinfo  {journal} {Annals Math.}\ }\textbf {\bibinfo {volume}
  {39}},\ \bibinfo {pages} {65--100} (\bibinfo {year} {1938})}\BibitemShut
  {NoStop}%
\bibitem [{\citenamefont {Eddington}\ and\ \citenamefont
  {Clark}(1938)}]{eddington1938problem}%
  \BibitemOpen
  \bibfield  {author} {\bibinfo {author} {\bibfnamefont {Arthur~Stanley}\
  \bibnamefont {Eddington}}\ and\ \bibinfo {author} {\bibfnamefont
  {Gordon~Leonard}\ \bibnamefont {Clark}},\ }\bibfield  {title} {\enquote
  {\bibinfo {title} {The problem of n bodies in general relativity theory},}\
  }\href@noop {} {\bibfield  {journal} {\bibinfo  {journal} {Proceedings of the
  Royal Society of London. Series A. Mathematical and Physical Sciences}\
  }\textbf {\bibinfo {volume} {166}},\ \bibinfo {pages} {465--475} (\bibinfo
  {year} {1938})}\BibitemShut {NoStop}%
\bibitem [{\citenamefont {Landau}(1987)}]{Landau:1987gn}%
  \BibitemOpen
  \bibfield  {author} {\bibinfo {author} {\bibfnamefont {L.D.}\ \bibnamefont
  {Landau}},\ }\href@noop {} {\emph {\bibinfo {title} {{Textbook on Theoretical
  Physics. Vol.2: Classical Field Theory (In German)}}}},\ edited by\ \bibinfo
  {editor} {\bibfnamefont {E.M.}\ \bibnamefont {Lifshitz}}, \bibinfo {editor}
  {\bibfnamefont {H.G.}\ \bibnamefont {Schopf}}, \ and\ \bibinfo {editor}
  {\bibfnamefont {P.}~\bibnamefont {Ziesche}}\ (\bibinfo {year}
  {1987})\BibitemShut {NoStop}%
\bibitem [{\citenamefont {Gultekin}\ \emph {et~al.}(2006)\citenamefont
  {Gultekin}, \citenamefont {Coleman~Miller},\ and\ \citenamefont
  {Hamilton}}]{Gultekin:2005fd}%
  \BibitemOpen
  \bibfield  {author} {\bibinfo {author} {\bibfnamefont {Kayhan}\ \bibnamefont
  {Gultekin}}, \bibinfo {author} {\bibfnamefont {M.}~\bibnamefont
  {Coleman~Miller}}, \ and\ \bibinfo {author} {\bibfnamefont {Douglas~P.}\
  \bibnamefont {Hamilton}},\ }\bibfield  {title} {\enquote {\bibinfo {title}
  {{Three-body dynamics with gravitational wave emission}},}\ }\href {\doibase
  10.1086/499917} {\bibfield  {journal} {\bibinfo  {journal} {Astrophys. J.}\
  }\textbf {\bibinfo {volume} {640}},\ \bibinfo {pages} {156--166} (\bibinfo
  {year} {2006})},\ \Eprint {http://arxiv.org/abs/astro-ph/0509885}
  {arXiv:astro-ph/0509885} \BibitemShut {NoStop}%
\bibitem [{\citenamefont {Iwasawa}\ \emph {et~al.}(2006)\citenamefont
  {Iwasawa}, \citenamefont {Funato},\ and\ \citenamefont
  {Makino}}]{Iwasawa:2005zh}%
  \BibitemOpen
  \bibfield  {author} {\bibinfo {author} {\bibfnamefont {Masaki}\ \bibnamefont
  {Iwasawa}}, \bibinfo {author} {\bibfnamefont {Yoko}\ \bibnamefont {Funato}},
  \ and\ \bibinfo {author} {\bibfnamefont {Junichiro}\ \bibnamefont {Makino}},\
  }\bibfield  {title} {\enquote {\bibinfo {title} {{Evolution of massive
  blackhole triples. 1. equal-mass binary-single systems}},}\ }\href {\doibase
  10.1086/507473} {\bibfield  {journal} {\bibinfo  {journal} {Astrophys. J.}\
  }\textbf {\bibinfo {volume} {651}},\ \bibinfo {pages} {1059--1067} (\bibinfo
  {year} {2006})},\ \Eprint {http://arxiv.org/abs/astro-ph/0511391}
  {arXiv:astro-ph/0511391} \BibitemShut {NoStop}%
\bibitem [{\citenamefont {Hoffman}\ and\ \citenamefont
  {Loeb}(2007)}]{Hoffman:2006iq}%
  \BibitemOpen
  \bibfield  {author} {\bibinfo {author} {\bibfnamefont {Loren}\ \bibnamefont
  {Hoffman}}\ and\ \bibinfo {author} {\bibfnamefont {Abraham}\ \bibnamefont
  {Loeb}},\ }\bibfield  {title} {\enquote {\bibinfo {title} {{Dynamics of
  triple black hole systems in hierarchically merging massive galaxies}},}\
  }\href {\doibase 10.1111/j.1365-2966.2007.11694.x} {\bibfield  {journal}
  {\bibinfo  {journal} {Mon. Not. Roy. Astron. Soc.}\ }\textbf {\bibinfo
  {volume} {377}},\ \bibinfo {pages} {957--976} (\bibinfo {year} {2007})},\
  \Eprint {http://arxiv.org/abs/astro-ph/0612517} {arXiv:astro-ph/0612517}
  \BibitemShut {NoStop}%
\bibitem [{\citenamefont {Gupta}\ \emph {et~al.}(2020)\citenamefont {Gupta},
  \citenamefont {Suzuki}, \citenamefont {Okawa},\ and\ \citenamefont
  {Maeda}}]{Gupta:2019unn}%
  \BibitemOpen
  \bibfield  {author} {\bibinfo {author} {\bibfnamefont {Priti}\ \bibnamefont
  {Gupta}}, \bibinfo {author} {\bibfnamefont {Haruka}\ \bibnamefont {Suzuki}},
  \bibinfo {author} {\bibfnamefont {Hirotada}\ \bibnamefont {Okawa}}, \ and\
  \bibinfo {author} {\bibfnamefont {Kei-ichi}\ \bibnamefont {Maeda}},\
  }\bibfield  {title} {\enquote {\bibinfo {title} {{Gravitational Waves from
  Hierarchical Triple Systems with Kozai-Lidov Oscillation}},}\ }\href
  {\doibase 10.1103/PhysRevD.101.104053} {\bibfield  {journal} {\bibinfo
  {journal} {Phys. Rev. D}\ }\textbf {\bibinfo {volume} {101}},\ \bibinfo
  {pages} {104053} (\bibinfo {year} {2020})},\ \Eprint
  {http://arxiv.org/abs/1911.11318} {arXiv:1911.11318 [gr-qc]} \BibitemShut
  {NoStop}%
\bibitem [{\citenamefont {Lousto}\ and\ \citenamefont
  {Nakano}(2008)}]{Lousto:2007ji}%
  \BibitemOpen
  \bibfield  {author} {\bibinfo {author} {\bibfnamefont {Carlos~O.}\
  \bibnamefont {Lousto}}\ and\ \bibinfo {author} {\bibfnamefont {Hiroyuki}\
  \bibnamefont {Nakano}},\ }\bibfield  {title} {\enquote {\bibinfo {title}
  {{Three-body equations of motion in successive post-Newtonian
  approximations}},}\ }\href {\doibase 10.1088/0264-9381/25/19/195019}
  {\bibfield  {journal} {\bibinfo  {journal} {Class. Quant. Grav.}\ }\textbf
  {\bibinfo {volume} {25}},\ \bibinfo {pages} {195019} (\bibinfo {year}
  {2008})},\ \Eprint {http://arxiv.org/abs/0710.5542} {arXiv:0710.5542 [gr-qc]}
  \BibitemShut {NoStop}%
\bibitem [{\citenamefont {Galaviz}(2011)}]{Galaviz:2011qb}%
  \BibitemOpen
  \bibfield  {author} {\bibinfo {author} {\bibfnamefont {Pablo}\ \bibnamefont
  {Galaviz}},\ }\bibfield  {title} {\enquote {\bibinfo {title} {{Stability and
  chaos of hierarchical three black hole configurations}},}\ }\href {\doibase
  10.1103/PhysRevD.84.104038} {\bibfield  {journal} {\bibinfo  {journal} {Phys.
  Rev. D}\ }\textbf {\bibinfo {volume} {84}},\ \bibinfo {pages} {104038}
  (\bibinfo {year} {2011})},\ \Eprint {http://arxiv.org/abs/1108.4485}
  {arXiv:1108.4485 [gr-qc]} \BibitemShut {NoStop}%
\bibitem [{\citenamefont {Naoz}\ \emph {et~al.}(2013)\citenamefont {Naoz},
  \citenamefont {Kocsis}, \citenamefont {Loeb},\ and\ \citenamefont
  {Yunes}}]{Naoz:2012bx}%
  \BibitemOpen
  \bibfield  {author} {\bibinfo {author} {\bibfnamefont {Smadar}\ \bibnamefont
  {Naoz}}, \bibinfo {author} {\bibfnamefont {Bence}\ \bibnamefont {Kocsis}},
  \bibinfo {author} {\bibfnamefont {Abraham}\ \bibnamefont {Loeb}}, \ and\
  \bibinfo {author} {\bibfnamefont {Nicolas}\ \bibnamefont {Yunes}},\
  }\bibfield  {title} {\enquote {\bibinfo {title} {{Resonant Post-Newtonian
  Eccentricity Excitation in Hierarchical Three-body Systems}},}\ }\href
  {\doibase 10.1088/0004-637X/773/2/187} {\bibfield  {journal} {\bibinfo
  {journal} {Astrophys. J.}\ }\textbf {\bibinfo {volume} {773}},\ \bibinfo
  {pages} {187} (\bibinfo {year} {2013})},\ \Eprint
  {http://arxiv.org/abs/1206.4316} {arXiv:1206.4316 [astro-ph.SR]} \BibitemShut
  {NoStop}%
\bibitem [{\citenamefont {Bonetti}\ \emph {et~al.}(2016)\citenamefont
  {Bonetti}, \citenamefont {Haardt}, \citenamefont {Sesana},\ and\
  \citenamefont {Barausse}}]{Bonetti:2016eif}%
  \BibitemOpen
  \bibfield  {author} {\bibinfo {author} {\bibfnamefont {Matteo}\ \bibnamefont
  {Bonetti}}, \bibinfo {author} {\bibfnamefont {Francesco}\ \bibnamefont
  {Haardt}}, \bibinfo {author} {\bibfnamefont {Alberto}\ \bibnamefont
  {Sesana}}, \ and\ \bibinfo {author} {\bibfnamefont {Enrico}\ \bibnamefont
  {Barausse}},\ }\bibfield  {title} {\enquote {\bibinfo {title}
  {{Post-Newtonian evolution of massive black hole triplets in galactic nuclei
  \textendash{} I. Numerical implementation and tests}},}\ }\href {\doibase
  10.1093/mnras/stw1590} {\bibfield  {journal} {\bibinfo  {journal} {Mon. Not.
  Roy. Astron. Soc.}\ }\textbf {\bibinfo {volume} {461}},\ \bibinfo {pages}
  {4419--4434} (\bibinfo {year} {2016})},\ \Eprint
  {http://arxiv.org/abs/1604.08770} {arXiv:1604.08770 [astro-ph.GA]}
  \BibitemShut {NoStop}%
\bibitem [{\citenamefont {Campanelli}\ \emph {et~al.}(2008)\citenamefont
  {Campanelli}, \citenamefont {Lousto},\ and\ \citenamefont
  {Zlochower}}]{Campanelli:2007ea}%
  \BibitemOpen
  \bibfield  {author} {\bibinfo {author} {\bibfnamefont {Manuela}\ \bibnamefont
  {Campanelli}}, \bibinfo {author} {\bibfnamefont {Carlos~O.}\ \bibnamefont
  {Lousto}}, \ and\ \bibinfo {author} {\bibfnamefont {Yosef}\ \bibnamefont
  {Zlochower}},\ }\bibfield  {title} {\enquote {\bibinfo {title} {{Close
  encounters of three black holes}},}\ }\href {\doibase
  10.1103/PhysRevD.77.101501} {\bibfield  {journal} {\bibinfo  {journal} {Phys.
  Rev. D}\ }\textbf {\bibinfo {volume} {77}},\ \bibinfo {pages} {101501}
  (\bibinfo {year} {2008})},\ \Eprint {http://arxiv.org/abs/0710.0879}
  {arXiv:0710.0879 [gr-qc]} \BibitemShut {NoStop}%
\bibitem [{\citenamefont {Lousto}\ and\ \citenamefont
  {Zlochower}(2008)}]{Lousto:2007rj}%
  \BibitemOpen
  \bibfield  {author} {\bibinfo {author} {\bibfnamefont {Carlos~O.}\
  \bibnamefont {Lousto}}\ and\ \bibinfo {author} {\bibfnamefont {Yosef}\
  \bibnamefont {Zlochower}},\ }\bibfield  {title} {\enquote {\bibinfo {title}
  {{Foundations of multiple black hole evolutions}},}\ }\href {\doibase
  10.1103/PhysRevD.77.024034} {\bibfield  {journal} {\bibinfo  {journal} {Phys.
  Rev. D}\ }\textbf {\bibinfo {volume} {77}},\ \bibinfo {pages} {024034}
  (\bibinfo {year} {2008})},\ \Eprint {http://arxiv.org/abs/0711.1165}
  {arXiv:0711.1165 [gr-qc]} \BibitemShut {NoStop}%
\bibitem [{\citenamefont {Galaviz}\ \emph {et~al.}(2010)\citenamefont
  {Galaviz}, \citenamefont {Bruegmann},\ and\ \citenamefont
  {Cao}}]{Galaviz:2010mx}%
  \BibitemOpen
  \bibfield  {author} {\bibinfo {author} {\bibfnamefont {Pablo}\ \bibnamefont
  {Galaviz}}, \bibinfo {author} {\bibfnamefont {Bernd}\ \bibnamefont
  {Bruegmann}}, \ and\ \bibinfo {author} {\bibfnamefont {Zhoujian}\
  \bibnamefont {Cao}},\ }\bibfield  {title} {\enquote {\bibinfo {title}
  {{Numerical evolution of multiple black holes with accurate initial data}},}\
  }\href {\doibase 10.1103/PhysRevD.82.024005} {\bibfield  {journal} {\bibinfo
  {journal} {Phys. Rev. D}\ }\textbf {\bibinfo {volume} {82}},\ \bibinfo
  {pages} {024005} (\bibinfo {year} {2010})},\ \Eprint
  {http://arxiv.org/abs/1004.1353} {arXiv:1004.1353 [gr-qc]} \BibitemShut
  {NoStop}%
\bibitem [{\citenamefont {Damour}\ \emph {et~al.}(2014)\citenamefont {Damour},
  \citenamefont {Jaranowski},\ and\ \citenamefont
  {Sch\"afer}}]{Damour:2014jta}%
  \BibitemOpen
  \bibfield  {author} {\bibinfo {author} {\bibfnamefont {Thibault}\
  \bibnamefont {Damour}}, \bibinfo {author} {\bibfnamefont {Piotr}\
  \bibnamefont {Jaranowski}}, \ and\ \bibinfo {author} {\bibfnamefont
  {Gerhard}\ \bibnamefont {Sch\"afer}},\ }\bibfield  {title} {\enquote
  {\bibinfo {title} {{Nonlocal-in-time action for the fourth post-Newtonian
  conservative dynamics of two-body systems}},}\ }\href {\doibase
  10.1103/PhysRevD.89.064058} {\bibfield  {journal} {\bibinfo  {journal} {Phys.
  Rev. D}\ }\textbf {\bibinfo {volume} {89}},\ \bibinfo {pages} {064058}
  (\bibinfo {year} {2014})},\ \Eprint {http://arxiv.org/abs/1401.4548}
  {arXiv:1401.4548 [gr-qc]} \BibitemShut {NoStop}%
\bibitem [{\citenamefont {Bernard}\ \emph {et~al.}(2016)\citenamefont
  {Bernard}, \citenamefont {Blanchet}, \citenamefont {Boh\'e}, \citenamefont
  {Faye},\ and\ \citenamefont {Marsat}}]{Bernard:2015njp}%
  \BibitemOpen
  \bibfield  {author} {\bibinfo {author} {\bibfnamefont {Laura}\ \bibnamefont
  {Bernard}}, \bibinfo {author} {\bibfnamefont {Luc}\ \bibnamefont {Blanchet}},
  \bibinfo {author} {\bibfnamefont {Alejandro}\ \bibnamefont {Boh\'e}},
  \bibinfo {author} {\bibfnamefont {Guillaume}\ \bibnamefont {Faye}}, \ and\
  \bibinfo {author} {\bibfnamefont {Sylvain}\ \bibnamefont {Marsat}},\
  }\bibfield  {title} {\enquote {\bibinfo {title} {{Fokker action of
  nonspinning compact binaries at the fourth post-Newtonian approximation}},}\
  }\href {\doibase 10.1103/PhysRevD.93.084037} {\bibfield  {journal} {\bibinfo
  {journal} {Phys. Rev. D}\ }\textbf {\bibinfo {volume} {93}},\ \bibinfo
  {pages} {084037} (\bibinfo {year} {2016})},\ \Eprint
  {http://arxiv.org/abs/1512.02876} {arXiv:1512.02876 [gr-qc]} \BibitemShut
  {NoStop}%
\bibitem [{\citenamefont {Damour}\ \emph {et~al.}(2016)\citenamefont {Damour},
  \citenamefont {Jaranowski},\ and\ \citenamefont
  {Sch\"afer}}]{Damour:2016abl}%
  \BibitemOpen
  \bibfield  {author} {\bibinfo {author} {\bibfnamefont {Thibault}\
  \bibnamefont {Damour}}, \bibinfo {author} {\bibfnamefont {Piotr}\
  \bibnamefont {Jaranowski}}, \ and\ \bibinfo {author} {\bibfnamefont
  {Gerhard}\ \bibnamefont {Sch\"afer}},\ }\bibfield  {title} {\enquote
  {\bibinfo {title} {{Conservative dynamics of two-body systems at the fourth
  post-Newtonian approximation of general relativity}},}\ }\href {\doibase
  10.1103/PhysRevD.93.084014} {\bibfield  {journal} {\bibinfo  {journal} {Phys.
  Rev. D}\ }\textbf {\bibinfo {volume} {93}},\ \bibinfo {pages} {084014}
  (\bibinfo {year} {2016})},\ \Eprint {http://arxiv.org/abs/1601.01283}
  {arXiv:1601.01283 [gr-qc]} \BibitemShut {NoStop}%
\bibitem [{\citenamefont {Bernard}\ \emph {et~al.}(2017)\citenamefont
  {Bernard}, \citenamefont {Blanchet}, \citenamefont {Boh\'e}, \citenamefont
  {Faye},\ and\ \citenamefont {Marsat}}]{Bernard:2016wrg}%
  \BibitemOpen
  \bibfield  {author} {\bibinfo {author} {\bibfnamefont {Laura}\ \bibnamefont
  {Bernard}}, \bibinfo {author} {\bibfnamefont {Luc}\ \bibnamefont {Blanchet}},
  \bibinfo {author} {\bibfnamefont {Alejandro}\ \bibnamefont {Boh\'e}},
  \bibinfo {author} {\bibfnamefont {Guillaume}\ \bibnamefont {Faye}}, \ and\
  \bibinfo {author} {\bibfnamefont {Sylvain}\ \bibnamefont {Marsat}},\
  }\bibfield  {title} {\enquote {\bibinfo {title} {{Energy and periastron
  advance of compact binaries on circular orbits at the fourth post-Newtonian
  order}},}\ }\href {\doibase 10.1103/PhysRevD.95.044026} {\bibfield  {journal}
  {\bibinfo  {journal} {Phys. Rev. D}\ }\textbf {\bibinfo {volume} {95}},\
  \bibinfo {pages} {044026} (\bibinfo {year} {2017})},\ \Eprint
  {http://arxiv.org/abs/1610.07934} {arXiv:1610.07934 [gr-qc]} \BibitemShut
  {NoStop}%
\bibitem [{\citenamefont {Foffa}\ \emph {et~al.}(2017)\citenamefont {Foffa},
  \citenamefont {Mastrolia}, \citenamefont {Sturani},\ and\ \citenamefont
  {Sturm}}]{Foffa:2016rgu}%
  \BibitemOpen
  \bibfield  {author} {\bibinfo {author} {\bibfnamefont {Stefano}\ \bibnamefont
  {Foffa}}, \bibinfo {author} {\bibfnamefont {Pierpaolo}\ \bibnamefont
  {Mastrolia}}, \bibinfo {author} {\bibfnamefont {Riccardo}\ \bibnamefont
  {Sturani}}, \ and\ \bibinfo {author} {\bibfnamefont {Christian}\ \bibnamefont
  {Sturm}},\ }\bibfield  {title} {\enquote {\bibinfo {title} {{Effective field
  theory approach to the gravitational two-body dynamics, at fourth
  post-Newtonian order and quintic in the Newton constant}},}\ }\href {\doibase
  10.1103/PhysRevD.95.104009} {\bibfield  {journal} {\bibinfo  {journal} {Phys.
  Rev. D}\ }\textbf {\bibinfo {volume} {95}},\ \bibinfo {pages} {104009}
  (\bibinfo {year} {2017})},\ \Eprint {http://arxiv.org/abs/1612.00482}
  {arXiv:1612.00482 [gr-qc]} \BibitemShut {NoStop}%
\bibitem [{\citenamefont {Porto}\ and\ \citenamefont
  {Rothstein}(2017)}]{Porto:2017dgs}%
  \BibitemOpen
  \bibfield  {author} {\bibinfo {author} {\bibfnamefont {Rafael~A.}\
  \bibnamefont {Porto}}\ and\ \bibinfo {author} {\bibfnamefont {Ira~Z.}\
  \bibnamefont {Rothstein}},\ }\bibfield  {title} {\enquote {\bibinfo {title}
  {{Apparent ambiguities in the post-Newtonian expansion for binary
  systems}},}\ }\href {\doibase 10.1103/PhysRevD.96.024062} {\bibfield
  {journal} {\bibinfo  {journal} {Phys. Rev. D}\ }\textbf {\bibinfo {volume}
  {96}},\ \bibinfo {pages} {024062} (\bibinfo {year} {2017})},\ \Eprint
  {http://arxiv.org/abs/1703.06433} {arXiv:1703.06433 [gr-qc]} \BibitemShut
  {NoStop}%
\bibitem [{\citenamefont {Marchand}\ \emph {et~al.}(2018)\citenamefont
  {Marchand}, \citenamefont {Bernard}, \citenamefont {Blanchet},\ and\
  \citenamefont {Faye}}]{Marchand:2017pir}%
  \BibitemOpen
  \bibfield  {author} {\bibinfo {author} {\bibfnamefont {Tanguy}\ \bibnamefont
  {Marchand}}, \bibinfo {author} {\bibfnamefont {Laura}\ \bibnamefont
  {Bernard}}, \bibinfo {author} {\bibfnamefont {Luc}\ \bibnamefont {Blanchet}},
  \ and\ \bibinfo {author} {\bibfnamefont {Guillaume}\ \bibnamefont {Faye}},\
  }\bibfield  {title} {\enquote {\bibinfo {title} {{Ambiguity-Free Completion
  of the Equations of Motion of Compact Binary Systems at the Fourth
  Post-Newtonian Order}},}\ }\href {\doibase 10.1103/PhysRevD.97.044023}
  {\bibfield  {journal} {\bibinfo  {journal} {Phys. Rev. D}\ }\textbf {\bibinfo
  {volume} {97}},\ \bibinfo {pages} {044023} (\bibinfo {year} {2018})},\
  \Eprint {http://arxiv.org/abs/1707.09289} {arXiv:1707.09289 [gr-qc]}
  \BibitemShut {NoStop}%
\bibitem [{\citenamefont {Damour}\ and\ \citenamefont
  {Jaranowski}(2017)}]{Damour:2017ced}%
  \BibitemOpen
  \bibfield  {author} {\bibinfo {author} {\bibfnamefont {Thibault}\
  \bibnamefont {Damour}}\ and\ \bibinfo {author} {\bibfnamefont {Piotr}\
  \bibnamefont {Jaranowski}},\ }\bibfield  {title} {\enquote {\bibinfo {title}
  {{Four-loop static contribution to the gravitational interaction potential of
  two point masses}},}\ }\href {\doibase 10.1103/PhysRevD.95.084005} {\bibfield
   {journal} {\bibinfo  {journal} {Phys. Rev. D}\ }\textbf {\bibinfo {volume}
  {95}},\ \bibinfo {pages} {084005} (\bibinfo {year} {2017})},\ \Eprint
  {http://arxiv.org/abs/1701.02645} {arXiv:1701.02645 [gr-qc]} \BibitemShut
  {NoStop}%
\bibitem [{\citenamefont {Foffa}\ and\ \citenamefont
  {Sturani}(2019)}]{Foffa:2019rdf}%
  \BibitemOpen
  \bibfield  {author} {\bibinfo {author} {\bibfnamefont {Stefano}\ \bibnamefont
  {Foffa}}\ and\ \bibinfo {author} {\bibfnamefont {Riccardo}\ \bibnamefont
  {Sturani}},\ }\bibfield  {title} {\enquote {\bibinfo {title} {{Conservative
  dynamics of binary systems to fourth Post-Newtonian order in the EFT approach
  I: Regularized Lagrangian}},}\ }\href {\doibase 10.1103/PhysRevD.100.024047}
  {\bibfield  {journal} {\bibinfo  {journal} {Phys. Rev. D}\ }\textbf {\bibinfo
  {volume} {100}},\ \bibinfo {pages} {024047} (\bibinfo {year} {2019})},\
  \Eprint {http://arxiv.org/abs/1903.05113} {arXiv:1903.05113 [gr-qc]}
  \BibitemShut {NoStop}%
\bibitem [{\citenamefont {Foffa}\ \emph
  {et~al.}(2019{\natexlab{a}})\citenamefont {Foffa}, \citenamefont {Porto},
  \citenamefont {Rothstein},\ and\ \citenamefont {Sturani}}]{Foffa:2019yfl}%
  \BibitemOpen
  \bibfield  {author} {\bibinfo {author} {\bibfnamefont {Stefano}\ \bibnamefont
  {Foffa}}, \bibinfo {author} {\bibfnamefont {Rafael~A.}\ \bibnamefont
  {Porto}}, \bibinfo {author} {\bibfnamefont {Ira}\ \bibnamefont {Rothstein}},
  \ and\ \bibinfo {author} {\bibfnamefont {Riccardo}\ \bibnamefont {Sturani}},\
  }\bibfield  {title} {\enquote {\bibinfo {title} {{Conservative dynamics of
  binary systems to fourth Post-Newtonian order in the EFT approach II:
  Renormalized Lagrangian}},}\ }\href {\doibase 10.1103/PhysRevD.100.024048}
  {\bibfield  {journal} {\bibinfo  {journal} {Phys. Rev. D}\ }\textbf {\bibinfo
  {volume} {100}},\ \bibinfo {pages} {024048} (\bibinfo {year}
  {2019}{\natexlab{a}})},\ \Eprint {http://arxiv.org/abs/1903.05118}
  {arXiv:1903.05118 [gr-qc]} \BibitemShut {NoStop}%
\bibitem [{\citenamefont {Bl\"umlein}\ \emph
  {et~al.}(2020{\natexlab{a}})\citenamefont {Bl\"umlein}, \citenamefont
  {Maier}, \citenamefont {Marquard},\ and\ \citenamefont
  {Sch\"afer}}]{Blumlein:2020pog}%
  \BibitemOpen
  \bibfield  {author} {\bibinfo {author} {\bibfnamefont {J.}~\bibnamefont
  {Bl\"umlein}}, \bibinfo {author} {\bibfnamefont {A.}~\bibnamefont {Maier}},
  \bibinfo {author} {\bibfnamefont {P.}~\bibnamefont {Marquard}}, \ and\
  \bibinfo {author} {\bibfnamefont {G.}~\bibnamefont {Sch\"afer}},\ }\bibfield
  {title} {\enquote {\bibinfo {title} {{Fourth post-Newtonian Hamiltonian
  dynamics of two-body systems from an effective field theory approach}},}\
  }\href {\doibase 10.1016/j.nuclphysb.2020.115041} {\bibfield  {journal}
  {\bibinfo  {journal} {Nucl. Phys. B}\ }\textbf {\bibinfo {volume} {955}},\
  \bibinfo {pages} {115041} (\bibinfo {year} {2020}{\natexlab{a}})},\ \Eprint
  {http://arxiv.org/abs/2003.01692} {arXiv:2003.01692 [gr-qc]} \BibitemShut
  {NoStop}%
\bibitem [{\citenamefont {Galley}\ \emph {et~al.}(2016)\citenamefont {Galley},
  \citenamefont {Leibovich}, \citenamefont {Porto},\ and\ \citenamefont
  {Ross}}]{Galley:2015kus}%
  \BibitemOpen
  \bibfield  {author} {\bibinfo {author} {\bibfnamefont {Chad~R.}\ \bibnamefont
  {Galley}}, \bibinfo {author} {\bibfnamefont {Adam~K.}\ \bibnamefont
  {Leibovich}}, \bibinfo {author} {\bibfnamefont {Rafael~A.}\ \bibnamefont
  {Porto}}, \ and\ \bibinfo {author} {\bibfnamefont {Andreas}\ \bibnamefont
  {Ross}},\ }\bibfield  {title} {\enquote {\bibinfo {title} {{Tail effect in
  gravitational radiation reaction: Time nonlocality and renormalization group
  evolution}},}\ }\href {\doibase 10.1103/PhysRevD.93.124010} {\bibfield
  {journal} {\bibinfo  {journal} {Phys. Rev. D}\ }\textbf {\bibinfo {volume}
  {93}},\ \bibinfo {pages} {124010} (\bibinfo {year} {2016})},\ \Eprint
  {http://arxiv.org/abs/1511.07379} {arXiv:1511.07379 [gr-qc]} \BibitemShut
  {NoStop}%
\bibitem [{\citenamefont {Foffa}\ \emph
  {et~al.}(2019{\natexlab{b}})\citenamefont {Foffa}, \citenamefont {Mastrolia},
  \citenamefont {Sturani}, \citenamefont {Sturm},\ and\ \citenamefont
  {Torres~Bobadilla}}]{Foffa:2019hrb}%
  \BibitemOpen
  \bibfield  {author} {\bibinfo {author} {\bibfnamefont {Stefano}\ \bibnamefont
  {Foffa}}, \bibinfo {author} {\bibfnamefont {Pierpaolo}\ \bibnamefont
  {Mastrolia}}, \bibinfo {author} {\bibfnamefont {Riccardo}\ \bibnamefont
  {Sturani}}, \bibinfo {author} {\bibfnamefont {Christian}\ \bibnamefont
  {Sturm}}, \ and\ \bibinfo {author} {\bibfnamefont {William~J.}\ \bibnamefont
  {Torres~Bobadilla}},\ }\bibfield  {title} {\enquote {\bibinfo {title}
  {{Static two-body potential at fifth post-Newtonian order}},}\ }\href
  {\doibase 10.1103/PhysRevLett.122.241605} {\bibfield  {journal} {\bibinfo
  {journal} {Phys. Rev. Lett.}\ }\textbf {\bibinfo {volume} {122}},\ \bibinfo
  {pages} {241605} (\bibinfo {year} {2019}{\natexlab{b}})},\ \Eprint
  {http://arxiv.org/abs/1902.10571} {arXiv:1902.10571 [gr-qc]} \BibitemShut
  {NoStop}%
\bibitem [{\citenamefont {Bl\"umlein}\ \emph
  {et~al.}(2020{\natexlab{b}})\citenamefont {Bl\"umlein}, \citenamefont
  {Maier},\ and\ \citenamefont {Marquard}}]{Blumlein:2019zku}%
  \BibitemOpen
  \bibfield  {author} {\bibinfo {author} {\bibfnamefont {J.}~\bibnamefont
  {Bl\"umlein}}, \bibinfo {author} {\bibfnamefont {A.}~\bibnamefont {Maier}}, \
  and\ \bibinfo {author} {\bibfnamefont {P.}~\bibnamefont {Marquard}},\
  }\bibfield  {title} {\enquote {\bibinfo {title} {{Five-Loop Static
  Contribution to the Gravitational Interaction Potential of Two Point
  Masses}},}\ }\href {\doibase 10.1016/j.physletb.2019.135100} {\bibfield
  {journal} {\bibinfo  {journal} {Phys. Lett. B}\ }\textbf {\bibinfo {volume}
  {800}},\ \bibinfo {pages} {135100} (\bibinfo {year} {2020}{\natexlab{b}})},\
  \Eprint {http://arxiv.org/abs/1902.11180} {arXiv:1902.11180 [gr-qc]}
  \BibitemShut {NoStop}%
\bibitem [{\citenamefont {Bini}\ \emph {et~al.}(2019)\citenamefont {Bini},
  \citenamefont {Damour},\ and\ \citenamefont {Geralico}}]{Bini:2019nra}%
  \BibitemOpen
  \bibfield  {author} {\bibinfo {author} {\bibfnamefont {Donato}\ \bibnamefont
  {Bini}}, \bibinfo {author} {\bibfnamefont {Thibault}\ \bibnamefont {Damour}},
  \ and\ \bibinfo {author} {\bibfnamefont {Andrea}\ \bibnamefont {Geralico}},\
  }\bibfield  {title} {\enquote {\bibinfo {title} {{Novel approach to binary
  dynamics: application to the fifth post-Newtonian level}},}\ }\href {\doibase
  10.1103/PhysRevLett.123.231104} {\bibfield  {journal} {\bibinfo  {journal}
  {Phys. Rev. Lett.}\ }\textbf {\bibinfo {volume} {123}},\ \bibinfo {pages}
  {231104} (\bibinfo {year} {2019})},\ \Eprint
  {http://arxiv.org/abs/1909.02375} {arXiv:1909.02375 [gr-qc]} \BibitemShut
  {NoStop}%
\bibitem [{\citenamefont {Bl\"umlein}\ \emph
  {et~al.}(2020{\natexlab{c}})\citenamefont {Bl\"umlein}, \citenamefont
  {Maier}, \citenamefont {Marquard},\ and\ \citenamefont
  {Sch\"afer}}]{Blumlein:2020znm}%
  \BibitemOpen
  \bibfield  {author} {\bibinfo {author} {\bibfnamefont {J.}~\bibnamefont
  {Bl\"umlein}}, \bibinfo {author} {\bibfnamefont {A.}~\bibnamefont {Maier}},
  \bibinfo {author} {\bibfnamefont {P.}~\bibnamefont {Marquard}}, \ and\
  \bibinfo {author} {\bibfnamefont {G.}~\bibnamefont {Sch\"afer}},\ }\bibfield
  {title} {\enquote {\bibinfo {title} {{Testing binary dynamics in gravity at
  the sixth post-Newtonian level}},}\ }\href {\doibase
  10.1016/j.physletb.2020.135496} {\bibfield  {journal} {\bibinfo  {journal}
  {Phys. Lett. B}\ }\textbf {\bibinfo {volume} {807}},\ \bibinfo {pages}
  {135496} (\bibinfo {year} {2020}{\natexlab{c}})},\ \Eprint
  {http://arxiv.org/abs/2003.07145} {arXiv:2003.07145 [gr-qc]} \BibitemShut
  {NoStop}%
\bibitem [{\citenamefont {Cheung}\ and\ \citenamefont
  {Solon}(2020)}]{Cheung:2020gyp}%
  \BibitemOpen
  \bibfield  {author} {\bibinfo {author} {\bibfnamefont {Clifford}\
  \bibnamefont {Cheung}}\ and\ \bibinfo {author} {\bibfnamefont {Mikhail~P.}\
  \bibnamefont {Solon}},\ }\bibfield  {title} {\enquote {\bibinfo {title}
  {{Classical gravitational scattering at $ \mathcal{O} $(G$^{3}$) from Feynman
  diagrams}},}\ }\href {\doibase 10.1007/JHEP06(2020)144} {\bibfield  {journal}
  {\bibinfo  {journal} {JHEP}\ }\textbf {\bibinfo {volume} {06}},\ \bibinfo
  {pages} {144} (\bibinfo {year} {2020})},\ \Eprint
  {http://arxiv.org/abs/2003.08351} {arXiv:2003.08351 [hep-th]} \BibitemShut
  {NoStop}%
\bibitem [{\citenamefont {Bini}\ \emph
  {et~al.}(2020{\natexlab{a}})\citenamefont {Bini}, \citenamefont {Damour},\
  and\ \citenamefont {Geralico}}]{Bini:2020nsb}%
  \BibitemOpen
  \bibfield  {author} {\bibinfo {author} {\bibfnamefont {Donato}\ \bibnamefont
  {Bini}}, \bibinfo {author} {\bibfnamefont {Thibault}\ \bibnamefont {Damour}},
  \ and\ \bibinfo {author} {\bibfnamefont {Andrea}\ \bibnamefont {Geralico}},\
  }\bibfield  {title} {\enquote {\bibinfo {title} {{Sixth post-Newtonian
  local-in-time dynamics of binary systems}},}\ }\href {\doibase
  10.1103/PhysRevD.102.024061} {\bibfield  {journal} {\bibinfo  {journal}
  {Phys. Rev. D}\ }\textbf {\bibinfo {volume} {102}},\ \bibinfo {pages}
  {024061} (\bibinfo {year} {2020}{\natexlab{a}})},\ \Eprint
  {http://arxiv.org/abs/2004.05407} {arXiv:2004.05407 [gr-qc]} \BibitemShut
  {NoStop}%
\bibitem [{\citenamefont {Bini}\ \emph
  {et~al.}(2020{\natexlab{b}})\citenamefont {Bini}, \citenamefont {Damour},\
  and\ \citenamefont {Geralico}}]{Bini:2020wpo}%
  \BibitemOpen
  \bibfield  {author} {\bibinfo {author} {\bibfnamefont {Donato}\ \bibnamefont
  {Bini}}, \bibinfo {author} {\bibfnamefont {Thibault}\ \bibnamefont {Damour}},
  \ and\ \bibinfo {author} {\bibfnamefont {Andrea}\ \bibnamefont {Geralico}},\
  }\bibfield  {title} {\enquote {\bibinfo {title} {{Binary dynamics at the
  fifth and fifth-and-a-half post-Newtonian orders}},}\ }\href {\doibase
  10.1103/PhysRevD.102.024062} {\bibfield  {journal} {\bibinfo  {journal}
  {Phys. Rev. D}\ }\textbf {\bibinfo {volume} {102}},\ \bibinfo {pages}
  {024062} (\bibinfo {year} {2020}{\natexlab{b}})},\ \Eprint
  {http://arxiv.org/abs/2003.11891} {arXiv:2003.11891 [gr-qc]} \BibitemShut
  {NoStop}%
\bibitem [{\citenamefont {Bini}\ \emph
  {et~al.}(2020{\natexlab{c}})\citenamefont {Bini}, \citenamefont {Damour},\
  and\ \citenamefont {Geralico}}]{Bini:2020hmy}%
  \BibitemOpen
  \bibfield  {author} {\bibinfo {author} {\bibfnamefont {Donato}\ \bibnamefont
  {Bini}}, \bibinfo {author} {\bibfnamefont {Thibault}\ \bibnamefont {Damour}},
  \ and\ \bibinfo {author} {\bibfnamefont {Andrea}\ \bibnamefont {Geralico}},\
  }\bibfield  {title} {\enquote {\bibinfo {title} {{Sixth post-Newtonian
  nonlocal-in-time dynamics of binary systems}},}\ }\href {\doibase
  10.1103/PhysRevD.102.084047} {\bibfield  {journal} {\bibinfo  {journal}
  {Phys. Rev. D}\ }\textbf {\bibinfo {volume} {102}},\ \bibinfo {pages}
  {084047} (\bibinfo {year} {2020}{\natexlab{c}})},\ \Eprint
  {http://arxiv.org/abs/2007.11239} {arXiv:2007.11239 [gr-qc]} \BibitemShut
  {NoStop}%
\bibitem [{\citenamefont {Bini}\ \emph
  {et~al.}(2020{\natexlab{d}})\citenamefont {Bini}, \citenamefont {Damour},
  \citenamefont {Geralico}, \citenamefont {Laporta},\ and\ \citenamefont
  {Mastrolia}}]{Bini:2020uiq}%
  \BibitemOpen
  \bibfield  {author} {\bibinfo {author} {\bibfnamefont {Donato}\ \bibnamefont
  {Bini}}, \bibinfo {author} {\bibfnamefont {Thibault}\ \bibnamefont {Damour}},
  \bibinfo {author} {\bibfnamefont {Andrea}\ \bibnamefont {Geralico}}, \bibinfo
  {author} {\bibfnamefont {Stefano}\ \bibnamefont {Laporta}}, \ and\ \bibinfo
  {author} {\bibfnamefont {Pierpaolo}\ \bibnamefont {Mastrolia}},\ }\bibfield
  {title} {\enquote {\bibinfo {title} {{Gravitational dynamics at $O(G^6)$:
  perturbative gravitational scattering meets experimental mathematics}},}\
  }\href@noop {} {\  (\bibinfo {year} {2020}{\natexlab{d}})},\ \Eprint
  {http://arxiv.org/abs/2008.09389} {arXiv:2008.09389 [gr-qc]} \BibitemShut
  {NoStop}%
\bibitem [{\citenamefont {Ohta}\ \emph {et~al.}(1974)\citenamefont {Ohta},
  \citenamefont {Okamura}, \citenamefont {Hiida},\ and\ \citenamefont
  {Kimura}}]{Ohta:1974pq}%
  \BibitemOpen
  \bibfield  {author} {\bibinfo {author} {\bibfnamefont {T.}~\bibnamefont
  {Ohta}}, \bibinfo {author} {\bibfnamefont {H.}~\bibnamefont {Okamura}},
  \bibinfo {author} {\bibfnamefont {K.}~\bibnamefont {Hiida}}, \ and\ \bibinfo
  {author} {\bibfnamefont {T.}~\bibnamefont {Kimura}},\ }\bibfield  {title}
  {\enquote {\bibinfo {title} {{Higher order gravitational potential for
  many-body system}},}\ }\href {\doibase 10.1143/PTP.51.1220} {\bibfield
  {journal} {\bibinfo  {journal} {Prog. Theor. Phys.}\ }\textbf {\bibinfo
  {volume} {51}},\ \bibinfo {pages} {1220--1238} (\bibinfo {year}
  {1974})}\BibitemShut {NoStop}%
\bibitem [{\citenamefont {Damour}\ and\ \citenamefont
  {Sch{\"a}fer}(1985)}]{Damour:1985mt}%
  \BibitemOpen
  \bibfield  {author} {\bibinfo {author} {\bibfnamefont {Thibault}\
  \bibnamefont {Damour}}\ and\ \bibinfo {author} {\bibfnamefont {Gerhard}\
  \bibnamefont {Sch{\"a}fer}},\ }\bibfield  {title} {\enquote {\bibinfo {title}
  {Lagrangians from point masses at the second post-newtonian approximation of
  general relativity},}\ }\href@noop {} {\bibfield  {journal} {\bibinfo
  {journal} {General relativity and gravitation}\ }\textbf {\bibinfo {volume}
  {17}},\ \bibinfo {pages} {879--905} (\bibinfo {year} {1985})}\BibitemShut
  {NoStop}%
\bibitem [{\citenamefont {Sch{\"a}fer}(1987)}]{Schafer:1987}%
  \BibitemOpen
  \bibfield  {author} {\bibinfo {author} {\bibfnamefont {G.}~\bibnamefont
  {Sch{\"a}fer}},\ }\bibfield  {title} {\enquote {\bibinfo {title} {{Three-body
  hamiltonian in general relativity}},}\ }\href {\doibase
  doi.org/10.1016/0375-9601(87)90389-6} {\bibfield  {journal} {\bibinfo
  {journal} {Phys. Lett. A}\ }\textbf {\bibinfo {volume} {123}},\ \bibinfo
  {pages} {336} (\bibinfo {year} {1987})}\BibitemShut {NoStop}%
\bibitem [{\citenamefont {Chu}(2009)}]{Chu:2008xm}%
  \BibitemOpen
  \bibfield  {author} {\bibinfo {author} {\bibfnamefont {Yi-Zen}\ \bibnamefont
  {Chu}},\ }\bibfield  {title} {\enquote {\bibinfo {title} {{The n-body problem
  in General Relativity up to the second post-Newtonian order from perturbative
  field theory}},}\ }\href {\doibase 10.1103/PhysRevD.79.044031} {\bibfield
  {journal} {\bibinfo  {journal} {Phys. Rev. D}\ }\textbf {\bibinfo {volume}
  {79}},\ \bibinfo {pages} {044031} (\bibinfo {year} {2009})},\ \Eprint
  {http://arxiv.org/abs/0812.0012} {arXiv:0812.0012 [gr-qc]} \BibitemShut
  {NoStop}%
\bibitem [{\citenamefont {Cheung}\ \emph {et~al.}(2018)\citenamefont {Cheung},
  \citenamefont {Rothstein},\ and\ \citenamefont {Solon}}]{Cheung:2018wkq}%
  \BibitemOpen
  \bibfield  {author} {\bibinfo {author} {\bibfnamefont {Clifford}\
  \bibnamefont {Cheung}}, \bibinfo {author} {\bibfnamefont {Ira~Z.}\
  \bibnamefont {Rothstein}}, \ and\ \bibinfo {author} {\bibfnamefont
  {Mikhail~P.}\ \bibnamefont {Solon}},\ }\bibfield  {title} {\enquote {\bibinfo
  {title} {{From Scattering Amplitudes to Classical Potentials in the
  Post-Minkowskian Expansion}},}\ }\href {\doibase
  10.1103/PhysRevLett.121.251101} {\bibfield  {journal} {\bibinfo  {journal}
  {Phys. Rev. Lett.}\ }\textbf {\bibinfo {volume} {121}},\ \bibinfo {pages}
  {251101} (\bibinfo {year} {2018})},\ \Eprint
  {http://arxiv.org/abs/1808.02489} {arXiv:1808.02489 [hep-th]} \BibitemShut
  {NoStop}%
\bibitem [{\citenamefont {Cristofoli}\ \emph {et~al.}(2019)\citenamefont
  {Cristofoli}, \citenamefont {Bjerrum-Bohr}, \citenamefont {Damgaard},\ and\
  \citenamefont {Vanhove}}]{Cristofoli:2019neg}%
  \BibitemOpen
  \bibfield  {author} {\bibinfo {author} {\bibfnamefont {Andrea}\ \bibnamefont
  {Cristofoli}}, \bibinfo {author} {\bibfnamefont {N.E.J.}\ \bibnamefont
  {Bjerrum-Bohr}}, \bibinfo {author} {\bibfnamefont {Poul~H.}\ \bibnamefont
  {Damgaard}}, \ and\ \bibinfo {author} {\bibfnamefont {Pierre}\ \bibnamefont
  {Vanhove}},\ }\bibfield  {title} {\enquote {\bibinfo {title}
  {{Post-Minkowskian Hamiltonians in general relativity}},}\ }\href {\doibase
  10.1103/PhysRevD.100.084040} {\bibfield  {journal} {\bibinfo  {journal}
  {Phys. Rev. D}\ }\textbf {\bibinfo {volume} {100}},\ \bibinfo {pages}
  {084040} (\bibinfo {year} {2019})},\ \Eprint
  {http://arxiv.org/abs/1906.01579} {arXiv:1906.01579 [hep-th]} \BibitemShut
  {NoStop}%
\bibitem [{\citenamefont {Bern}\ \emph
  {et~al.}(2019{\natexlab{a}})\citenamefont {Bern}, \citenamefont {Cheung},
  \citenamefont {Roiban}, \citenamefont {Shen}, \citenamefont {Solon},\ and\
  \citenamefont {Zeng}}]{Bern:2019nnu}%
  \BibitemOpen
  \bibfield  {author} {\bibinfo {author} {\bibfnamefont {Zvi}\ \bibnamefont
  {Bern}}, \bibinfo {author} {\bibfnamefont {Clifford}\ \bibnamefont {Cheung}},
  \bibinfo {author} {\bibfnamefont {Radu}\ \bibnamefont {Roiban}}, \bibinfo
  {author} {\bibfnamefont {Chia-Hsien}\ \bibnamefont {Shen}}, \bibinfo {author}
  {\bibfnamefont {Mikhail~P.}\ \bibnamefont {Solon}}, \ and\ \bibinfo {author}
  {\bibfnamefont {Mao}\ \bibnamefont {Zeng}},\ }\bibfield  {title} {\enquote
  {\bibinfo {title} {{Scattering Amplitudes and the Conservative Hamiltonian
  for Binary Systems at Third Post-Minkowskian Order}},}\ }\href {\doibase
  10.1103/PhysRevLett.122.201603} {\bibfield  {journal} {\bibinfo  {journal}
  {Phys. Rev. Lett.}\ }\textbf {\bibinfo {volume} {122}},\ \bibinfo {pages}
  {201603} (\bibinfo {year} {2019}{\natexlab{a}})},\ \Eprint
  {http://arxiv.org/abs/1901.04424} {arXiv:1901.04424 [hep-th]} \BibitemShut
  {NoStop}%
\bibitem [{\citenamefont {Bern}\ \emph
  {et~al.}(2019{\natexlab{b}})\citenamefont {Bern}, \citenamefont {Cheung},
  \citenamefont {Roiban}, \citenamefont {Shen}, \citenamefont {Solon},\ and\
  \citenamefont {Zeng}}]{Bern:2019crd}%
  \BibitemOpen
  \bibfield  {author} {\bibinfo {author} {\bibfnamefont {Zvi}\ \bibnamefont
  {Bern}}, \bibinfo {author} {\bibfnamefont {Clifford}\ \bibnamefont {Cheung}},
  \bibinfo {author} {\bibfnamefont {Radu}\ \bibnamefont {Roiban}}, \bibinfo
  {author} {\bibfnamefont {Chia-Hsien}\ \bibnamefont {Shen}}, \bibinfo {author}
  {\bibfnamefont {Mikhail~P.}\ \bibnamefont {Solon}}, \ and\ \bibinfo {author}
  {\bibfnamefont {Mao}\ \bibnamefont {Zeng}},\ }\bibfield  {title} {\enquote
  {\bibinfo {title} {{Black Hole Binary Dynamics from the Double Copy and
  Effective Theory}},}\ }\href {\doibase 10.1007/JHEP10(2019)206} {\bibfield
  {journal} {\bibinfo  {journal} {JHEP}\ }\textbf {\bibinfo {volume} {10}},\
  \bibinfo {pages} {206} (\bibinfo {year} {2019}{\natexlab{b}})},\ \Eprint
  {http://arxiv.org/abs/1908.01493} {arXiv:1908.01493 [hep-th]} \BibitemShut
  {NoStop}%
\bibitem [{\citenamefont {Damour}(2020)}]{Damour:2020tta}%
  \BibitemOpen
  \bibfield  {author} {\bibinfo {author} {\bibfnamefont {Thibault}\
  \bibnamefont {Damour}},\ }\bibfield  {title} {\enquote {\bibinfo {title}
  {{Radiative contribution to classical gravitational scattering at the third
  order in $G$}},}\ }\href {\doibase 10.1103/PhysRevD.102.124008} {\bibfield
  {journal} {\bibinfo  {journal} {Phys. Rev. D}\ }\textbf {\bibinfo {volume}
  {102}},\ \bibinfo {pages} {124008} (\bibinfo {year} {2020})},\ \Eprint
  {http://arxiv.org/abs/2010.01641} {arXiv:2010.01641 [gr-qc]} \BibitemShut
  {NoStop}%
\bibitem [{\citenamefont {Di~Vecchia}\ \emph {et~al.}(2020)\citenamefont
  {Di~Vecchia}, \citenamefont {Heissenberg}, \citenamefont {Russo},\ and\
  \citenamefont {Veneziano}}]{DiVecchia:2020ymx}%
  \BibitemOpen
  \bibfield  {author} {\bibinfo {author} {\bibfnamefont {Paolo}\ \bibnamefont
  {Di~Vecchia}}, \bibinfo {author} {\bibfnamefont {Carlo}\ \bibnamefont
  {Heissenberg}}, \bibinfo {author} {\bibfnamefont {Rodolfo}\ \bibnamefont
  {Russo}}, \ and\ \bibinfo {author} {\bibfnamefont {Gabriele}\ \bibnamefont
  {Veneziano}},\ }\bibfield  {title} {\enquote {\bibinfo {title} {{Universality
  of ultra-relativistic gravitational scattering}},}\ }\href {\doibase
  10.1016/j.physletb.2020.135924} {\bibfield  {journal} {\bibinfo  {journal}
  {Phys. Lett. B}\ }\textbf {\bibinfo {volume} {811}},\ \bibinfo {pages}
  {135924} (\bibinfo {year} {2020})},\ \Eprint
  {http://arxiv.org/abs/2008.12743} {arXiv:2008.12743 [hep-th]} \BibitemShut
  {NoStop}%
\bibitem [{\citenamefont {K\"alin}\ and\ \citenamefont
  {Porto}(2020)}]{Kalin:2020mvi}%
  \BibitemOpen
  \bibfield  {author} {\bibinfo {author} {\bibfnamefont {Gregor}\ \bibnamefont
  {K\"alin}}\ and\ \bibinfo {author} {\bibfnamefont {Rafael~A.}\ \bibnamefont
  {Porto}},\ }\bibfield  {title} {\enquote {\bibinfo {title} {{Post-Minkowskian
  Effective Field Theory for Conservative Binary Dynamics}},}\ }\href {\doibase
  10.1007/JHEP11(2020)106} {\bibfield  {journal} {\bibinfo  {journal} {JHEP}\
  }\textbf {\bibinfo {volume} {11}},\ \bibinfo {pages} {106} (\bibinfo {year}
  {2020})},\ \Eprint {http://arxiv.org/abs/2006.01184} {arXiv:2006.01184
  [hep-th]} \BibitemShut {NoStop}%
\bibitem [{\citenamefont {K\"alin}\ \emph {et~al.}(2020)\citenamefont
  {K\"alin}, \citenamefont {Liu},\ and\ \citenamefont {Porto}}]{Kalin:2020fhe}%
  \BibitemOpen
  \bibfield  {author} {\bibinfo {author} {\bibfnamefont {Gregor}\ \bibnamefont
  {K\"alin}}, \bibinfo {author} {\bibfnamefont {Zhengwen}\ \bibnamefont {Liu}},
  \ and\ \bibinfo {author} {\bibfnamefont {Rafael~A.}\ \bibnamefont {Porto}},\
  }\bibfield  {title} {\enquote {\bibinfo {title} {{Conservative Dynamics of
  Binary Systems to Third Post-Minkowskian Order from the Effective Field
  Theory Approach}},}\ }\href@noop {} {\  (\bibinfo {year} {2020})},\ \Eprint
  {http://arxiv.org/abs/2007.04977} {arXiv:2007.04977 [hep-th]} \BibitemShut
  {NoStop}%
\bibitem [{\citenamefont {Westpfahl}(1985)}]{Westpfahl:1985tsl}%
  \BibitemOpen
  \bibfield  {author} {\bibinfo {author} {\bibfnamefont {Konradin}\
  \bibnamefont {Westpfahl}},\ }\bibfield  {title} {\enquote {\bibinfo {title}
  {{High-Speed Scattering of Charged and Uncharged Particles in General
  Relativity}},}\ }\href {\doibase 10.1002/prop.2190330802} {\bibfield
  {journal} {\bibinfo  {journal} {Fortsch. Phys.}\ }\textbf {\bibinfo {volume}
  {33}},\ \bibinfo {pages} {417--493} (\bibinfo {year} {1985})}\BibitemShut
  {NoStop}%
\bibitem [{\citenamefont {Bel}\ \emph {et~al.}(1981)\citenamefont {Bel},
  \citenamefont {Damour}, \citenamefont {Deruelle}, \citenamefont {Ibanez},\
  and\ \citenamefont {Martin}}]{Bel:1981be}%
  \BibitemOpen
  \bibfield  {author} {\bibinfo {author} {\bibfnamefont {LLuis}\ \bibnamefont
  {Bel}}, \bibinfo {author} {\bibfnamefont {T.}~\bibnamefont {Damour}},
  \bibinfo {author} {\bibfnamefont {N.}~\bibnamefont {Deruelle}}, \bibinfo
  {author} {\bibfnamefont {J.}~\bibnamefont {Ibanez}}, \ and\ \bibinfo {author}
  {\bibfnamefont {J.}~\bibnamefont {Martin}},\ }\bibfield  {title} {\enquote
  {\bibinfo {title} {{Poincar\'e-invariant gravitational field and equations of
  motion of two pointlike objects: The postlinear approximation of general
  relativity}},}\ }\href {\doibase 10.1007/BF00756073} {\bibfield  {journal}
  {\bibinfo  {journal} {Gen. Rel. Grav.}\ }\textbf {\bibinfo {volume} {13}},\
  \bibinfo {pages} {963--1004} (\bibinfo {year} {1981})}\BibitemShut {NoStop}%
\bibitem [{\citenamefont {Ledvinka}\ \emph {et~al.}(2008)\citenamefont
  {Ledvinka}, \citenamefont {Schaefer},\ and\ \citenamefont
  {Bicak}}]{Ledvinka:2008tk}%
  \BibitemOpen
  \bibfield  {author} {\bibinfo {author} {\bibfnamefont {Tomas}\ \bibnamefont
  {Ledvinka}}, \bibinfo {author} {\bibfnamefont {Gerhard}\ \bibnamefont
  {Schaefer}}, \ and\ \bibinfo {author} {\bibfnamefont {Jiri}\ \bibnamefont
  {Bicak}},\ }\bibfield  {title} {\enquote {\bibinfo {title} {{Relativistic
  Closed-Form Hamiltonian for Many-Body Gravitating Systems in the
  Post-Minkowskian Approximation}},}\ }\href {\doibase
  10.1103/PhysRevLett.100.251101} {\bibfield  {journal} {\bibinfo  {journal}
  {Phys. Rev. Lett.}\ }\textbf {\bibinfo {volume} {100}},\ \bibinfo {pages}
  {251101} (\bibinfo {year} {2008})},\ \Eprint {http://arxiv.org/abs/0807.0214}
  {arXiv:0807.0214 [gr-qc]} \BibitemShut {NoStop}%
\bibitem [{\citenamefont {Damour}(2016)}]{Damour:2016gwp}%
  \BibitemOpen
  \bibfield  {author} {\bibinfo {author} {\bibfnamefont {Thibault}\
  \bibnamefont {Damour}},\ }\bibfield  {title} {\enquote {\bibinfo {title}
  {{Gravitational scattering, post-Minkowskian approximation and Effective
  One-Body theory}},}\ }\href {\doibase 10.1103/PhysRevD.94.104015} {\bibfield
  {journal} {\bibinfo  {journal} {Phys. Rev. D}\ }\textbf {\bibinfo {volume}
  {94}},\ \bibinfo {pages} {104015} (\bibinfo {year} {2016})},\ \Eprint
  {http://arxiv.org/abs/1609.00354} {arXiv:1609.00354 [gr-qc]} \BibitemShut
  {NoStop}%
\bibitem [{\citenamefont {Blanchet}\ and\ \citenamefont
  {Fokas}(2018)}]{Blanchet:2018yvb}%
  \BibitemOpen
  \bibfield  {author} {\bibinfo {author} {\bibfnamefont {Luc}\ \bibnamefont
  {Blanchet}}\ and\ \bibinfo {author} {\bibfnamefont {Athanassios~S.}\
  \bibnamefont {Fokas}},\ }\bibfield  {title} {\enquote {\bibinfo {title}
  {{Equations of motion of self-gravitating $N$-body systems in the first
  post-Minkowskian approximation}},}\ }\href {\doibase
  10.1103/PhysRevD.98.084005} {\bibfield  {journal} {\bibinfo  {journal} {Phys.
  Rev. D}\ }\textbf {\bibinfo {volume} {98}},\ \bibinfo {pages} {084005}
  (\bibinfo {year} {2018})},\ \Eprint {http://arxiv.org/abs/1806.08347}
  {arXiv:1806.08347 [gr-qc]} \BibitemShut {NoStop}%
\bibitem [{\citenamefont {Mogull}\ \emph {et~al.}(2020)\citenamefont {Mogull},
  \citenamefont {Plefka},\ and\ \citenamefont {Steinhoff}}]{Mogull:2020sak}%
  \BibitemOpen
  \bibfield  {author} {\bibinfo {author} {\bibfnamefont {Gustav}\ \bibnamefont
  {Mogull}}, \bibinfo {author} {\bibfnamefont {Jan}\ \bibnamefont {Plefka}}, \
  and\ \bibinfo {author} {\bibfnamefont {Jan}\ \bibnamefont {Steinhoff}},\
  }\bibfield  {title} {\enquote {\bibinfo {title} {{Classical black hole
  scattering from a worldline quantum field theory}},}\ }\href@noop {} {\
  (\bibinfo {year} {2020})},\ \Eprint {http://arxiv.org/abs/2010.02865}
  {arXiv:2010.02865 [hep-th]} \BibitemShut {NoStop}%
\bibitem [{\citenamefont {Goldberger}\ and\ \citenamefont
  {Rothstein}(2006)}]{Goldberger:2004jt}%
  \BibitemOpen
  \bibfield  {author} {\bibinfo {author} {\bibfnamefont {Walter~D.}\
  \bibnamefont {Goldberger}}\ and\ \bibinfo {author} {\bibfnamefont {Ira~Z.}\
  \bibnamefont {Rothstein}},\ }\bibfield  {title} {\enquote {\bibinfo {title}
  {{An Effective field theory of gravity for extended objects}},}\ }\href
  {\doibase 10.1103/PhysRevD.73.104029} {\bibfield  {journal} {\bibinfo
  {journal} {Phys. Rev. D}\ }\textbf {\bibinfo {volume} {73}},\ \bibinfo
  {pages} {104029} (\bibinfo {year} {2006})},\ \Eprint
  {http://arxiv.org/abs/hep-th/0409156} {arXiv:hep-th/0409156} \BibitemShut
  {NoStop}%
\bibitem [{\citenamefont {Chicherin}\ \emph {et~al.}(2018)\citenamefont
  {Chicherin}, \citenamefont {Kazakov}, \citenamefont {Loebbert}, \citenamefont
  {M{\"u}ller},\ and\ \citenamefont {Zhong}}]{Chicherin:2017cns}%
  \BibitemOpen
  \bibfield  {author} {\bibinfo {author} {\bibfnamefont {Dmitry}\ \bibnamefont
  {Chicherin}}, \bibinfo {author} {\bibfnamefont {Vladimir}\ \bibnamefont
  {Kazakov}}, \bibinfo {author} {\bibfnamefont {Florian}\ \bibnamefont
  {Loebbert}}, \bibinfo {author} {\bibfnamefont {Dennis}\ \bibnamefont
  {M{\"u}ller}}, \ and\ \bibinfo {author} {\bibfnamefont {De-liang}\
  \bibnamefont {Zhong}},\ }\bibfield  {title} {\enquote {\bibinfo {title}
  {{Yangian Symmetry for Bi-Scalar Loop Amplitudes}},}\ }\href {\doibase
  10.1007/JHEP05(2018)003} {\bibfield  {journal} {\bibinfo  {journal} {JHEP}\
  }\textbf {\bibinfo {volume} {05}},\ \bibinfo {pages} {003} (\bibinfo {year}
  {2018})},\ \Eprint {http://arxiv.org/abs/1704.01967} {arXiv:1704.01967
  [hep-th]} \BibitemShut {NoStop}%
\bibitem [{\citenamefont {Loebbert}\ \emph
  {et~al.}(2020{\natexlab{a}})\citenamefont {Loebbert}, \citenamefont
  {Miczajka}, \citenamefont {M\"uller},\ and\ \citenamefont
  {M\"unkler}}]{Loebbert:2020hxk}%
  \BibitemOpen
  \bibfield  {author} {\bibinfo {author} {\bibfnamefont {Florian}\ \bibnamefont
  {Loebbert}}, \bibinfo {author} {\bibfnamefont {Julian}\ \bibnamefont
  {Miczajka}}, \bibinfo {author} {\bibfnamefont {Dennis}\ \bibnamefont
  {M\"uller}}, \ and\ \bibinfo {author} {\bibfnamefont {Hagen}\ \bibnamefont
  {M\"unkler}},\ }\bibfield  {title} {\enquote {\bibinfo {title} {{Massive
  Conformal Symmetry and Integrability for Feynman Integrals}},}\ }\href
  {\doibase 10.1103/PhysRevLett.125.091602} {\bibfield  {journal} {\bibinfo
  {journal} {Phys. Rev. Lett.}\ }\textbf {\bibinfo {volume} {125}},\ \bibinfo
  {pages} {091602} (\bibinfo {year} {2020}{\natexlab{a}})},\ \Eprint
  {http://arxiv.org/abs/2005.01735} {arXiv:2005.01735 [hep-th]} \BibitemShut
  {NoStop}%
\bibitem [{\citenamefont {Loebbert}\ \emph
  {et~al.}(2020{\natexlab{b}})\citenamefont {Loebbert}, \citenamefont
  {M{\"u}ller},\ and\ \citenamefont {M{\"u}nkler}}]{Loebbert:2019vcj}%
  \BibitemOpen
  \bibfield  {author} {\bibinfo {author} {\bibfnamefont {Florian}\ \bibnamefont
  {Loebbert}}, \bibinfo {author} {\bibfnamefont {Dennis}\ \bibnamefont
  {M{\"u}ller}}, \ and\ \bibinfo {author} {\bibfnamefont {Hagen}\ \bibnamefont
  {M{\"u}nkler}},\ }\bibfield  {title} {\enquote {\bibinfo {title} {{Yangian
  Bootstrap for Conformal Feynman Integrals}},}\ }\href {\doibase
  10.1103/PhysRevD.101.066006} {\bibfield  {journal} {\bibinfo  {journal}
  {Phys. Rev. D}\ }\textbf {\bibinfo {volume} {101}},\ \bibinfo {pages}
  {066006} (\bibinfo {year} {2020}{\natexlab{b}})},\ \Eprint
  {http://arxiv.org/abs/1912.05561} {arXiv:1912.05561 [hep-th]} \BibitemShut
  {NoStop}%
\bibitem [{\citenamefont {Corcoran}\ \emph {et~al.}(2020)\citenamefont
  {Corcoran}, \citenamefont {Loebbert}, \citenamefont {Miczajka},\ and\
  \citenamefont {Staudacher}}]{Corcoran:2020epz}%
  \BibitemOpen
  \bibfield  {author} {\bibinfo {author} {\bibfnamefont {Luke}\ \bibnamefont
  {Corcoran}}, \bibinfo {author} {\bibfnamefont {Florian}\ \bibnamefont
  {Loebbert}}, \bibinfo {author} {\bibfnamefont {Julian}\ \bibnamefont
  {Miczajka}}, \ and\ \bibinfo {author} {\bibfnamefont {Matthias}\ \bibnamefont
  {Staudacher}},\ }\bibfield  {title} {\enquote {\bibinfo {title} {{Minkowski
  Box from Yangian Bootstrap}},}\ }\href@noop {} {\  (\bibinfo {year}
  {2020})},\ \Eprint {http://arxiv.org/abs/2012.07852} {arXiv:2012.07852
  [hep-th]} \BibitemShut {NoStop}%
\bibitem [{\citenamefont {Coriano}\ \emph {et~al.}(2013)\citenamefont
  {Coriano}, \citenamefont {Delle~Rose}, \citenamefont {Mottola},\ and\
  \citenamefont {Serino}}]{Coriano:2013jba}%
  \BibitemOpen
  \bibfield  {author} {\bibinfo {author} {\bibfnamefont {Claudio}\ \bibnamefont
  {Coriano}}, \bibinfo {author} {\bibfnamefont {Luigi}\ \bibnamefont
  {Delle~Rose}}, \bibinfo {author} {\bibfnamefont {Emil}\ \bibnamefont
  {Mottola}}, \ and\ \bibinfo {author} {\bibfnamefont {Mirko}\ \bibnamefont
  {Serino}},\ }\bibfield  {title} {\enquote {\bibinfo {title} {{Solving the
  Conformal Constraints for Scalar Operators in Momentum Space and the
  Evaluation of Feynman's Master Integrals}},}\ }\href {\doibase
  10.1007/JHEP07(2013)011} {\bibfield  {journal} {\bibinfo  {journal} {JHEP}\
  }\textbf {\bibinfo {volume} {07}},\ \bibinfo {pages} {011} (\bibinfo {year}
  {2013})},\ \Eprint {http://arxiv.org/abs/1304.6944} {arXiv:1304.6944
  [hep-th]} \BibitemShut {NoStop}%
\bibitem [{\citenamefont {Bzowski}\ \emph {et~al.}(2014)\citenamefont
  {Bzowski}, \citenamefont {McFadden},\ and\ \citenamefont
  {Skenderis}}]{Bzowski:2013sza}%
  \BibitemOpen
  \bibfield  {author} {\bibinfo {author} {\bibfnamefont {Adam}\ \bibnamefont
  {Bzowski}}, \bibinfo {author} {\bibfnamefont {Paul}\ \bibnamefont
  {McFadden}}, \ and\ \bibinfo {author} {\bibfnamefont {Kostas}\ \bibnamefont
  {Skenderis}},\ }\bibfield  {title} {\enquote {\bibinfo {title} {{Implications
  of conformal invariance in momentum space}},}\ }\href {\doibase
  10.1007/JHEP03(2014)111} {\bibfield  {journal} {\bibinfo  {journal} {JHEP}\
  }\textbf {\bibinfo {volume} {03}},\ \bibinfo {pages} {111} (\bibinfo {year}
  {2014})},\ \Eprint {http://arxiv.org/abs/1304.7760} {arXiv:1304.7760
  [hep-th]} \BibitemShut {NoStop}%
\bibitem [{\citenamefont {Loebbert}\ \emph
  {et~al.}(2020{\natexlab{c}})\citenamefont {Loebbert}, \citenamefont
  {Miczajka}, \citenamefont {M\"uller},\ and\ \citenamefont
  {M\"unkler}}]{Loebbert:2020glj}%
  \BibitemOpen
  \bibfield  {author} {\bibinfo {author} {\bibfnamefont {Florian}\ \bibnamefont
  {Loebbert}}, \bibinfo {author} {\bibfnamefont {Julian}\ \bibnamefont
  {Miczajka}}, \bibinfo {author} {\bibfnamefont {Dennis}\ \bibnamefont
  {M\"uller}}, \ and\ \bibinfo {author} {\bibfnamefont {Hagen}\ \bibnamefont
  {M\"unkler}},\ }\bibfield  {title} {\enquote {\bibinfo {title} {{Yangian
  Bootstrap for Massive Feynman Integrals}},}\ }\href@noop {} {\  (\bibinfo
  {year} {2020}{\natexlab{c}})},\ \Eprint {http://arxiv.org/abs/2010.08552}
  {arXiv:2010.08552 [hep-th]} \BibitemShut {NoStop}%
\bibitem [{\citenamefont {York}(1972)}]{York:1972sj}%
  \BibitemOpen
  \bibfield  {author} {\bibinfo {author} {\bibfnamefont {James~W.}\
  \bibnamefont {York}, \bibfnamefont {Jr.}},\ }\bibfield  {title} {\enquote
  {\bibinfo {title} {{Role of conformal three geometry in the dynamics of
  gravitation}},}\ }\href {\doibase 10.1103/PhysRevLett.28.1082} {\bibfield
  {journal} {\bibinfo  {journal} {Phys. Rev. Lett.}\ }\textbf {\bibinfo
  {volume} {28}},\ \bibinfo {pages} {1082--1085} (\bibinfo {year}
  {1972})}\BibitemShut {NoStop}%
\bibitem [{\citenamefont {Gibbons}\ and\ \citenamefont
  {Hawking}(1977)}]{Gibbons:1976ue}%
  \BibitemOpen
  \bibfield  {author} {\bibinfo {author} {\bibfnamefont {G.~W.}\ \bibnamefont
  {Gibbons}}\ and\ \bibinfo {author} {\bibfnamefont {S.~W.}\ \bibnamefont
  {Hawking}},\ }\bibfield  {title} {\enquote {\bibinfo {title} {{Action
  Integrals and Partition Functions in Quantum Gravity}},}\ }\href {\doibase
  10.1103/PhysRevD.15.2752} {\bibfield  {journal} {\bibinfo  {journal} {Phys.
  Rev. D}\ }\textbf {\bibinfo {volume} {15}},\ \bibinfo {pages} {2752--2756}
  (\bibinfo {year} {1977})}\BibitemShut {NoStop}%
\bibitem [{\citenamefont {Sannan}(1986)}]{Sannan:1986tz}%
  \BibitemOpen
  \bibfield  {author} {\bibinfo {author} {\bibfnamefont {Sigurd}\ \bibnamefont
  {Sannan}},\ }\bibfield  {title} {\enquote {\bibinfo {title} {{Gravity as the
  Limit of the Type \{II\} Superstring Theory}},}\ }\href {\doibase
  10.1103/PhysRevD.34.1749} {\bibfield  {journal} {\bibinfo  {journal} {Phys.
  Rev. D}\ }\textbf {\bibinfo {volume} {34}},\ \bibinfo {pages} {1749}
  (\bibinfo {year} {1986})}\BibitemShut {NoStop}%
\bibitem [{\citenamefont {Damour}\ and\ \citenamefont
  {Schaefer}(1991)}]{Damour:1990jh}%
  \BibitemOpen
  \bibfield  {author} {\bibinfo {author} {\bibfnamefont {Thibault}\
  \bibnamefont {Damour}}\ and\ \bibinfo {author} {\bibfnamefont {Gerhard}\
  \bibnamefont {Schaefer}},\ }\bibfield  {title} {\enquote {\bibinfo {title}
  {{Redefinition of position variables and the reduction of higher order
  Lagrangians}},}\ }\href {\doibase 10.1063/1.529135} {\bibfield  {journal}
  {\bibinfo  {journal} {J. Math. Phys.}\ }\textbf {\bibinfo {volume} {32}},\
  \bibinfo {pages} {127--134} (\bibinfo {year} {1991})}\BibitemShut {NoStop}%
\bibitem [{\citenamefont {Chicherin}\ \emph {et~al.}(2017)\citenamefont
  {Chicherin}, \citenamefont {Kazakov}, \citenamefont {Loebbert}, \citenamefont
  {M\"uller},\ and\ \citenamefont {Zhong}}]{Chicherin:2017frs}%
  \BibitemOpen
  \bibfield  {author} {\bibinfo {author} {\bibfnamefont {Dmitry}\ \bibnamefont
  {Chicherin}}, \bibinfo {author} {\bibfnamefont {Vladimir}\ \bibnamefont
  {Kazakov}}, \bibinfo {author} {\bibfnamefont {Florian}\ \bibnamefont
  {Loebbert}}, \bibinfo {author} {\bibfnamefont {Dennis}\ \bibnamefont
  {M\"uller}}, \ and\ \bibinfo {author} {\bibfnamefont {De-liang}\ \bibnamefont
  {Zhong}},\ }\bibfield  {title} {\enquote {\bibinfo {title} {{Yangian Symmetry
  for Fishnet Feynman Graphs}},}\ }\href {\doibase 10.1103/PhysRevD.96.121901}
  {\bibfield  {journal} {\bibinfo  {journal} {Phys. Rev. D}\ }\textbf {\bibinfo
  {volume} {96}},\ \bibinfo {pages} {121901} (\bibinfo {year} {2017})},\
  \Eprint {http://arxiv.org/abs/1708.00007} {arXiv:1708.00007 [hep-th]}
  \BibitemShut {NoStop}%
\bibitem [{\citenamefont {Boos}\ and\ \citenamefont
  {Davydychev}(1991)}]{Boos:1990rg}%
  \BibitemOpen
  \bibfield  {author} {\bibinfo {author} {\bibfnamefont {E.E.}\ \bibnamefont
  {Boos}}\ and\ \bibinfo {author} {\bibfnamefont {Andrei~I.}\ \bibnamefont
  {Davydychev}},\ }\bibfield  {title} {\enquote {\bibinfo {title} {{A Method of
  evaluating massive Feynman integrals}},}\ }\href {\doibase
  10.1007/BF01016805} {\bibfield  {journal} {\bibinfo  {journal} {Theor. Math.
  Phys.}\ }\textbf {\bibinfo {volume} {89}},\ \bibinfo {pages} {1052--1063}
  (\bibinfo {year} {1991})}\BibitemShut {NoStop}%
\bibitem [{\citenamefont {Ohta}\ \emph {et~al.}(1973)\citenamefont {Ohta},
  \citenamefont {Okamura}, \citenamefont {Kimura},\ and\ \citenamefont
  {Hiida}}]{Ohta:1973je}%
  \BibitemOpen
  \bibfield  {author} {\bibinfo {author} {\bibfnamefont {T.}~\bibnamefont
  {Ohta}}, \bibinfo {author} {\bibfnamefont {H.}~\bibnamefont {Okamura}},
  \bibinfo {author} {\bibfnamefont {T.}~\bibnamefont {Kimura}}, \ and\ \bibinfo
  {author} {\bibfnamefont {K.}~\bibnamefont {Hiida}},\ }\bibfield  {title}
  {\enquote {\bibinfo {title} {{Physically acceptable solution of einstein's
  equation for many-body system}},}\ }\href {\doibase 10.1143/PTP.50.492}
  {\bibfield  {journal} {\bibinfo  {journal} {Prog. Theor. Phys.}\ }\textbf
  {\bibinfo {volume} {50}},\ \bibinfo {pages} {492--514} (\bibinfo {year}
  {1973})}\BibitemShut {NoStop}%
\bibitem [{\citenamefont {Isaev}(2008)}]{Isaev:2007uy}%
  \BibitemOpen
  \bibfield  {author} {\bibinfo {author} {\bibfnamefont {A.P.}\ \bibnamefont
  {Isaev}},\ }\bibfield  {title} {\enquote {\bibinfo {title} {{Operator
  approach to analytical evaluation of Feynman diagrams}},}\ }\href {\doibase
  10.1134/S1063778808050219} {\bibfield  {journal} {\bibinfo  {journal} {Phys.
  Atom. Nucl.}\ }\textbf {\bibinfo {volume} {71}},\ \bibinfo {pages} {914--924}
  (\bibinfo {year} {2008})},\ \Eprint {http://arxiv.org/abs/0709.0419}
  {arXiv:0709.0419 [hep-th]} \BibitemShut {NoStop}%
\bibitem [{\citenamefont {Loebbert}\ \emph {et~al.}(2018)\citenamefont
  {Loebbert}, \citenamefont {Mojaza},\ and\ \citenamefont
  {Plefka}}]{Loebbert:2018xce}%
  \BibitemOpen
  \bibfield  {author} {\bibinfo {author} {\bibfnamefont {Florian}\ \bibnamefont
  {Loebbert}}, \bibinfo {author} {\bibfnamefont {Matin}\ \bibnamefont
  {Mojaza}}, \ and\ \bibinfo {author} {\bibfnamefont {Jan}\ \bibnamefont
  {Plefka}},\ }\bibfield  {title} {\enquote {\bibinfo {title} {{Hidden
  Conformal Symmetry in Tree-Level Graviton Scattering}},}\ }\href {\doibase
  10.1007/JHEP05(2018)208} {\bibfield  {journal} {\bibinfo  {journal} {JHEP}\
  }\textbf {\bibinfo {volume} {05}},\ \bibinfo {pages} {208} (\bibinfo {year}
  {2018})},\ \Eprint {http://arxiv.org/abs/1802.05999} {arXiv:1802.05999
  [hep-th]} \BibitemShut {NoStop}%
\end{thebibliography}%

\end{document}